\PassOptionsToPackage{table}{xcolor}
\documentclass[acmsmall]{acmart}

\usepackage{graphicx}
\usepackage[table]{xcolor}
\usepackage{cprotect}
\usepackage{verbatim}
\usepackage{listings}
\usepackage{rotating}
\usepackage{amsfonts}
\usepackage{todonotes}
\usepackage{amsmath}
\usepackage{subcaption}
\usepackage{algorithmicx,algorithm,algpseudocode}
\usepackage{multirow}
\usepackage{ifthen,color}
\usepackage[export]{adjustbox}
\usepackage{booktabs}
\usepackage{nicematrix} %
\usepackage{enumitem}
\usepackage{url}
\usepackage{arydshln}
\usepackage{lscape}

\newif\ifminted
\mintedtrue
\ifminted
\usepackage[frozencache,cachedir=.]{minted}
\usepackage{tcolorbox}

\newcommand{\bcolorbox}{\begin{tcolorbox}[colback=white!15,arc=0pt,outer arc=0pt,
	left=5pt,right=5pt,top=1.5pt,bottom=1.5pt, boxrule=0.5pt]}
\newcommand{\ecolorbox}{\end{tcolorbox}}

\setminted{fontsize=\footnotesize,baselinestretch=.97,linenos,xleftmargin=6pt,numbersep=3pt,mathescape=true,escapeinside=||,bgcolor=bg}
\definecolor{bg}{rgb}{1,1,1}
\newminted{cpp}{linenos}
\newenvironment{cplus}{\VerbatimEnvironment\begin{cppcode}}{\end{cppcode}}
\newmintinline[cplusinl]{cpp}{}
\newminted{python3}{}

\newmintinline[pytinl]{python3}{fontsize=\small,breaklines}
\else
\lstnewenvironment{cplus}[1][\footnotesize]{\lstset{language=C++custom,%
    style=numbers,basicstyle=\color{lightblue}#1\ttfamily,%
    keywordstyle=#1\ttfamily,%
    style=bold-keywords,style=frametb}}{}

\lstnewenvironment{cplusuf}[1][\small]{\lstset{language=C++custom,%
    style=numbers,basicstyle=\color{lightblue}#1\ttfamily,%
    keywordstyle=#1\ttfamily,%
    style=bold-keywords,frame=t%
  }}{}

\lstnewenvironment{mjava}[1][\small]{\lstset{language=Java,%
    escapeinside={/*@}{@*/},style=numbers,basicstyle=\color{lightblue}#1\ttfamily,%
    keywordstyle=#1\ttfamily,%
    style=bold-keywords,style=frametb}}{}

\lstnewenvironment{cplusnln}[1][\small]{\lstset{language=C++custom,%
    xleftmargin=8pt,xrightmargin=8pt,basicstyle=\color{lightblue}#1\ttfamily,%
    keywordstyle=#1\ttfamily,%
    style=bold-keywords,style=frametb}}{}

\makeatletter
\lstnewenvironment{code}[1][\footnotesize]{\lstset{language=C++custom,columns=fullflexible,%
    xleftmargin=8pt,xrightmargin=8pt,basicstyle=\color{lightblue}#1,%
    keywordstyle=#1,style=numbers,%
    style=bold-keywords,mathescape=true}}{\@endparenv}
\makeatother

\lstnewenvironment{python}[1][\footnotesize]{\lstset{language=Pythoncustom,%
    style=numbers,basicstyle=\color{lightblue}#1\ttfamily,%
    keywordstyle=#1\ttfamily,%
    style=bold-keywords,style=frametb}}{}

\lstnewenvironment{pythonuf}[1][\small]{\lstset{language=Pythoncustom,%
    style=numbers,basicstyle=\color{lightblue}#1\ttfamily,%
    keywordstyle=#1\ttfamily,%
    style=bold-keywords,frame=t%
  }}{}

\lstnewenvironment{pythonln}[1][\small]{\lstset{language=Pythoncustom,%
    xleftmargin=8pt,xrightmargin=8pt,basicstyle=\color{lightblue}#1\ttfamily,%
    keywordstyle=#1\ttfamily,%
    style=bold-keywords,style=frametb}}{}

\makeatletter
\lstnewenvironment{pcode}[1][\footnotesize]{\lstset{language=Pythoncustom,columns=fullflexible,%
    xleftmargin=8pt,xrightmargin=8pt,basicstyle=\color{lightblue}#1,%
    keywordstyle=#1,style=numbers,%
    style=bold-keywords,mathescape=true}}{\@endparenv}
\makeatother\fi

\providecommand{\codeinl}[2][\small] %
{{\lstinline[language=C++custom,breaklines=false,columns=fullflexible,%
basicstyle=\color{lightblue}#1\sffamily%
,mathescape=true,keywordstyle=#1\sffamily]@#2@}}%

\providecommand{\cplusinl}[2][\small] %
{{\lstinline[language=C++custom,breaklines=false,columns=fullflexible,%
basicstyle=\color{lightblue}#1\ttfamily%
,keywordstyle=#1\ttfamily]@#2@}}%

\providecommand{\pytinl}[2][\small] %
{{\lstinline[language=Pythoncustom,breaklines=false,columns=fullflexible,%
basicstyle=\color{lightblue}#1\ttfamily%
,keywordstyle=#1\ttfamily]@#2@}}%

\lstdefinestyle{cppmarkers}{rangeprefix=/*\#\ ,%
includerangemarker=false,%
rangesuffix=\ \#*/}%

\lstdefinelanguage{Haskell-custom}
{%
escapeinside={--@}{@--},breaklines=true,breakatwhitespace=true%
language=Haskell,basicstyle=\color{lightblue}\ttfamily,keywordstyle=\ttfamily,%
morekeywords={class,instance,type,newtype,data,where,deriving,import},%
lineskip=-.1\baselineskip,morekeywords={concept,requires,concept_map},
literate={+}{{$+$}}1 {/}{{$/$}}1 {*}{{$*$}}1 {=}{{$=$}}1
               {>}{{$>$}}1 {<}{{$<$}}1 {\\}{{$\lambda$}}1
               {\\\\}{{\char`\\\char`\\}}1
               {->}{{$\rightarrow$}}2 {>=}{{$\geq$}}2 {<-}{{$\leftarrow$}}2
               {=>}{{$\Rightarrow$}}2
               {\ .}{{$\circ$}}2 {\ .\ }{{$\circ$}}2
               {>>}{{>>}}2 {>>=}{{>>=}}2
               {|}{{$\mid$}}1
             }

\lstnewenvironment{hask}[1][\small]{\lstset{language=Haskell-custom,%
    style=numbers,basicstyle=\color{lightblue}#1\ttfamily,keywordstyle=#1\ttfamily,%
    style=bold-keywords,style=frametb}}{}

\lstdefinestyle{markers}{rangeprefix=\{-\:\ ,%
includerangemarker=false,%
rangesuffix=\ \:-\}}%

\lstdefinestyle{C++}
{
  language=C++1y,
  columns=fullflexible,
  breaklines=true,
}

\lstdefinelanguage{CASL}{
  escapechar=@,
  breakatwhitespace=true,
  morekeywords = {
    spec, sort, then, op, ops, var, vars, pred, end
  }
}

\lstnewenvironment{casl}[1][\small]
{
  \lstset{
    language=CASL,
    basicstyle=#1\sffamily,
    keywordstyle=#1\sffamily\bfseries,
    commentstyle=#1\sffamily,
    columns=fullflexible,
    breaklines=true,
  }
}
{}

\usepackage{ifthen,color}
\definecolor{darkred}{rgb} {0.5,0.0,0.0}
\definecolor{darkgreen}{rgb} {0.0,0.5,0.0}
\definecolor{darkblue}{rgb} {0.0,0.0,0.5}
\definecolor{lightgray}{gray}{0.94}

\newcommand{\John}   [1]{{{\color{darkgreen}(John) #1}}}
\newcommand{\Aydin}   [1]{{{\color{violet}(Aydin) #1}}}
\newcommand{\Carl}   [1]{{{\color{darkblue}(Carl) #1}}}
\newcommand{\final}{0}
\newcommand{\revision}   [1]{{{\color{black}#1}}}
\newcommand{\cmd}[1]{\texttt{#1}}

\ifthenelse{\equal{\final}{1}}{\renewcommand{\John}[1]{}}{}
\ifthenelse{\equal{\final}{1}}{\renewcommand{\Aydin}[1]{}}{}
\ifthenelse{\equal{\final}{1}}{\renewcommand{\Carl}[1]{}}{}

\newcommand{\mM}{\mathbf{M}}
\newcommand{\mA}{\mathbf{A}}
\newcommand{\mB}{\mathbf{B}}
\newcommand{\mC}{\mathbf{C}}
\newcommand{\mU}{\mathbf{u}}
\newcommand{\mV}{\mathbf{v}}
\newcommand{\mX}{\mathbf{x}}
\newcommand{\mW}{\mathbf{w}}
\newcommand{\mR}{\mathbb{R}}

\newcommand{\mD}{\mathbb{D}}
\newcommand{\mI}{\mathbb{I}}

\newcommand{\dnnz}{\textit{nnz}}
\newcommand{\flops}{\mathrm{flops}}
\newcommand{\transpose}     {^{\mbox{\scriptsize \sf T}}}
\def\checkmark{\tikz\fill[scale=0.4](0,.35) -- (.25,0) -- (1,.7) -- (.25,.15) -- cycle;}

\makeatletter
\ifcase \@ptsize \relax%
  \newcommand{\miniscule}{\@setfontsize\miniscule{4}{5}}%
\or%
  \newcommand{\miniscule}{\@setfontsize\miniscule{5}{6}}%
\or%
  \newcommand{\miniscule}{\@setfontsize\miniscule{5}{6}}%
\fi
\makeatother

\setcopyright{rightsretained}
\acmJournal{TOMS}
\acmYear{2021} \acmVolume{1} \acmNumber{1} \acmArticle{1} \acmMonth{1} \acmPrice{}\acmDOI{10.1145/3466795}

\begin{document}

\title{GraphBLAST\@: A High-Performance Linear Algebra-based Graph Framework on the GPU}

        \author{Carl Yang}
        \orcid{0000-0002-4357-0906}
        \affiliation{
                \institution{University of California, Davis and Lawrence Berkeley National Laboratory}
                \streetaddress{Dept.\ of Electrical and Computer Engineering, 1 Shields Avenue}
                \city{Davis}
                \state{California}
                \postcode{95616}}
        \email{ctcyang@ucdavis.edu}

        \author{Ayd\i{}n Bulu{\c{c}}}
        \orcid{0000-0001-7253-9038}
        \affiliation{
                \institution{Lawrence Berkeley National Laboratory and University of California, Berkeley}
                \streetaddress{1 Cyclotron Road}
                \city{Berkeley}
                \state{California}
                \postcode{94720}}
        \email{abuluc@lbl.gov}

        \author{John D. Owens}
        \orcid{0000-0001-6582-8237}
        \affiliation{
                \institution{University of California, Davis}
                \streetaddress{Dept.\ of Electrical and Computer Engineering, 1 Shields Avenue}
                \city{Davis}
                \state{California}
                \postcode{95616}}
        \email{jowens@ece.ucdavis.edu}

        \renewcommand{\shortauthors}{C. Yang, A. Bulu{\c{c}}, J.D. Owens}%

\maketitle

        \label{sec:abstract}
        High-performance implementations of graph algorithms are challenging to implement on new parallel hardware such as GPUs because of three challenges: (1)~the difficulty of coming up with graph building blocks, (2)~load imbalance on parallel hardware, and (3)~graph problems having low arithmetic intensity. To address some of these challenges, GraphBLAS is an
innovative, on-going effort by the graph analytics community to propose building blocks based on sparse linear algebra, which allow graph algorithms to be expressed in a performant, succinct, composable, and portable manner. In this paper, we examine the performance challenges of a linear-algebra-based approach to building graph frameworks and describe new design principles for overcoming these bottlenecks. Among the new design principles is \emph{exploiting input sparsity}, which allows users to write graph algorithms without specifying push and pull direction. \emph{Exploiting output sparsity} allows users to tell the backend which values of the output in a single vectorized computation they do not want computed. \emph{Load-balancing} is an important feature for balancing work amongst parallel workers. We describe the important load-balancing features for handling graphs with different characteristics. The design principles described in this paper have been implemented in ``GraphBLAST'', \revision{the first high-performance linear algebra-based graph framework on NVIDIA GPUs that is open-source}. The results show that on a single GPU, GraphBLAST has on average at least an order of magnitude speedup over previous GraphBLAS implementations SuiteSparse and GBTL, comparable performance to the fastest GPU hardwired primitives and shared-memory graph frameworks Ligra and Gunrock, and better performance than any other GPU graph framework, while offering a simpler and more concise programming model.

\section{Introduction}
\label{sec:intro}
Graphs are a representation that naturally emerges when solving problems in domains including  bioinformatics~\cite{Georganas:2018:ESD}, social network analysis~\cite{Ching:2015:OTE}, molecular synthesis~\cite{Jin:2018:JTV}, and route planning~\cite{Delling:2009:ERP}. Graphs may contain billions of vertices and edges, so parallelization has become a must.

The past two decades have seen the rise of parallel processors into a commodity product---both general-purpose processors in the form of graphic processor units (GPUs), as well as domain-specific processors such as tensor processor units (TPUs) and the graph processors being developed under the DARPA HIVE program. \revision{Research into developing parallel hardware and initiatives such as the DIMACS and HPEC graph challenges~\cite{Johnson:1993:NFA,Samsi:2018:GRT} have succeeded in speeding up graph algorithms~\cite{Shun:2013:LLG,Wang:2017:GGG}.} However, the improvement in graph performance has come at the cost of a more challenging programming model. The result has been a mismatch between the high-level languages that users and graph algorithm designers would prefer to program in (e.g., Python) and  programming languages for parallel hardware (e.g., C++, CUDA, OpenMP, or MPI).

\begin{figure*}[t]
  \centering
  \includegraphics[width=0.8\textwidth]{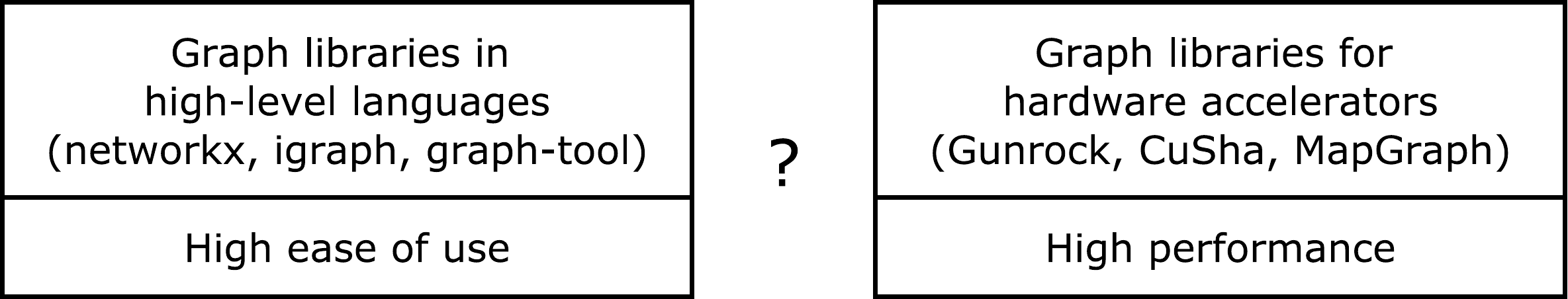}
  \caption{Mismatch between existing frameworks targeting high-level languages and hardware accelerators.\label{fig:gap}}
\end{figure*}

To address this mismatch, many initiatives, including NVIDIA's RAPIDS effort~\cite{RAPIDS:2018}, have been launched in order to provide an open-source Python-based ecosystem for data science and graphs on GPUs. One such initiative, GraphBLAS, is an open standard~\cite{Buluc:2017:TGC} for graph frameworks. It promises standard building blocks for expressing graph algorithms in the language of linear algebra. Such a standard attempts to solve the following problems:

\begin{enumerate}
  \item \emph{Performance portability}: Graph algorithms need no modification to have high performance across hardware.
  \item \emph{Concise expression}: Graph algorithms are expressed in few lines of code.
  \item \emph{High-performance}: Graph algorithms achieve state-of-the-art performance.
  \item \emph{Scalability}: An implementation is effective at both small-scale and exascale.
\end{enumerate}

Goal 1 (\emph{performance portability}) is central to the GraphBLAS philosophy, and it has made inroads with several implementations already being developed using this common interface~\cite{Davis:2018:SGG,Zhang:2016:GCG,Moreira:2018:IBM}. Regarding Goal 2 (\emph{concise expression}), GraphBLAS encourages users to think in a vectorized manner, which yields an order-of-magnitude reduction in lines of code as shown in Table~\ref{tab:loc}. Before Goal 4 (\emph{scalability}) can be achieved, Goal 3 (\emph{high-performance}) on the small scale must first be demonstrated.

\begin{table}
  \centering
  \rowcolors{2}{lightgray}{white}
  \begin{tabular}{lcccccccc}
    \toprule
    & This & \multicolumn{7}{c}{Framework}     \\
    Algorithm     & Work &  CS     & GL        & GR  & LI & MG & GB      & SS \\ \midrule
    Breadth-first-search & 22 & 76 & 353 & 1161 & 45 & 140 & 22 & 29 \\
    Single-source shortest-path & 24 & 78 & 440 & 465 & 60 & 184 & 25 & N/A \\
    PageRank & 32 & 84 & 342 & 805 & 68 & 144 & 47 & 31 \\
    Connected components & 50 & N/A & 595 & 1435 & 61 & 153 & N/A & 132 \\
    Triangle counting & 8 & N/A & 283 & 297 & 60 & N/A & 17 & 15 \\ \bottomrule
  \end{tabular}
  \caption{Comparison of lines of C or C++ application code for seven graph frameworks and this work. The graph frameworks we compared with are CuSha (CS)~\cite{Khorasani:2014:CVG}, Galois (GL)~\cite{Nguyen:2013:ALI}, Gunrock (GR)~\cite{Wang:2017:GGG}, Ligra (LI)~\cite{Shun:2013:LLG}, Mapgraph (MG)~\cite{Fu:2014:MAH}, GBTL (GB)~\cite{Zhang:2016:GCG}, and SuiteSparse (SS)~\cite{Davis:2018:SGG}.\label{tab:loc}}
\end{table}

However to date, GraphBLAS has lacked high-performance implementations for GPUs. The GraphBLAS Template Library (GBTL)~\cite{Zhang:2016:GCG} is a GraphBLAS-inspired GPU graph framework. The architecture of GBTL is C++-based and maintains a separation of concerns between a top-level interface defined by the GraphBLAS C API specification and the low-level backend. However, since it was intended as a proof-of-concept in programming language research, it is an order of magnitude slower than state-of-the-art graph frameworks on the GPU in terms of performance.

We identify several reasons graph frameworks are \emph{challenging} to implement on the GPU\@:

\begin{description}
  \item [Generalizability of optimization] While many graph algorithms share similarities, the optimizations found in high-performance graph frameworks often seem ad hoc and difficult to reconcile with the goal of a clean and simple interface. What are the \emph{optimizations} most deserving of attention when designing a high-performance graph framework on the GPU\@?
  \item[Load imbalance] Graph problems have irregular memory access patterns making it hard to extract parallelism. On parallel systems such as GPUs, this is further complicated by the challenge of balancing work amongst parallel compute units. How should this problem of \emph{load-balancing} be addressed?
  \item[Low compute-to-memory access ratio] Graph problems typically require multiple memory accesses on unstructured data rather than many floating-point computations. Therefore, graph problems are often memory-bound rather than compute-bound. What can be done to reduce the \emph{number of memory accesses}?
\end{description}

What are the design principles required to build a GPU implementation based on linear algebra that matches the state-of-the-art graph frameworks in performance? Towards that end, we have designed GraphBLAST\footnote{\url{https://github.com/gunrock/graphblast}}: the first high-performance implementation of GraphBLAS for the GPU\@. Our implementation is for a single GPU, but given the similarity between the GraphBLAS interface we are adhering to and the CombBLAS interface~\cite{Buluc:2011:TCB}, which is a graph framework for distributed CPUs, we are confident the design we propose here will allow us to extend it to a distributed implementation with future work.

In order to perform a comprehensive evaluation of our system, we compare our framework against state-of-the-art graph frameworks on the CPU and GPU, as well as hardwired GPU implementations, which are problem-specific GPU implementations that developers have hand-tuned for performance. The state-of-the-art graph frameworks against which we will be comparing are Ligra~\cite{Shun:2013:LLG}, Gunrock~\cite{Wang:2017:GGG}, CuSha~\cite{Khorasani:2014:CVG}, Galois~\cite{Nguyen:2013:ALI}, Mapgraph~\cite{Fu:2014:MAH}, GBTL~\cite{Zhang:2016:GCG}, and SuiteSparse~\cite{Davis:2018:SGG}, which we will describe in greater detail in Section~\ref{sec:previous}. The hardwired implementations will be Enterprise (BFS)~\protect\cite{Liu:2015:EBG}, delta-stepping SSSP~\protect\cite{Davidson:2014:WPG}, pull-based PR~\protect\cite{Khorasani:2014:CVG}, hooking and pointer-jumping CC~\protect\cite{Soman:2010:AFG}, and bitmap-based triangle counting~\protect\cite{Bisson:2017:HPE}. The \revision{five} graph algorithms on which we will be evaluating our system are:

\begin{itemize}
  \item Breadth-first-search (BFS)
  \item Single-source shortest-path (SSSP)
  \item PageRank (PR)
  \revision{\item Connected components (CC)}
  \item Triangle counting (TC)
\end{itemize}

GraphBLAST has also been used for graph coloring~\cite{Osama:2019:GCO} as well as DARPA HIVE graph algorithms on the GPU~\cite{Owens:2018:HYR}, including graph projections, local graph clustering, and seeded graph matching.

Our contributions in this paper are as follows:
\begin{enumerate}
  \item We briefly categorize parallel graph frameworks (Section~\ref{sec:related}) and give a short introduction to GraphBLAS's computation model (Section~\ref{sec:graphblast}).
  \item We demonstrate the importance of exploiting \emph{input sparsity}, which means picking the algorithm based on a cost model that selects between an algorithm that exploits the input vector's sparsity and another algorithm that is more efficient for denser input vectors. One of the consequences is direction optimization (Section~\ref{sec:diropt}).
  \item We show the importance of exploiting \emph{output sparsity}, which is implemented as masking and can be used to reduce the number of memory accesses of several graph algorithms (Section~\ref{sec:masking}).
  \item \revision{We explain the design considerations required for high-performance on the GPU, which are avoiding CPU-to-GPU memory copies, supporting generalized semiring operators, and load-balancing (Section~\ref{sec:loadbalancing})}.
  \item We review how common graph algorithms are expressed in GraphBLAST (Section~\ref{sec:app}).
  \item \revision{We show that, enabled by the optimizations \emph{exploiting sparsity}, \emph{masking}, and \emph{proper load-balancing}, GraphBLAST gets $43.51\times$ geomean (i.e., geometric mean) and $1268\times$ peak over SuiteSparse GraphBLAS for multi-threaded CPUs. Compared to state-of-the-art graph frameworks on the CPU and GPU on five graph algorithms running on scale-free graphs, GraphBLAST gets $2.31\times$ geomean ($10.97\times$ peak) and $1.14\times$ ($5.24\times$ peak) speed-up (Section~\ref{sec:results})}.
\end{enumerate}

\begin{table}
  \centering
  \begin{NiceTabular}{llccccl}%
  [code-before = \rowcolor{white}{1-2}\rowcolors{3}{lightgray}{}\columncolor{white}{1}]
    \toprule
    & & \multicolumn{5}{c}{Application}     \\
    Major Feature & Component     & BFS     & SSSP        & PR      &     CC & TC  \\ \midrule
    \multirow{3}{*}{Exploit input sparsity}
    & Generalized direction optimization      & \checkmark & \checkmark  & \checkmark & \checkmark &    \\
    & Boolean semiring & \checkmark &  & & &   \\
    & Avoid sparse-to-dense conversion      & \checkmark  & \checkmark & &  &  \\ \midrule
    \multirow{1}{*}{Exploit output sparsity}
    & Masking  & \checkmark & \checkmark & & \checkmark & \checkmark   \\ \midrule
    \multirow{2}{*}{Load-balancing}
    & Row split & \checkmark & \checkmark  & \checkmark& \checkmark & \checkmark    \\
    & Merge-based  & \checkmark & \checkmark  & \checkmark & \checkmark &   \\ \bottomrule
  \end{NiceTabular}
  \caption{Applicability of design principles.\label{tab:applicability}}
\end{table}

Over the next three sections, we will discuss the most important design principles for making this code performant, which are exploiting input sparsity and output sparsity, and making good decisions for considerations specific to the GPU\@. Table~\ref{tab:applicability} shows  which of the five graph algorithms discussed in this paper our optimizations apply to.

\section{Background \& Motivation}
\label{sec:related}
We begin by describing related literature in the field of graph frameworks on parallel hardware (Section~\ref{sec:relatedwork}), and move to discussing the limitations of previous systems that inspired ours (Section~\ref{sec:previous}). Further, we review the connection between graph algorithms and linear algebra (Section~\ref{sec:graphmatrix}). \revision{For a broader survey of parallel graph frameworks, refer to Doekemeijer and Varbanescu's 2014 work~\cite{Doekemeijer:2014:ASO}.}

\subsection{Related work}
\label{sec:relatedwork}

\revision{Large-scale graph frameworks on multi-threaded CPUs, distributed-memory CPU systems (surveyed by Batarfi et al.~\cite{Batarfi:2015:LSG}), and massively parallel GPUs (surveyed by Shi et al.~\cite{Shi:2018:GPO}) fall into three broad categories: vertex-centric (surveyed by McCune, Weninger and Madey~\cite{McCune:2015:TLA}), edge-centric, and linear-algebra-based. In this section, we will explain this categorization and the influential graph frameworks from each category.}

\subsubsection{Vertex-centric}
Introduced by Pregel~\cite{Malewicz:2010:PAS}, vertex-centric frameworks are based on parallelizing over vertices. The computation in Pregel is inspired by the distributed CPU programming model of MapReduce~\cite{Dean:2008:MSD} and is based on message passing. At the beginning of the algorithm, all vertices are active. Pregel follows the bulk synchronous programming model (BSP) consisting of global synchronization barriers called \emph{supersteps}. At the end of a superstep, the runtime receives the messages from each sending vertex and computes the set of active vertices for the superstep. Computation continues until convergence or a user-defined condition is reached.

Pregel's programming model is good for scalability and fault tolerance. However, standard graph algorithms in most Pregel-like graph processing systems suffer from slow convergence on large-diameter graphs and load imbalance on scale-free graphs. Apache Giraph~\cite{Ching:2015:OTE} is an open-source implementation of Google's Pregel. It is a popular graph computation engine in the Hadoop ecosystem initially open-sourced by Yahoo!. Han et al.~\cite{Han:2014:AEC} provide a full survey of Pregel-like frameworks. %

Galois~\cite{Nguyen:2013:ALI} is a graph system for shared memory based on a different operator abstraction that supports priority scheduling and dynamic graphs and processes on subsets of vertices called active elements. However, their model does not abstract implementation details of the loop from the user. Users have to generate the active elements set directly for different graph algorithms.

\revision{First introduced by PowerGraph~\cite{Gonzalez:2012:PDG}, the Gather-Apply-Scatter (GAS) model is a concrete implementation of the vertex-centric model designed to address the slow convergence of vertex-centric models on power law graphs. For the load imbalance problem, it uses vertex-cut to split high-degree vertices into equal degree-sized redundant vertices. This exposes greater parallelism in real-world graphs. It supports both BSP and asynchronous execution. Like Pregel, PowerGraph is a distributed CPU framework. For flexibility, PowerGraph also offers a vertex-centric programming model, which is efficient on non-power law graphs.

MapGraph~\cite{Fu:2014:MAH} is a similar GAS framework and integrates both Baxter's load-balanced search~\cite{Baxter:2015:MGL} and Merrill, Garland, and Grimshaw's dynamic grouping workload mapping strategy~\cite{Merrill:2012:SGG} to increase its performance. CuSha~\cite{Khorasani:2014:CVG} is also a GAS model-based GPU graph analytics system. It solves the load imbalance and GPU underutilization problem with a GPU adoption of the parallel sliding window technique. They call this preprocessing step ``G-Shard'' and combine it with a concatenated window method to group edges from the same source indices.

Noteworthy systems for processing dynamic graphs are STINGER~\cite{ediger2012stinger}, Hornet~\cite{Busato:2018:HAE}, Kineograph~\cite{cheng2012kineograph}, Aspen~\cite{dhulipala2019low}, and Terrace~\cite{terracesigmod21}. The systems all avoid using the popular CSR data structure for storing the graph. For example, Hornet stores the adjacency list in arrays of memory blocks, using a vectorized bit tree to find the next available memory block, and leveraging B+ trees for managing memory blocks~\cite{Beyer:1972:OAM}.

HavoqGT is a distributed graph framework built for high performance~\cite{Pearce:2014:FPT,Pearce:2019:OQT}. Its novelty is a new algorithmic technique called \emph{vertex delegates} in its programming model, which both load-balances and performs asynchronous broadcast and reduction operations for the high-degree vertices. This has the impact of performing much better than a simple 1D partitioning strategy on distributed systems.}

\subsubsection{Edge-centric}
\revision{First introduced by X-Stream~\cite{Roy:2013:XEG}, the edge-centric model treats edges rather than vertices as the first-class graph entities. There, authors build an out-of-core engine that relies on streaming unordered edge lists. Even though updates to vertices must still be random access, the updates to edges can have sequential access. Roy et al.~\cite{Roy:2013:XEG} contend that this takes advantage of storage media (main memory, solid-state disk, and magnetic disk) having superior sequential access performance vs.\ random memory access.}

\subsubsection{Linear algebra-based}
Linear algebra-based graph frameworks were pioneered by the Combinatorial BLAS (CombBLAS)~\cite{Buluc:2011:TCB}, a distributed memory CPU-based graph framework. Algebra-based graph frameworks rely on the fact that graph traversal can be described as a matrix-vector product. CombBLAS offers a small but powerful set of linear algebra primitives. Combined with algebraic semirings, this small set of primitives can describe a broad set of graph algorithms.  The advantage of CombBLAS is that it is the only framework that can express a 2D partitioning of the adjacency matrix, which is helpful in scaling to large-scale graphs.

In the context of bridging the gap between vertex-centric and linear algebra-based frameworks, GraphMat~\cite{Sundaram:2015:GHP} is a groundbreaking work. Traditionally, linear algebra-based frameworks have found difficulty gaining adoption, because they rely on users' understanding how to express graph algorithms in terms of linear algebra. GraphMat addresses this problem by exposing a vertex-centric interface to the user, automatically converting such a program to a generalized sparse matrix-vector multiply, and then performing the computation on a linear-algebra-based backend.

nvGRAPH~\cite{Eaton:2016:NVG} is a high-performance GPU graph analytics library developed by NVIDIA\@. It views graph analytics problems from the perspective of linear algebra and matrix computations~\cite{Kepner:2016:MFG}, and uses semiring matrix-vector multiply operations to present graph algorithms. As of version 10.1, it supports five algorithms: PageRank, single-source shortest-path (SSSP), triangle counting,  single-source widest-path, and spectral clustering. SuiteSparse~\cite{Davis:2018:SGG} is notable for being the first GraphBLAS-compliant library. We compare against the multithreaded CPU implementation of SuiteSparse. GBTL~\cite{Zhang:2016:GCG} is a GraphBLAS-like framework on the GPU\@. Rather than high performance, its implementation focused on programming language research and a separation of concerns between the interface and the backend.

\subsubsection{Implementation challenges on GPUs}

\revision{Whether vertex-centric, edge-centric, or linear-algebra-based, GPU implementations of graph frameworks face several common challenges in achieving high performance.

\paragraph{Fine-grained load imbalance} The most straightforward form of parallelism in graph problems is parallelizing across vertices. However, in many graphs, particularly scale-free graphs, the number of outbound edges at each vertex may vary dramatically. Consequently, the amount of work per vertex varies in the same way. Thus, a GPU implementation that assigns vertices to neighboring threads results in significant fine-grained load imbalance across those neighboring threads. This imbalance is identical to the imbalance in sparse matrix operations with a variable amount of non-zero elements per row that choose to assign a thread per matrix row~\cite{Bell:2009:ISM}.

We describe this problem from the linear algebra perspective in Section~\ref{sec:loadbalancing}. Native graph frameworks typically address this problem through a variety of techniques, including dynamically binning vertices by the size of their workload and processing like-sized bins with an appropriately sized grain of computation~\cite{Merrill:2012:SGG} or converting parallelism over vertices to parallelism over edges using prefix-sum-like methods~\cite{Davidson:2014:WPG}. The challenges are choosing the right load-balance method and balancing the cost of load balance vs.\ its performance benefits.

\paragraph{Minimizing overhead} \label{sec:kernel-launch} \label{sec:kernel-fusion} GPU kernels that run on large, load-balanced datasets with a large amount of work per input element achieve their peak throughput. However, in the course of a graph computation, a GPU framework may often face situations where its runtime is not dominated by processing time but instead by overheads. One form of overhead occurs when the GPU does not have enough work to keep the entire GPU busy; in such a case, that kernel's runtime is dominated by the cost of the kernel launch rather than the cost of the work. A related form of overhead for bulk-synchronous operations is the requirement for global synchronization at the end of each kernel; to first order, the entire GPU must wait until the last element processed by a kernel is complete before starting the next kernel.

Another form of overhead is when multiple kernels are used to perform a computation that could be combined into a single kernel (``kernel fusion''), one that  potentially exploits producer-consumer locality within a kernel. The performance gap between ``hardwired'' graph algorithm implementations that are tailored to a single algorithm and more general programmable graph frameworks is often a result of this additional overhead for the programmable framework~\cite[Section~3.1]{Wang:2017:GGG} because frameworks are generally built from smaller, modular operations and cannot automatically perform kernel fusion and exploit producer-consumer locality as hardwired implementations do.

The overhead of a kernel launch is on the order of several microseconds~\cite{Zhang:2019:UTO}. So, for graph computations that require many iterations and have many kernel launches per iteration, the aggregate cost of kernel launches is significant. Thus, minimizing kernel launches is an important goal of any high-performance graph framework.}

\subsection{Previous systems}
\label{sec:previous}

Two systems that directly inspired our work are Gunrock and Ligra.

\subsubsection{Gunrock}
Gunrock~\cite{Wang:2017:GGG} is a state-of-the-art GPU-based graph processing framework. It is notable for being the only high-level GPU-based graph analytics system with support for both vertex-centric and edge-centric operations, as well as fine-grained runtime load balancing strategies, without requiring any preprocessing of input datasets. Indeed,  Table~\ref{tab:comparison} shows Gunrock has the most performance optimizations out of all graph frameworks, but this comes at a cost of increasing the complexity and amount of user application code. In our work, we want the performance Gunrock optimizations provide while moving more work to the backend. In other words, we want to adhere to GraphBLAS's compact and easy-to-use user interface, while maintaining state-of-the-art performance.

\begin{table}
  \centering
  \rowcolors{2}{lightgray}{white}
  \begin{tabular}{lcccccccc}
    \toprule
    & This & \multicolumn{7}{c}{Framework}     \\
    Component     & Work &  CS     & GL        & GR  & LI & MG & GB      & SS \\ \midrule
    Programming model & LA & GA & GA & GA & GA & GA & LA & LA \\
    Backend & GPU & GPU & CPU & GPU & CPU & GPU & GPU & CPU \\
    Preprocessing & no & yes & no & no & no & no & no & no \\
    BFS lines of code & 22 & 76 & 353 & 1161 & 45 & 140 & 22 & 29 \\ \midrule
    Direction optimization      & \checkmark &  & \checkmark & \checkmark & \checkmark & & & \checkmark \\
    Generalized direction optimization & \checkmark & & & & \checkmark & & & \\
    Early-exit optimization~\cite{Yang:2018:IPE} & \checkmark & & & \checkmark & & & & \checkmark \\
    Structure-only optimization~\cite{Yang:2018:IPE} & \checkmark & \checkmark & \checkmark & \checkmark & \checkmark & \checkmark & \checkmark & \checkmark \\
    Avoid sparse-to-dense conversion~\cite{Yang:2018:IPE}      & \checkmark  &  & & \checkmark & & &  & \\
    Masking (kernel fusion) & \checkmark &  & &   \checkmark & & & & \checkmark \\
    Static mapping (vertex-centric) & \checkmark & \checkmark & \checkmark & \checkmark & \checkmark & \checkmark & \checkmark & \checkmark \\
    Dynamic mapping (edge-centric) & \checkmark &  & & \checkmark & \checkmark & & & \checkmark \\ \bottomrule
  \end{tabular}
  \caption{Detailed comparison of different parallel graph frameworks on the CPU and GPU\@. LA indicates a linear algebra-based model and GA indicates a native graph abstraction composed of vertices and edges. The five graph abstraction-based frameworks we compared with are CuSha (CS)~\cite{Khorasani:2014:CVG}, Galois (GL)~\cite{Nguyen:2013:ALI}, Gunrock (GR)~\cite{Wang:2017:GGG}, Ligra (LI)~\cite{Shun:2013:LLG}, and MapGraph (MG)~\cite{Fu:2014:MAH}. The two linear-algebra-based frameworks we compared with are GBTL (GB)~\cite{Zhang:2016:GCG} and SuiteSparse (SS)~\cite{Davis:2018:SGG}. Note that part of load balancing work in CuSha is done during the (offline) G-shard generation process. The difference between direction optimization and generalized direction optimization is that the former indicates the framework supports this optimization, while the latter indicates the selection of push and pull is automated and generalized to graph algorithms besides BFS\@. Early-exit, structure-only and avoid sparse-to-dense conversion optimizations are discussed in previous work~\cite{Yang:2018:IPE}.\label{tab:comparison}}
\end{table}

\subsubsection{Ligra}
Ligra~\cite{Shun:2013:LLG} is a CPU-based graph processing framework for shared memory. Its lightweight implementation is targeted at shared memory architectures and uses CilkPlus for its multi-threading implementation. It is notable for being the first graph processing framework to generalize Beamer, Asanovi\'{c} and Patterson's direction-optimized BFS~\cite{Beamer:2012:DOB} to many graph-traversal-based algorithms. However, Ligra does not support multi-source graph traversals. In our framework, multi-source graph traversals find natural expression as sparse BLAS 3 operations (matrix-matrix multiplications).

\subsection{Graph traversal vs.\ matrix-vector multiply}
\label{sec:graphmatrix}

\begin{figure}[t]
        \centering
          \centering
          \begin{subfigure}[t]{0.2\textwidth}
                \centering
                \includegraphics[width=1\linewidth]{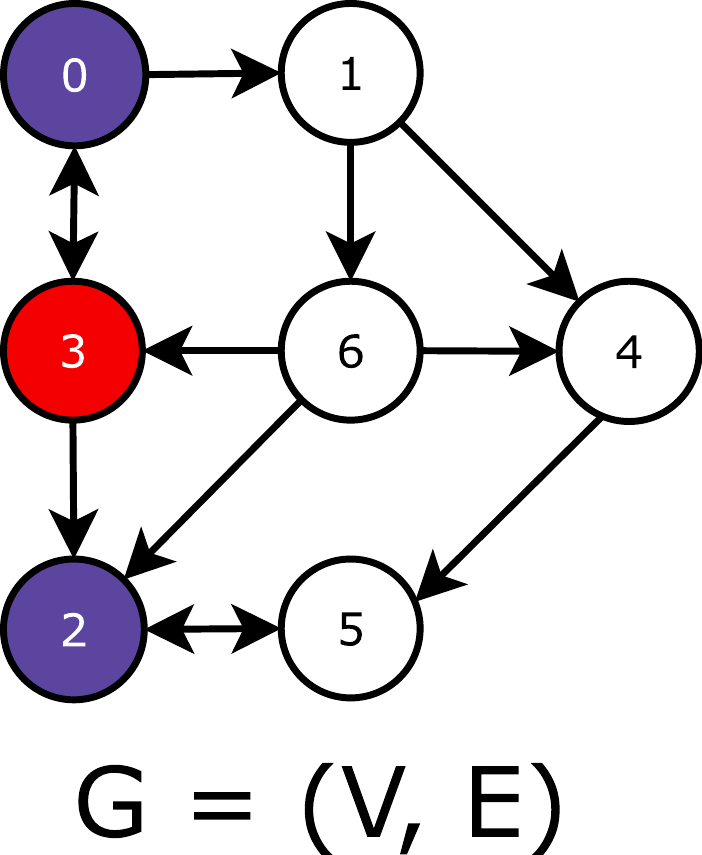}
                \caption{graph traversal\label{fig:duality-graph}}
          \end{subfigure}
          \hspace{5em}
          \begin{subfigure}[t]{0.34\textwidth}
                \centering
                \includegraphics[width=1\textwidth]{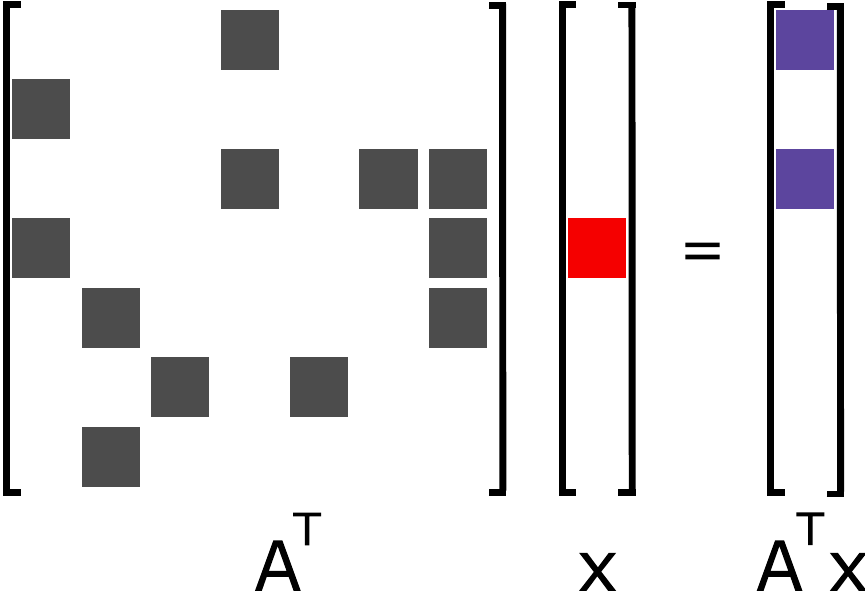}
                \caption{matrix-vector multiply\label{fig:duality-matrix}}
          \end{subfigure}
          \caption{The adjacency matrix $\mathbf{A}$ is one representation of graph $G=(V,E)$ with its set of vertices $V$ and set of edges $E$. The matrix-vector multiply $\mathbf{A}\transpose \mathbf{x}$  is one representation of the BFS graph traversal where the sparse vector $\mathbf{x}$ represents the current active frontier of vertices.}
          \label{fig:bfsmatrix}
\end{figure}

The connection between graph traversal and linear algebra was noted by Denes K\"{o}nig~\cite{Konig:1931:KGM} in the early days of graph theory. Since then the connection between graphs and matrices has been established by the popular representation of a graph as an adjacency matrix. More specifically, it has become popular to represent a vector-matrix multiply as being equivalent to one iteration of breadth-first-search traversal (see Figure~\ref{fig:bfsmatrix}). \revision{Some seemingly non-traversal graph algorithms such as triangle counting, which can be solved efficiently using a masked SpGEMM, can also be thought in terms of traversals. Multiplying the lower triangle of the adjacency matrix with its transpose, as we do in Section~\ref{sec:tricount}, simply does a 2-hop traversal of the graph from every vertex. The use of lower triangle and its transpose as opposed to the whole adjacency matrix ensures that identical paths are not explored redundantly. The masking checks whether such 2-hop paths, also called \emph{wedges} or \emph{triads} in the literature, close to form triangles.}

\section{GraphBLAS Concepts}
\label{sec:graphblast}
The following section introduces GraphBLAS's model of computation. A full treatment of GraphBLAS is beyond the scope of this paper; we give a brief introduction to the reader, so that he or she can better follow our contributions in later sections. We refer the interested reader to the GraphBLAS C API specification~\cite{Buluc:2017:TGC} and selected papers~\cite{Kepner:2016:MFG,Buluc:2017:DGA,Mattson:2017:GCA} for a full treatment. At the end of this section, we give a running example (Section~\ref{sec:runningexample}). In later sections, we will show how taking advantage of \emph{input} and \emph{output sparsity} will, even in the small running example, allow computation to complete with fewer memory accesses.

\revision{GraphBLAS's model of computation includes the following concepts:

  \begin{enumerate}
    \item Abstract algebraic constructs: Matrix, Vector, Semiring, and Monoid
    \item Programming constructs: Masking and Descriptor
    \item Compute using constructs: Operation
  \end{enumerate}

  As the name may suggest, the abstract algebraic constructs (Matrix, Vector, Semiring, and Monoid) come directly from abstract algebra and have precise mathematical definitions from that community. The programming constructs (Masking and Descriptor) are used to provide expressibility and performance by changing the abstract algebraic constructs or operations slightly. The Operations are the functions used to carry out computation over the algebraic and programming constructs.}

\subsection{Abstract algebraic constructs}

\subsubsection{Matrix}
A Matrix in GraphBLAST is a general $M$-by-$N$ matrix but it is often used to represent the adjacency matrix of a graph. A full list of methods used to interact with Matrix objects is shown in Table~\ref{tab:operation}. When referring to matrices in mathematical notation, we will indicate them with uppercase boldface, i.e., $\mA$\@. \revision{We index the $(i,j)$-th element of the matrix with $\mA(i,j)$, the $i$-th row with $\mA(i,:)$, and the $j$-th column with  $\mA(:,j)$.}

\begin{table}[t]
  \centering
  \begin{tabular}{lll}
    \toprule
    Operation & Description & Graph application \\ \midrule
    \cmd{Matrix} & matrix constructor & create graph\\
    \cmd{Vector} & vector constructor & create vertex set\\ \midrule
    \cmd{dup} & copy assignment & copy graph or vertex set\\
    \cmd{clear} & empty vector or matrix & empty graph or vertex set\\ \midrule
    \cmd{size} & no.\ of elements (vector only) & no.\ of vertices\\
    \cmd{nrows} & no.\ of rows (matrix only) & no.\ of vertices\\
    \cmd{ncols} & no.\ of columns (matrix only) & no.\ of vertices\\
    \cmd{nvals} & no.\ of stored elements & no.\ of active vertices or edges\\ \midrule
    \cmd{build} & build sparse vector or matrix & build vertex set or graph from tuples\\
    \cmd{buildDense}$^\dag$ & build dense vector or matrix & build vertex set or graph from tuples\\
    \cmd{fill}$^\dag$ & build dense vector or matrix & build vertex set or graph from constant\\ \midrule
    \cmd{setElement} & set single element & modify single vertex or edge\\
    \cmd{extractElement} & extract single element & read value of single vertex or edge\\
    \cmd{extractTuples} & extract tuples & read values of vertices or edges\\ \bottomrule
  \end{tabular}
  \caption{A list of Matrix and Vector operations in GraphBLAST\@.\\$^\dag$: These are convenience operations not found in the GraphBLAS specification, but were added by the authors for GraphBLAST\@.\label{tab:matrixop}}
\end{table}

\subsubsection{Vector}
A Vector is the set of vertices in a graph that are currently actively involved in the graph computation. We call these vertices \emph{active}. The list of methods used to interact with Vector objects overlaps heavily with the one for Matrix objects. When referring to vectors in mathematical notation, we will indicate them with lowercase boldface, i.e., $\mathbf{x}$.
\revision{We index the $i$-th element of the vector with $\mX(i)$.}

\subsubsection{Semiring}
A semiring encapsulates computation on vertices and edges of the graph. In classical matrix multiplication, the semiring used is the $(+, \times, \mR, 0)$ arithmetic semiring. However, this can be generalized to $(\oplus, \otimes, \mD, \mI)$ in order to vary what operations are performed during the graph search. $(\oplus, \otimes, \mD, \mI)$ represents the following:

\begin{itemize}
  \item $\otimes$: Semiring multiply
  \item $\oplus$: Semiring add
  \item $\mD$: Semiring domain
  \item $\mI$: Additive identity
\end{itemize}

Here is an example using the \cmd{MinPlus} semiring (also known as the tropical semiring) $(\oplus, \otimes, \mD, \mI) =  \{\text{min}, +, \mR \cup \{+\infty\}, +\infty\}$, which can be used for shortest-path calculation:

\begin{itemize}
  \item $\otimes$: In \cmd{MinPlus}, $\otimes = +$. The vector represents currently known shortest distances between a source vertex $s$ and vertices whose distance from $s$ we want to update, say $v$. During the multiplication $\otimes = +$, we want to add up distances from parents of $v$ whose distance from $s$ is finite. This gives distances from $s \rightarrow u \rightarrow v$, potentially via many parent vertices $u$.
  \item $\oplus$: In \cmd{MinPlus}, $\oplus = \text{min}$. This operation chooses choosing the distance from $s \rightarrow u \rightarrow v$ such that the distance is a minimum for all intermediate vertices $u$.
  \item $\mD$: In \cmd{MinPlus}, $\mD = \mR \cup \{+\infty\}$, which is the set of real numbers augmented by infinity (indicating unreachability).
  \item $\mI$: In \cmd{MinPlus}, $\mI = +\infty$, representing that doing the reduction $\oplus$ if there are no elements to be reduced---there is no parent $u$ that is reachable from $s$---the default output should be infinity, indicating $v$ is unreachable from $s$ as well.
\end{itemize}

The most frequently used semirings are shown in Table~\ref{tab:semiring}.

\begin{table}
  \centering
  \begin{tabular}{lll}
    \toprule
    Name & Semiring    & Application   \\ \midrule
    \cmd{PlusMultiplies} & $\{+, \times, \mR, 0\}$    & Classical linear algebra \\
    \cmd{LogicalOrAnd} & $\{||, \&\&, \{0,1\}, 0\}$        & Graph connectivity  \\
    \cmd{MinPlus} & $\{\text{min}, +, \mR \cup \{+\infty\}, +\infty\}$ & Shortest path  \\
    \cmd{MaxPlus} & $\{\text{max}, +, \mR, -\infty\}$ & Graph matching                     \\
    \cmd{MinMultiplies} &$ \{\text{min}, \times, \mR, +\infty\}$ & Maximal independent set \\ \midrule
    Name & Monoid    & Application   \\ \midrule
    \cmd{PlusMonoid} & $\{+, 0\}$ & Sum-reduce \\
    \cmd{MultipliesMonoid} & $\{\times, 1\}$ & Times-reduce \\
    \cmd{MinimumMonoid} & $\{\text{min}, +\infty\}$ & Min-reduce \\
    \cmd{MaximumMonoid} & $\{\text{max}, -\infty\}$ & Max-reduce \\
    \cmd{LogicalOrMonoid} & $\{||, 0\}$ & Or-reduce \\
    \cmd{LogicalAndMonoid} & $\{\&\&, 1\}$ & And-reduce \\
    \bottomrule
  \end{tabular}
  \caption{A list of commonly used semirings and monoids in GraphBLAST\@.\label{tab:semiring}}
\end{table}

\subsubsection{Monoid}

A monoid is similar to a semiring, but it only has one operation, which must be associative and have an identity. A monoid should be passed in to GraphBLAS operations that only need one operation instead of two. As a rule of thumb, the only operations that require two operations (i.e., a semiring) are \cmd{mxm}, \cmd{mxv}, and \cmd{vxm}. This means that for GraphBLAS operations \cmd{eWiseMult}, \cmd{eWiseAdd}, and \cmd{reduce}, a monoid should be passed in. A list of frequently used monoids is shown in Table~\ref{tab:semiring}.

\subsection{Programing constructs}

\subsubsection{Masking}
\label{subsubsec:masking}
Masking is an important tool in GraphBLAST that lets a user mark the indices where the result of any operation in Table~\ref{tab:operation} should be written to the output. This set of indices is called the \emph{mask} and must be in the form of a \cmd{Vector} or \cmd{Matrix} object. The masking semantic is:

\begin{quote}
  \emph{For a given pair of indices $(i, j)$, if the mask matrix $\mM(i,j)$ has a value 0, then the output at location $(i,j)$ will not be written to $\mC(i,j)$. However, if $\mM(i, j)$ is not equal to 0, then the output at location $(i,j)$ will be written to $\mC(i,j)$.}
\end{quote}

Sometimes, the user may want the opposite to happen: when the mask matrix has a value 0 at $\mM(i,j)$, then it will be written to the output matrix $\mC(i,j)$. Likewise, if the mask matrix has a non-zero, then it will not be written. This construction is called the \emph{structural complement} of the mask.

To represent masking, we borrow the elementwise multiplication operation from MATLAB\@. Given we are trying to multiply matrices $\mA$, $\mB$ into matrix $\mC$ and the operation is masked by matrix $\mM$, our operation is $\mC \leftarrow \mA \mB\, .\!* \mM$.

\subsubsection{Descriptor}
A descriptor is an object passed into all operations listed in Table~\ref{tab:operation} that can be used to modify the operation. For example, a mask can be set to use the structural complement using a method \cmd{Descriptor::set(GrB\_MASK, GrB\_SCMP)}. The other operations we include are listed in Table~\ref{tab:descriptor}.

\begin{table}
  \centering
  \begin{tabular}{lll}
    \toprule
    Field & Value    & Behavior   \\ \midrule
    \cmd{GrB\_MASK} & (default) &  Mask \\
          & \cmd{GrB\_SCMP}   & Structural complement of mask \\
    \cmd{GrB\_INP0}  & (default) & Do not transpose first input parameter\\
          & \cmd{GrB\_TRAN}  & Transpose first input parameter  \\
    \cmd{GrB\_INP1}  & (default) & Do not transpose second input parameter \\
          & \cmd{GrB\_TRAN} & Transpose second input parameter  \\ \midrule
    \cmd{GrB\_OUTP} & (default) & Do not clear output before writing to masked indices \\
          & \cmd{GrB\_REPLACE} & Clear output before writing to masked indices \\ \bottomrule
  \end{tabular}
  \caption{A list of descriptor settings in GraphBLAST\@. Below the line are variants that are in the GraphBLAS API specification that we do not currently support. \label{tab:descriptor}}
\end{table}

\subsection{Operation}

\begin{figure}[t]
  \centering
  \includegraphics[width=0.6\textwidth]{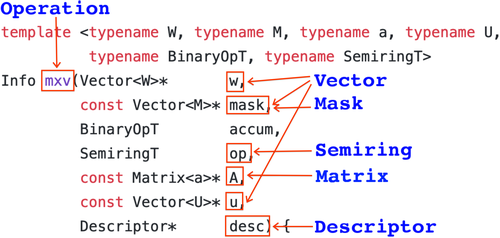}
  \caption{\revision{Calling the operation \cmd{mxv} (matrix-vector multiply) performs $\mW = \mA \mU .* \mathbf{mask}$ over the Semiring \cmd{op}. The template parameters can be used to do compile-time type-checking. \cmd{Info} is an error type that is returned according to the C API specification~\cite{Buluc:2017:TGC}. \cmd{accum} is an optional parameter for controlling whether the output of the calculation overwrites $\mW$ or whether it is accumulated to $\mW$. The Descriptor can be used to control whether the matrix should be transposed or not, and whether the mask or the structural complement of the mask is used (see Section~\ref{subsubsec:masking}).}}
  \label{fig:mxv-concept}
\end{figure}

\begin{figure}[t]
  \centering
  \includegraphics[width=0.6\textwidth]{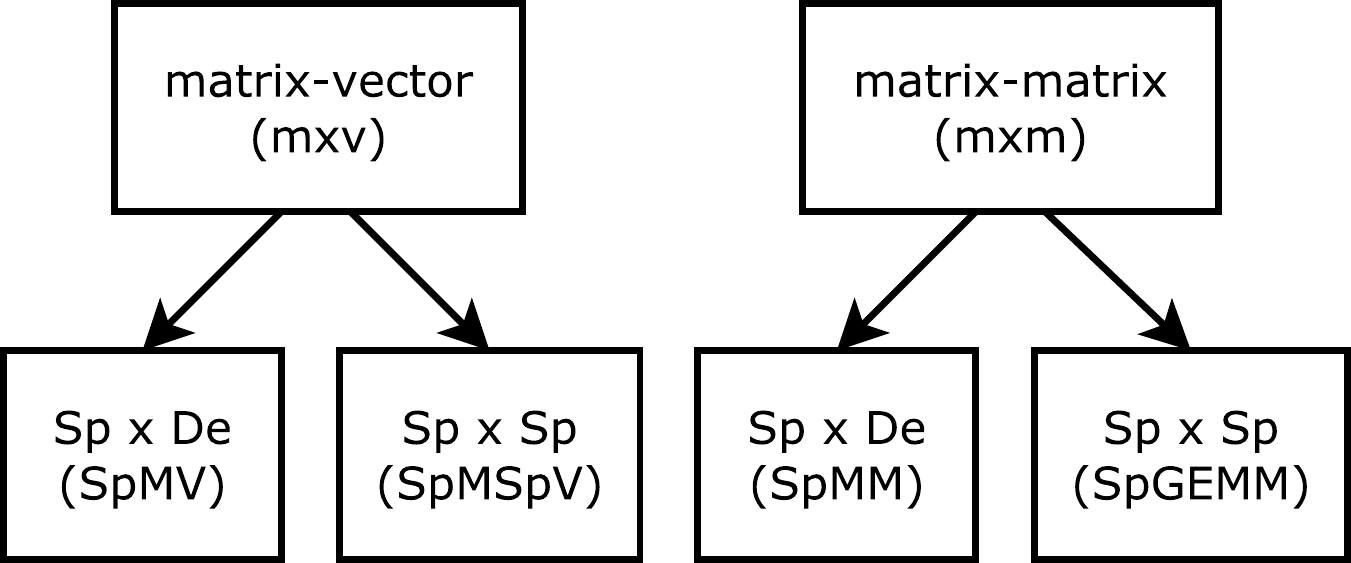}
  \caption{Decomposition of key GraphBLAS operations. Note that vxm is the same as mxv and setting the matrix to be transposed, so it is not shown.}
  \label{fig:decomposition}
\end{figure}

\revision{An operation is a commonly used linear algebra operation. A full list of operations is shown in Table~\ref{tab:operation}.  Of the operations in Table~\ref{tab:operation}, the most computationally intensive and useful operations are \cmd{mxm}, \cmd{mxv}, and \cmd{vxm}. We find empirically that these operations take over 90\% of graph algorithm runtime. For these operations, we decompose them into constituent parts in order to better optimize their performance (see Figure~\ref{fig:decomposition}). Intuitively, \cmd{mxv} and \cmd{vxm} represent a graph traversal from a single-nodeset (one single set of active nodes $\mU(i)$ for all $i$). On ther other hand, \cmd{mxm} represents a multi-nodeset, that is to say, a graph-traversal from multiple, independent sets of active nodes at the same time (for all independent sets of active nodes $B(:,j)$, start a traversal for all active nodes $i$ in that set).

  An operation is used to tie all the objects in GraphBLAS together.  Figure~\ref{fig:mxv-concept} shows a matrix-vector multiply, which represents a graph traversal from all active nodes $\mU(i)$ to their neighbor nodes, applying the semiring multiply $\otimes$ to get a temporary $\mathbf{c}(i,j) = \mA(i, j) \otimes \mU(i)$ between the edge value $\mA(i,j)$ and the active node's value $\mU(i)$ and finally doing a reduction over the temporary output $\mW(i) = \bigoplus_j \mathbf{c}(i,j)$ using the semiring add operation $\oplus$ and the semiring identity $\mathrm{I}$\@. Optionally, the user has the option of applying a boolean mask to the output so that the output is effectively $\mW(i) = \mathbf{mask}(i) \times \mW(i)$ for all $i$. The descriptor can further mutate this operation to account for matrix transposition and whether the mask or its structural complement should be used.}

\begin{table}
  \centering
  \begin{tabular}{llll}
    \toprule
    Operation & Math Equivalent    & Description & Graph application \\ \midrule
    \cmd{mxm} & $\mC = \mA \mB$ & matrix-matrix mult. & multi-nodeset traversal\\
    \cmd{mxv} & $\mW = \mA \mU$ & matrix-vector mult. & single-nodeset traversal \\
    \cmd{vxm} & $\mW = \mU \mA$ & vector-matrix mult. & single-nodeset traversal \\ \midrule
    \cmd{eWiseMult} & $\mC = \mA\text{ }.* \mB$ & element-wise mult. & graph intersection\\
              & $\mW = \mU\text{ }.* \mV$ & & vertex intersection \\ \midrule
    \cmd{eWiseAdd} &$\mC = \mA + \mB$ & element-wise add & graph union \\
              &$\mW = \mU + \mV$ & & vertex union \\ \midrule
    \cmd{extract} & $\mC = \mA(\mathbf{i}, \mathbf{j})$ & extract submatrix & extract subgraph\\
              & $\mW = \mU(\mathbf(i))$ & extract subvector & extract subset of vertices\\ \midrule
    \cmd{assign} &$\mC(\mathbf{i}, \mathbf{j}) = \mA$& assign to submatrix & assign to subgraph \\
              &$\mW(\mathbf{i}) = \mU$ & assign to subvector & assign to subset of vertices\\ \midrule
    \cmd{apply} & $\mC = f(\mA)$ & apply unary op & apply function to each edge \\
              & $\mW = f(\mU)$ & & apply function to each vertex \\ \midrule
    \cmd{reduce} & $\mW = \sum_i \mA(i, :)$ & reduce to vector & compute out-degrees \\
              & $\mW = \sum_j \mA(:, j)$ & reduce to vector & compute in-degrees \\
              & $w = \sum \mW$ & reduce to scalar & \\ \midrule
    \cmd{transpose} & $\mC = \mA^T$ & transpose & reverse edges in graph \\
    \bottomrule
  \end{tabular}
  \caption{A list of operations in GraphBLAST\@.\label{tab:operation}}
\end{table}

\subsection{Running example}
\label{sec:runningexample}

\begin{figure}
  \centering
  \captionsetup[subfigure]{justification=centering}
  \begin{subfigure}[t]{0.4\textwidth}
    \centering
    \includegraphics[width=1\textwidth]{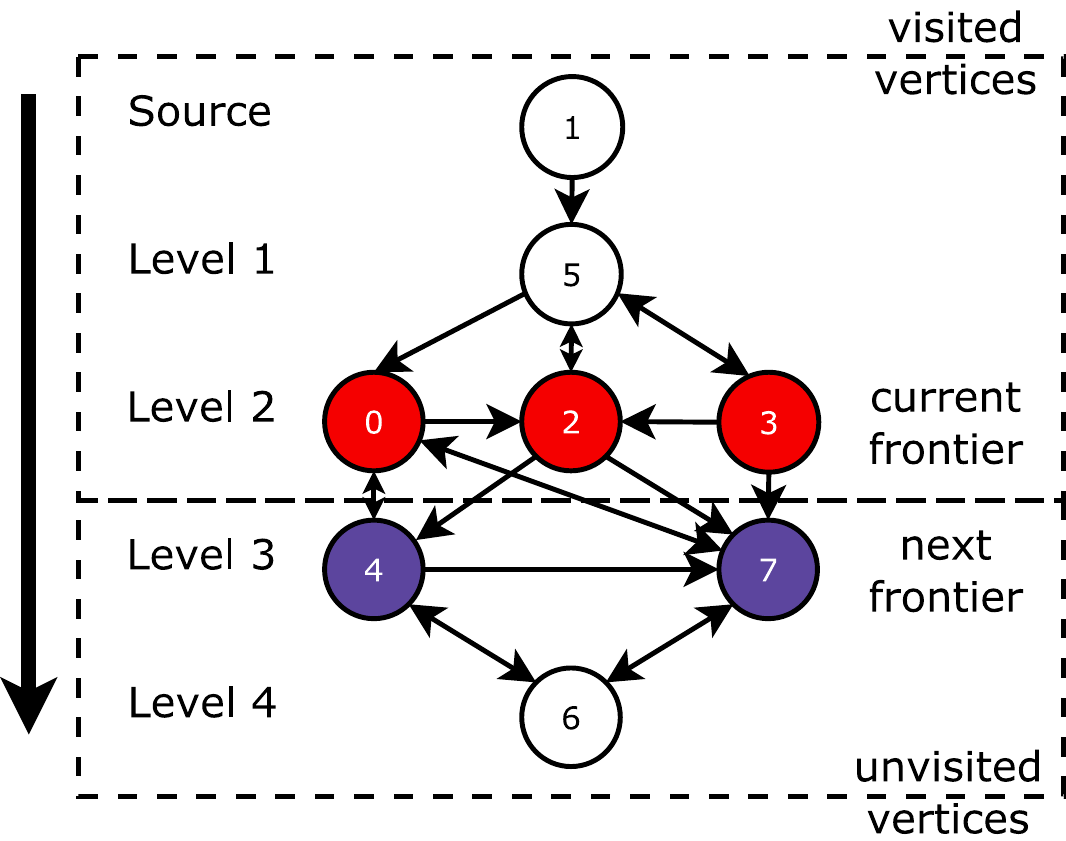}
    \caption{Graph representation\label{fig:graph-journal}}
    \vspace{0.5em}
  \end{subfigure}
  \hfill
  \begin{subfigure}[t]{0.55\textwidth}
    \centering
    \includegraphics[width=1\textwidth]{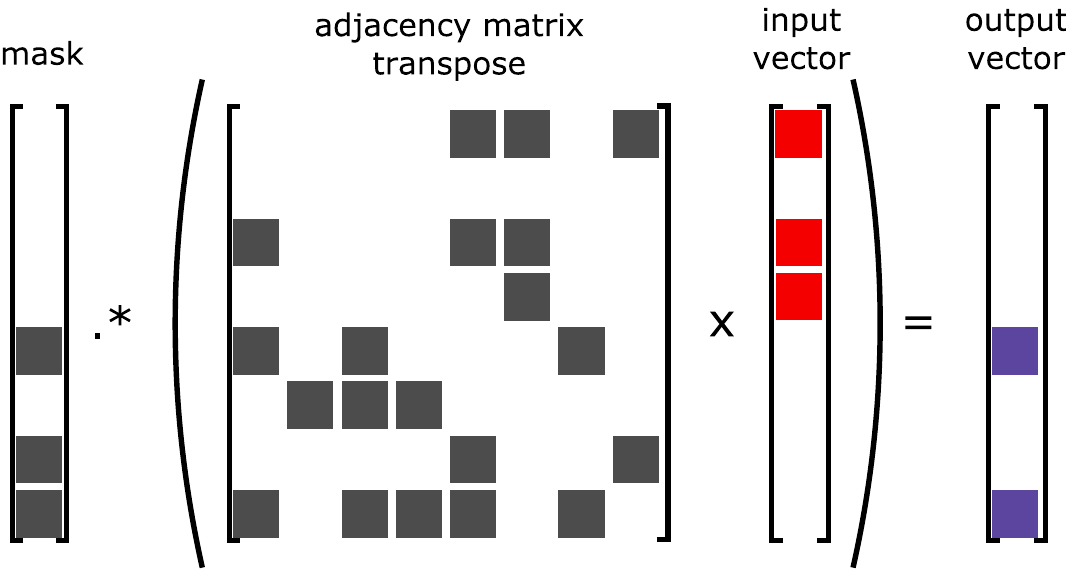}
    \caption{Linear algebraic representation\label{fig:spmv}}
    \vspace{0.5em}
  \end{subfigure}
  \begin{subfigure}[t]{\textwidth}
    \centering
    \includegraphics[width=0.8\textwidth]{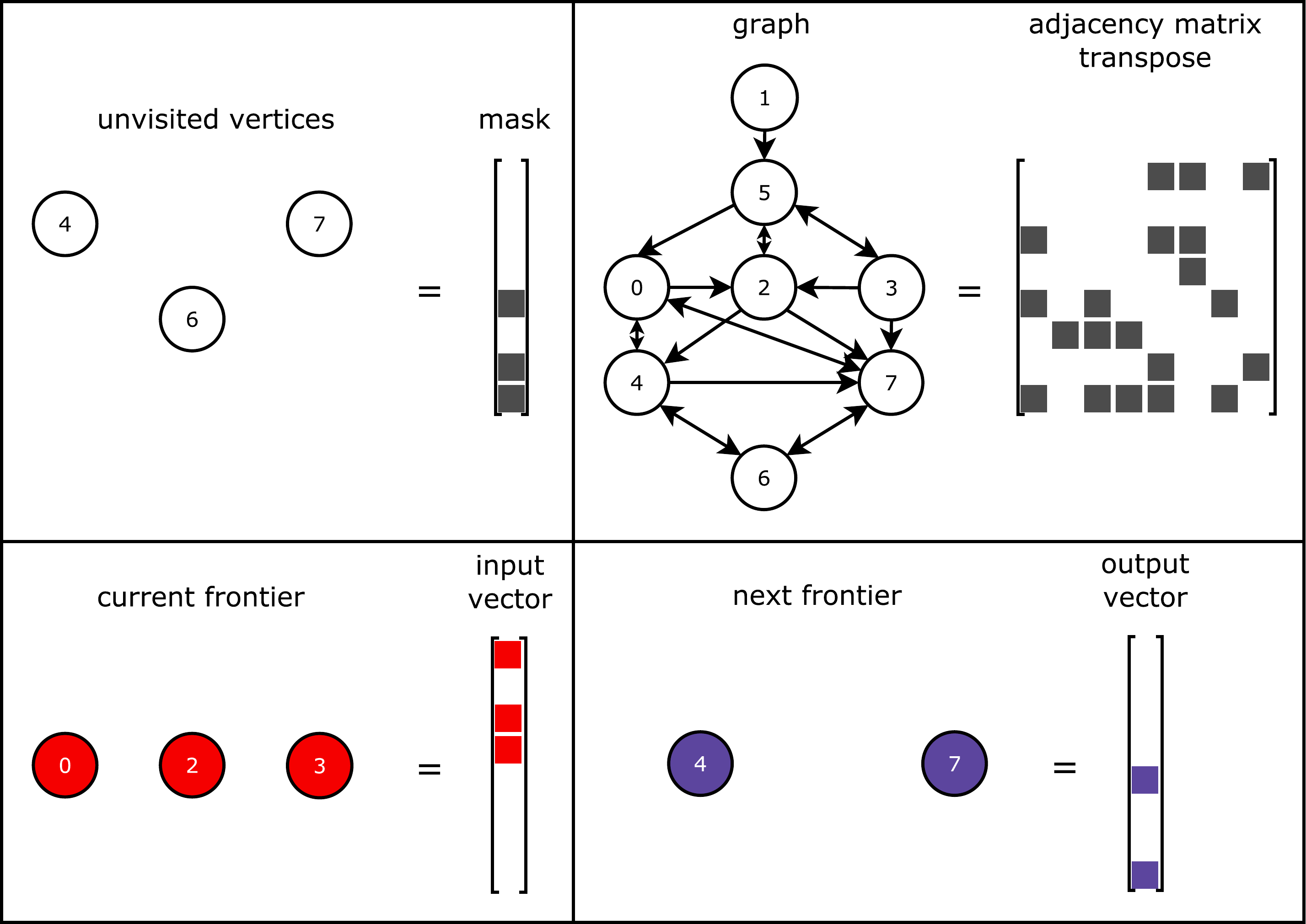}
    \caption{Graph concepts and linear algebraic equivalents\label{fig:graph-spmv}}
  \end{subfigure}
  \caption{Running example of breadth-first-search from source node 1. Currently, we are on level 2 and trying to get to level 3. To do so we need to do a graph traversal from the current frontier (vertices 0, 2, 3) to their neighbors (vertices 2, 4, 5, 7). This corresponds to the multiplication $\mA^T \mathbf{f}$. This is followed by filtering out visited vertices (vertices 2, 5), leaving us with the next frontier (vertices 4, 7). This corresponds to the elementwise multiply $\neg \mathbf{v} \,.\!* (\mA^T \mathbf{f})$.\label{fig:runningexample}}
\end{figure}

As a running example in this paper, we discuss sparse-matrix multiplication by dense-vector (SpMV) and sparse-vector (SpMSpV) with a direct dependence on graph traversal. The key step we will be discussing is Line~\ref{line:multiplyline} of Algorithm~\ref{alg:matrixbfs}, which is the matrix formulation of parallel breadth-first-search. Illustrated in Figure~\ref{fig:spmv}, this problem consists of a matrix-vector multiply followed by an elementwise multiplication between two vectors: one is the output of the matrix-vector multiply and the other is the negation (or \emph{structural complement}) of the visited vector.

Using the standard dense matrix-vector multiplication algorithm (GEMV), we would require $8 \times 8 = 64$ loads and store instructions. However, if we instead treat the matrix not as a dense matrix but as a sparse matrix in order to take advantage of input matrix sparsity, we can perform the same computation in a mere 20 loads and stores into the sparse matrix. This number comes from counting the number of nonzeroes in the sparse matrix, which is equivalent to the number of edges in the graph. Using this as the baseline, we will show in later sections how optimizations such as exploiting the input vector and output vector sparsity can further reduce the number of loads and stores required.

\subsection{Code example}

Having described the different components of GraphBLAST, we show a short code example of how to do breadth-first-search using the GraphBLAST interface alongside the linear algebra in Algorithm~\ref{alg:matrixbfs}. Before the while-loop, the vectors $\mathbf{f}$ and $\mathbf{v}$ (representing the vertices currently active in the traversal and the set of previously visited vertices) are initialized.

Then in each iteration of the while-loop, the following steps take place: (1)~vertices currently active are added to the visited vertex vector, marked by the iteration $d$ where they were first encountered; (2)~the active vertices are traversed to find the next set of active vertices, and then elementwise-multiplied by the negation of the set of active vertices (filtering out previously visited vertices); (3)~the number of active vertices of the next iterations is reduced to variable $c$; (4)~the iteration number is incremented. This while-loop continues until there are no more active vertices ($c$ reaches 0).

As demonstrated in the code example, GraphBLAS has the advantage of being concise. Developing new graph algorithms in GraphBLAS requires modifying a single file and writing simple C++ code. Provided a GraphBLAS implementation exists for a particular processor, GraphBLAS code can be used with minimal changes. Over the next three sections, we will discuss the most important design principles for making this code performant: exploiting input sparsity, exploiting output sparsity, and good load-balancing.

\bcolorbox
\begin{minipage}[t]{\textwidth}
  \begin{algorithmic}[1]
    \Procedure{MatrixBFS}{Graph $\mA$, Vector $\mathbf{v}$, Source $s$}
    \State{}Initialize $d \leftarrow 1$
    \State{}Initialize $\mathbf{f}(i) \leftarrow
    \begin{cases}
      1, \text{   if }i = s\\
      0, \text{   if }i \neq s
    \end{cases}$
    \State{}Initialize $\mathbf{v} \leftarrow [0, 0, \ldots, 0]$
    \State{}Initialize $c \leftarrow 1$
    \While{ $c>0$ }
    \State{}Update $\mathbf{v} \leftarrow d\mathbf{f} + \mathbf{v}$ %
    \State{}Update $\mathbf{f} \leftarrow \mathbf{A^Tf} \,.\!* \mathbf{\neg v}$\label{line:multiplyline}
    \Comment{using Boolean semiring (see Table~\ref{tab:semiring})}
    \State{}Compute $c \leftarrow \sum_{i = 0}^n \mathbf{f}(i)$
    \Comment{using standard plus monoid (see Table~\ref{tab:semiring})}
    \State{}Update $d \leftarrow d+1$
    \EndWhile{}
    \EndProcedure{}
  \end{algorithmic}
\end{minipage}
\ecolorbox
\bcolorbox
\begin{minipage}[]{\textwidth}
\begin{cplus}
#include <graphblas/graphblas.hpp>

void bfs(Vector<float>*       v,
         const Matrix<float>* A,
         Index                s) {
  Descriptor desc;
  Index A_nrows;
  A->nrows(&A_nrows);
  float d = 1.f;

  Vector<float> f1(A_nrows);
  Vector<float> f2(A_nrows);
  std::vector<Index> indices(1, s);
  std::vector<float> values(1, 1.f);
  f1.build(&indices, &values, 1, GrB_NULL);

  v->fill(0.f);
  float c = 1.f;
  while (c > 0) {
    // Assign level d at indices f1 to visited vector v
    graphblas::assign(v, &f1, GrB_NULL, d, GrB_ALL, A_nrows, &desc);
    // Set mask to use structural complement (negation)
    desc->toggle(GrB_MASK);
    // Multiply frontier f1 by transpose of matrix A using visited vector v as mask
    // Semiring: Boolean semiring (see Table 4)
    graphblas::vxm(&f2, v, GrB_NULL, LogicalOrAndSemiring<float>(), &f1, A, &desc);
    // Set mask to not use structural complement (negation)
    desc->toggle(GrB_MASK);
    f2.swap(&f1);
    // Check how many vertices of frontier f1 are active, stop when number reaches 0
    // Monoid: Standard addition (see Table 4)
    graphblas::reduce(&c, GrB_NULL, PlusMonoid<float>(), &f1, &desc);
    d++;
  }
}
\end{cplus}
\end{minipage}
\ecolorbox
{\captionof{algorithm}{Matrix formulation of BFS (top) and example GraphBLAST code (bottom).\label{alg:matrixbfs}}} %

\subsection{Differences with GraphBLAS C API standard}

\revision{We have tried to make our framework interface adhere to the GraphBLAS C API as close as possible, but since we decided to take advantage of certain C++ features (templates and functors, in particular) in our framework, we have departed from the strict API standard. We observe that some differences are motivation for the design of a GraphBLAS C++ API~\cite{Brock:2020:ARF}. Some major differences in the interface are listed below:

  \begin{enumerate}
    \item  We use C++ templates and functors to implement GraphBLAS Semirings. For a more in-depth discussion, see Section~\ref{subsec:operators}.
    \item We require passing in a template parameter specifying the type in place of: (i) passing a datatype of \cmd{GrB\_Type} to \cmd{Matrix} and \cmd{Vector} declaration, (ii) specifying types used in the semiring. This allows more compile-time type-checking to ensure that the types are correct, which is not possible as a compiled binary, but has the disadvantage of the user having to do more work than necessary (i.e., despite the backend knowing what type it is in the \cmd{Matrix} declaration).
    \item We provide a header-only C++ library rather than a shared object library. This design choice carries all the advantages and disadvantages header-only libraries have compared to shared object libraries, in addition to the advantages and disadvantages of the first two differences. In our design, this is mainly a consequence of our choice of using C++ templates and functors rather than code generation (see Section~\ref{subsec:operators}).
  \end{enumerate}

  A few minor interface differences follow:

  \begin{enumerate}
    \item We require \cmd{Matrix::build} and \cmd{Vector::build} to use \cmd{std::vector} rather than C-style arrays. However, it would be a simple addition to maintain compatibility with GraphBLAS C API specification by allowing C-style arrays too.
    \item We use a \cmd{graphblas} namespace instead of prefixing our methods  with \cmd{GrB\_}.
    \item We provide convenience methods \cmd{Vector::fill} and \cmd{Descriptor::toggle} that are extensions to the GraphBLAS C API specification.
    \item We choose not to include the \cmd{GrB\_REPLACE} descriptor setting. This is motivated by our design principle of choosing not to implement what can be composed by a few simpler operations. In this case, if desired, the user can reproduce the \cmd{GrB\_REPLACE} behavior by first calling \cmd{Matrix::clear()} or \cmd{Vector::clear()} and then calling the operation they wanted to modify with \cmd{GrB\_REPLACE}.
    \item We choose to ignore the \cmd{accum} input parameter, which is responsible for choosing to accumulate results into the output Vector or Matrix. Our motivation is the same as the above decision. This accumulation can also be done by following up the initial operation by an elementwise addition or elementwise multiply.
    \item We have matrix-vector, matrix-scalar, and vector-scalar variants of elementwise addition and multiplication for convenience and performance. These variants are called rank promotion~\cite{Mattson:2017:GCA} or Numpy-style broadcasting~\cite{harris2020array}.
  \end{enumerate}}

\section{Exploiting Input Sparsity (Direction-Optimization)}
\label{sec:diropt}
In this section, we discuss our design philosophy of making exploiting input sparsity and one of its consequences, direction optimization, a first-class feature of our implementation. Since the matrix represents a graph, the matrix $\mA$ will be assumed to be stored in sparse format. In traversal-based graph algorithms, the operation we care most about is:

\[\mathbf{y} \gets \mA \mathbf{x}\,.\!* \neg \mathbf{m}\]

Here, the matrix $\mA$ represents the graph, the input vector $\mathbf{x}$ represents the current frontier, the output vector $\mathbf{y}$ represents the next iteration frontier, and the mask vector $\mathbf{m}$ represents the set of visited vertices. The negation $\neg$ converts this set of visited vertices to unvisited vertices. For more discussion on masks, see Section~\ref{sec:masking}.

In this section, we try to limit our discussion to \emph{input sparsity}, by which we are referring to the input vector $\mathbf{x}$ being sparse. We exploit this fact to reduce the number of operations. We provide quantitative data to support our conclusion that doing so is of the foremost importance in building a high-performance graph framework. We present three seemingly unrelated challenges with implementing a linear-algebra-based graph framework based on the GraphBLAS specification, but which we will show are actually facets of the same problem:

\begin{enumerate}
        \item Previous work~\cite{Beamer:2012:DOB,Shun:2013:LLG} has shown that direction optimization is critical to achieving state-of-the-art performance on breadth-first-search. However, direction optimization has been notably absent in linear-algebra-based graph frameworks and assumed only possible for traditional, vertex-centric graph frameworks. How can direction optimization be implemented as matrix-vector multiplication in a linear-algebra-based framework like GraphBLAS\@?
        \item The GraphBLAS definition for \cmd{mxv} operation is intentionally underspecified. As Figure~\ref{fig:decomposition} shows, there are two ways to implement \cmd{mxv}. How should it be implemented?
        \item The GraphBLAS definition for \cmd{Matrix} and \cmd{Vector} objects are intentionally underspecified. What should the underlying data structure for these objects look like?
\end{enumerate}

\revision{The results of this section show that the GraphBLAST library would automatically discover the direction optimization idea while only having access to \cmd{mxv} function parameters, without having any knowledge of the semantics of the computation (e.g., graph traversals). Because GraphBLAST changes direction based on the sparsity of inputs alone, which are abstract vector and matrix objects, it has the potential to discover more opportunities for push-pull like optimization on domains beyond graph processing.}

\subsection{Two roads to matrix-vector multiplication}

Before we address the above challenges, we draw a distinction between two different ways the matrix-vector multiply $\mathbf{y} \gets \mathbf{A} \mathbf{x}$ can be computed. We distinguish between multiplying a sparse-matrix by a dense-vector (SpMV) and by a sparse-vector (SpMSpV). There is extensive literature focusing on SpMV for GPUs (including a comprehensive survey~\cite{Filippone:2017:SMV}). However, we concentrate on SpMSpV, because it is more relevant to graph search algorithms where the vector represents the subset of vertices that are currently active and is typically sparse.

\begin{figure}[t]
  \centering
  \begin{subfigure}[t]{0.45\textwidth}
    \centering
    \includegraphics[width=1\linewidth]{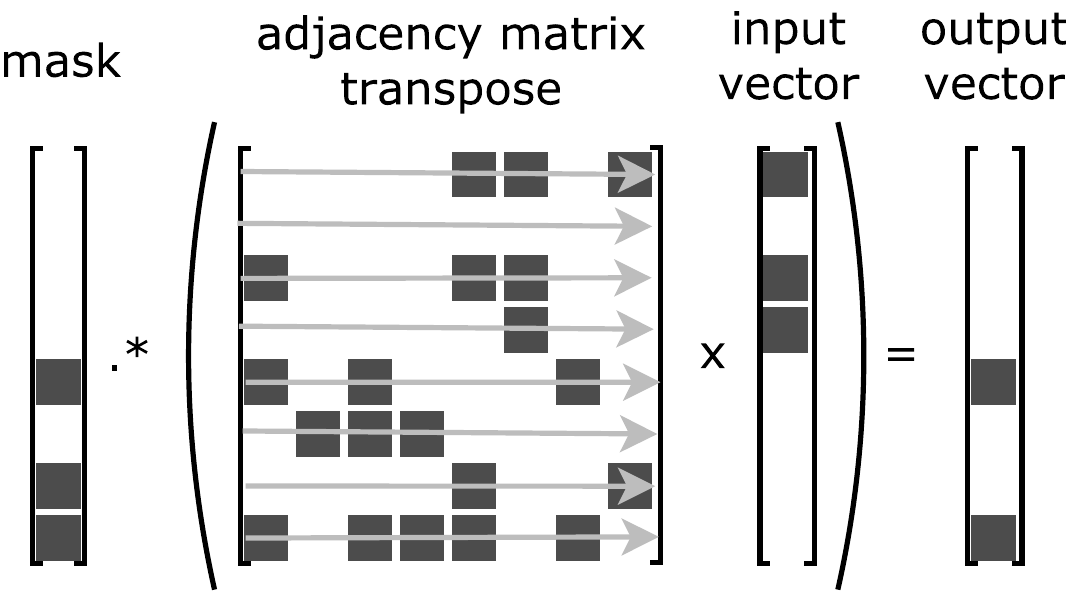}
    \caption{SpMV (20 loads and stores)\label{fig:spmv2}}
  \end{subfigure}
  \hfill
  \begin{subfigure}[t]{0.45\textwidth}
    \centering
    \includegraphics[width=1\textwidth]{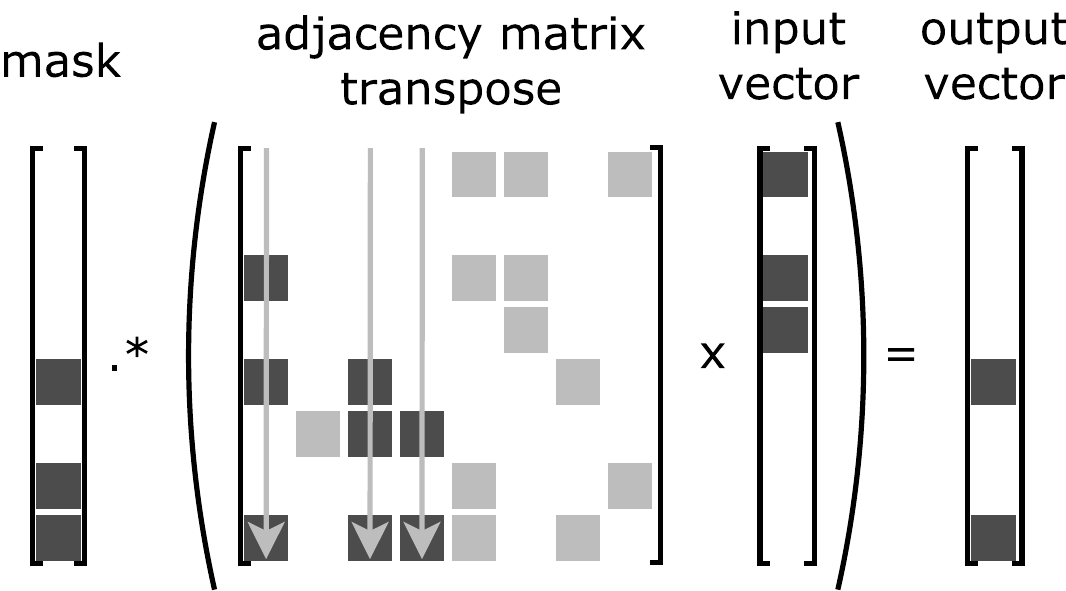}
    \caption{SpMSpV (8 loads and stores)\label{fig:spmspv2}}
  \end{subfigure}
  \begin{subfigure}[t]{0.45\textwidth}
    \centering
    \includegraphics[width=1\linewidth]{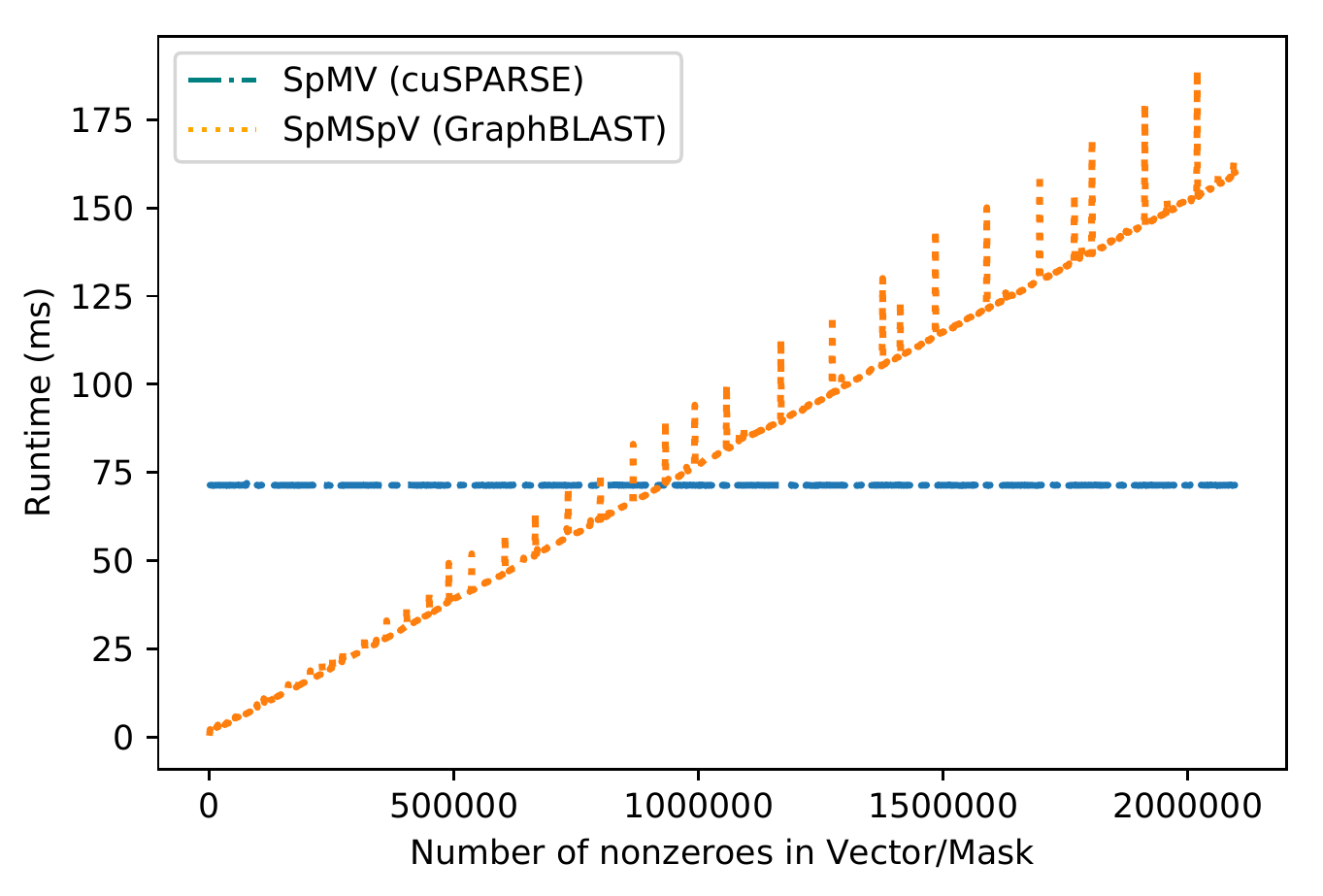}
    \caption{Algorithmic complexity of SpMV and SpMSpV as a function of vector sparsity.\label{fig:vsdensecomplexity}}
  \end{subfigure}
  \hfill
  \begin{subfigure}[t]{0.45\textwidth}
    \centering
    \includegraphics[width=1\textwidth]{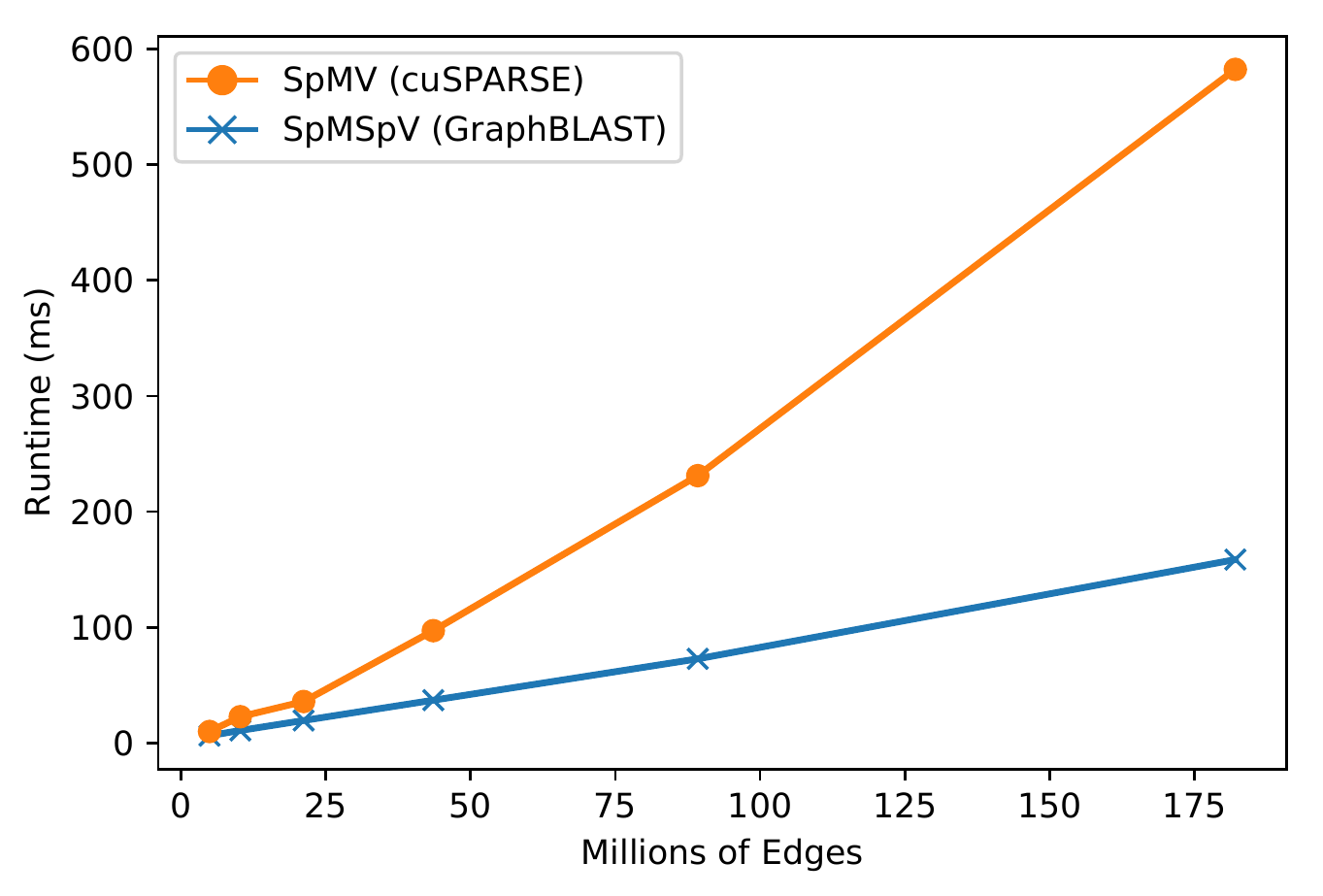}
    \caption{SpMV and SpMSpV runtime on Kronecker scale-16--21 graphs.\label{fig:vs_cusparse}}
  \end{subfigure}
  \caption{Comparison of SpMV and SpMSpV\@.\label{fig:sparsevsdense}}
\end{figure}

Recall in the running example in Section~\ref{sec:runningexample} that by exploiting matrix sparsity (SpMV) in favor of dense matrix-vector multiplication (GEMV), we were able to bring the number of \revision{load and store instructions} down from GEMV's 64 to SpMV's 20. A natural question to ask is whether it is possible to decrease the number of \revision{load and store instructions} further when the input vector is sparse. Indeed, when we exploit input sparsity (SpMSpV) to get the situation in Figure~\ref{fig:spmspv2}, we can reduce the number of \revision{loads and stores} from 20 to 8. Similar to how our move from GEMV to SpMV involved changing the matrix storage format from dense to sparse, moving from SpMV to SpMSpV motivates storing the vector in sparse format. It is worth noting that the sparse vectors are assumed to be implemented as lists of indices and values. A summary is shown in Table~\ref{tab:complexity-journal}.

\begin{table}
\centering
  \rowcolors{2}{lightgray}{white}
\begin{tabular}{lclccc}
        \toprule
        & & & Matrix & Input Vector & Output Vector \\
        Operation & Mask          & Complexity & Sparsity ($\mA$) & Sparsity ($\mathbf{x}$) & Sparsity ($\mathbf{m}$)\\ \midrule
        GEMV & no & $O(MN)$ & & & \\
        SpMV (pull)    & no         & $O(dM)$ & \checkmark & & \\
        SpMSpV (push) & no & $O(d\text{ }nnz(\mathbf{x}))$ & \checkmark & \checkmark & \\ \midrule
        GEMV & yes & $O(N\text{ }nnz(\mathbf{m}))$ & & & \checkmark \\
        SpMV (pull) & yes   & $O(d\text{ }nnz(\mathbf{m}))$ & \checkmark & & \checkmark \\
        SpMSpV (push) & yes    &   $O(d\text{ }nnz(\mathbf{x}))$ & \checkmark & \checkmark & \\ \bottomrule
\end{tabular}
\caption{Computational complexity of matrix-vector multiplication where $d$ is the average number of nonzeroes per row or column, and $\mA$ is an $M$-by-$N$ matrix. The top three rows indicate the standard case $\mathbf{y} \leftarrow \mA \mathbf{x}$, while the bottom three rows represent the masked case $\mathbf{y} \leftarrow \mA \mathbf{x} \,.\!* \mathbf{m}$. Checkmarks indicate which form of sparsity each operation exploits. The notation $nnz(\mathbf{a}))$ refers to the number of nonzero entries in a vector $\mathbf{a}$. \label{tab:complexity-journal}}
\end{table}

\subsection{Related work}

Mirroring the dichotomy between SpMSpV and SpMV, there are two methods to perform one iteration of graph traversal, which are called \emph{push} and \emph{pull}.\footnote{To the authors' best knowledge, the terminology of ``push'' and ``pull'' was first introduced by Karp et al.~\cite{Karp:2000:RRS} in the context of updates to distributed copies of a database.} They can be used to describe graph traversals in a variety of graph traversal-based algorithms such as breadth-first-search, single-source shortest-path, and PageRank.

In the case of breadth-first-search, \emph{push} begins with the current frontier (the set of vertices from which we are traversing the graph) and looks for children of this set of vertices. Then, from this set of children, the previously visited vertices must be filtered out to generate the output frontier (the frontier we use as input on the next iteration). In contrast, \emph{pull} starts from the set of \emph{unvisited} vertices and looks back to find their parents. If a node in the unvisited-vertex set has a parent in the current frontier, we add it to the output frontier. Beamer, Asanovi\'{c}, and Patterson~\cite{Beamer:2012:DOB} observed that in the middle iterations of a BFS on scale-free graphs, the frontier becomes large and each neighbor is found many times, leading to redundant work. They show that for optimal performance, in these intermediate iterations, they should switch to \emph{pull}, and then in later iterations, back to \emph{push}.

Many graph algorithms such as breadth-first-search, single-source shortest-path, and PageRank involve multiple iterations of graph traversals. Switching between \emph{push} and \emph{pull} in different iterations applied to the specific algorithm of breadth-first-search is called \emph{direction optimization} or \emph{direction-optimized BFS}, which was also described by Beamer, Asanovi\'{c}, and Patterson~\cite{Beamer:2012:DOB}. This approach is also termed \emph{push-pull}. Building on this work, Shun and Blelloch~\cite{Shun:2013:LLG} generalized direction optimization to graph traversal algorithms beyond BFS\@. To avoid confusion with the BFS-specific instance, we refer to Shun and Blelloch's contribution as \emph{generalized direction optimization}.

Beamer, Asanovi\'{c} and Patterson later studied matrix-vector multiplication in the context of SpMV- and SpMSpV-based implementations for PageRank~\cite{Beamer:2017:RPC}. In both their work and that of Besta et al.~\cite{Besta:2017:TPT}, the authors noted that switching between push/pull is the same as switching between SpMSpV/SpMV\@. In both works, authors show a one-to-one correspondence between push and SpMSpV, and between pull and SpMV; they are two ways of thinking about the same concept.

Our work differs from Beamer, Asanovi\'{c}, and Patterson and Besta et al.\ in three ways: (1)~they emphasize graph algorithm research, whereas we focus on building a graph framework, (2)~their work targets multithreaded CPUs, while ours targets GPUs, and (3)~their interface is vertex-centric, but ours is linear-algebra-based.

The work we present here builds on our earlier work and is first to extend the \emph{generalized direction optimization} technique to linear-algebra-based frameworks based on the GraphBLAS specification. In contrast, previous implementations to the GraphBLAS specification, such as GBTL~\cite{Zhang:2016:GCG} and SuiteSparse~\cite{Davis:2018:SGG}, do not support \emph{generalized direction optimization} and as a consequence, trail state-of-art graph frameworks in performance.

In both implementations, the operation \cmd{mxv} is implemented as a special case of \cmd{mxm} when one of the matrix dimensions is 1 (i.e., is a \cmd{Vector}). The \cmd{mxm} implementation is a variant of Gustavson's algorithm~\cite{Gustavson:1978:TFA}, which takes advantage of both matrix sparsity and input vector sparsity, so it has a similar performance characteristic as SpMSpV\@. Therefore, it shares SpMSpV's poor performance when either: (1)~there are more elements in the input vector or (2)~when there are fewer elements in the mask (representing fewer operations that need to be performed). In other words, neither GBTL and SuiteSparse automatically switch to \emph{pull} when the input vector becomes large in the middle iterations of graph traversal algorithms like BFS, and perform \emph{push} throughout the entire BFS\@. In comparison, our graph framework balances exploiting input vector sparsity (SpMSpV) with the efficiency of SpMV during iterations of high input vector sparsity. This helps us match or exceed the performance of existing graph frameworks (Section~\ref{sec:results}).

\subsection{Implementation}
\label{sec:sparsevsdense}

In this subsection, we revisit the three challenges we claimed boil down to different facets of the same challenge: exploiting input sparsity.

\begin{description}
  \item[Direction optimization] Our backend automatically handles direction optimization when \cmd{mxv} is called, by consulting an empirical cost model and calling either the SpMV or SpMSpV routine that we expect will result in the fewest memory accesses.
  \item[mxv: SpMV or SpMSpV] Both routines are necessary for an efficient implementation of \cmd{mxv} in a graph framework.
  \item[Matrix and Vector storage format] For \cmd{Matrix}, store both CSR and CSC, but give users the option to save memory by  only storing one of these two representations. The result is a memory-efficient, performance-inefficient solution. For \cmd{Vector}, since both dense vector and sparse vector are required for the two different routines SpMV and SpMSpV respectively, we give the backend the responsibility to switch between dense and sparse vector representations. We allow the user to specify the initial storage format of the \cmd{Matrix} and \cmd{Vector} objects.
\end{description}

\begin{table}
  \centering
  \begin{tabular}{llll}
    \toprule
    Work & Direction & Criteria & Application \\ \midrule
    Beamer et al.~\cite{Beamer:2012:DOB} & push $\rightarrow$ pull & $|E_f| > |E_u|/14$ and increasing & BFS only \\
         & push $\leftarrow$ pull & $|V_f| < |V|/24$ and decreasing & BFS only\\
    Ligra~\cite{Shun:2013:LLG} & push $\rightarrow$ pull & $|E_f| > |E|/20$ & generalized  \\
         & push $\leftarrow$ pull & $|E_f| < |E|/20$ & generalized \\
    Gunrock~\cite{Wang:2017:GGG} & push $\rightarrow$ pull & $|E_f^*| > |E_u^*|/1000$ & BFS only \\
         & push $\leftarrow$ pull & $|E_f^*| < |E_u^*|/5$ & BFS only \\
    This work & push $\rightarrow$ pull & $|E_f^*| > |E|/10$ & generalized \\
         & push $\leftarrow$ pull & $|E_f^*| < |E|/10$ & generalized \\ \bottomrule
  \end{tabular}
  \caption{Direction-optimization criteria for four different works. $|V_f|$ indicates the number of nonzeroes in the frontier $\mathbf{f}$. $|E_f|$ indicates the number of neighbors from the frontier $\mathbf{f}$. $|E_u|$ indicates the number of neighbors from unvisited vertices. Superscript $^*$ indicates the value is approximated rather than precisely calculated.\label{tab:switchpoint}}
\end{table}

\subsubsection{Direction optimization}

When a user calls \cmd{mxv}, our backend chooses either the SpMV or SpMSpV routine, using an empirical cost model to select the one with fewer memory accesses. Table~\ref{tab:switchpoint} shows how our decision to change directions compares with existing literature. We make the following simplifying assumptions:

\begin{enumerate}
  \item On GPUs, computing the precise number of neighbors $|E_f|$ for a given frontier $\mathbf{f}$ requires prefix-sum computations. To avoid what Beamer et al.\ called a non-significant amount of overhead, we instead approximate the precise number of neighbors using the number of nonzeroes in the vector by assuming that in expectation, each element in the vector has around the same number of neighbors, i.e., $d |V_f| \approx |E_f|$. Gunrock also makes this assumption.
  \item When the mask (representing the unvisited vertices) is dense, counting the number of unvisited vertices $|V_u|$ requires an additional GPU kernel launch, which represents significant overhead (Section~\ref{sec:kernel-launch}). Therefore, we make the assumption that the number of unvisited vertices is all vertices, i.e., $|V_u| \approx |V|$ so $|E_u| \approx |E|$. We find this is a reasonable assumption to make, because for scale-free graphs the optimal time to switch from push to pull is very early on, so $|V_u| \approx |V|$. Ligra also makes this assumption.
\end{enumerate}

\subsubsection{mxv: SpMV or SpMSpV}
\label{sec:spmv-spmspv-input-sparsity-benchmark}

Following our earlier work~\cite{Yang:2015:FSM}, which showed that SpMV is not performant enough for graph traversal and that SpMSpV is necessary, we run our own microbenchmark regarding GraphBLAS\@. In our microbenchmark, we benchmarked \cmd{graphblas::mxv} implemented with two variants---\cmd{SparseVector} and  \cmd{DenseVector}---as a function of \cmd{Vector} sparsity for a synthetic undirected Kronecker graph with 2M vertices and 182M edges. For more details of this experiment, see Section~\ref{sec:results}.

As our microbenchmark in Figure~\ref{fig:sparsevsdense} illustrates, the performance of the SpMSpV variant of \cmd{graphblas::mxv} is proportional to the sparsity of the input vector. However, the SpMV variant is constant. This matches the theoretical complexity shown in Table~\ref{tab:complexity-journal}, which shows that SpMV scales with $O(dM)$, which is independent of input vector sparsity. However, SpMSpV is able to factor in the sparsity of the input vector (i.e., $nnz(\mathbf{x})$) into the computational cost. For more details, see Section~\ref{subsubsec:spmspv}.

\subsubsection{Matrix and Vector storage format}

One of the most important design choices for an implementer is whether \cmd{Matrix} and \cmd{Vector} objects ought to be stored in dense or sparse storage, and if sparse, which type of sparse matrix or vector storage?

For \cmd{Matrix} objects, the decision is clear-cut. Since graphs tend to have more than 99.9\% sparsity and upwards of millions of vertices, storing them in dense format would be wasteful and in some cases impossible because of the limitation of available device memory. We use the popular CSR (Compressed Sparse Row) format, because it is standard in graph analytics and supports the fast row access required by SpMV\@. Similarly, since we also need to support SpMSpV and fast column access, we also support the CSC data structure (Figure~\ref{fig:sparsevsdense}).

For \cmd{Vector} objects, we support both dense and sparse storage formats. The dense storage is a flat array of values. The sparse storage is a list of sorted indices and values for all nonzero elements in the vector. Through additional \cmd{Vector} object methods \cmd{Vector::buildDense} and \cmd{Vector::fill} (shown in Table~\ref{tab:matrixop}), we allow users to give the backend hints on whether they want the object to initially be stored in dense or sparse storage.

\subsection{Direction optimization insights}

Exploiting input sparsity is a useful and important strategy for high-performance in graph traversals. We believe that the GraphBLAS interface decision where users do not have to specify whether or not they want to exploit input sparsity is a good one; we showed that instead, users  must only write code once using the \cmd{mxv} interface and both forms of SpMV and SpMSpV code can be automatically generated for them by GraphBLAST\@. In the next section, we will show how the number of memory accesses can also be reduced by exploiting output sparsity.

\section{Exploiting Output Sparsity (Masking)}
\label{sec:masking}
The previous section discussed the importance of reducing the number of load and store instructions using input vector sparsity. This section deals with the mirror situation, which is \emph{output vector sparsity} (or \emph{output sparsity}). Output vector sparsity can also be referred to as an output mask or \emph{masking} for short.

Masking allows GraphBLAS users to tell the framework they are planning to follow a matrix-vector or matrix-matrix multiply with an elementwise product. This allows the backend to implement the fused mask optimization, which in some cases may reduce the number of computations needed. Alongside exploiting input sparsity, our design philosophy was to make exploiting output sparsity a first-class feature in GraphBLAST with highly efficient implementations of masking. Masking raises the following implementation challenges.

\begin{enumerate}
  \item Masking is a novel concept introduced by the GraphBLAS API to allow users to decide which output indices they do and do not care about computing. How can masking be implemented efficiently?
  \item When should the mask be accessed before the computation in out-of-order fashion and when should it be processed after the computation?
\end{enumerate}

\subsection{Motivation and applications of masking}
Following the brief introduction to masking in Section~\ref{sec:masking}, the reader may wonder why such an operation is necessary. Masking can be thought of in two ways: (i)~masking is a way to fuse an element-wise operation with another operation from Table~\ref{tab:operation}; and (ii)~masking allows the user to express for which indices they do and do not require a value before the actual computation is performed. We define this as \emph{output sparsity}. The former means that masking is a way for the user to tell the framework there is an opportunity for kernel fusion, while the latter is an intuitive way to understand why masking can reduce the number of computations in graph algorithms.

There are several graph algorithms where exploiting \emph{output sparsity} can be used to reduce the number of computations:

\begin{enumerate}
  \item In breadth-first-search~\cite{Buluc:2017:TGC,Yang:2018:IPE}, the mask \cmd{Vector} represents the \emph{visited} set of vertices. Since in a breadth-first-search each vertex only needs to be visited once, the user can let the software know that the output need not include any vertices from the \emph{visited} set.
  \item In single-source shortest-path~\cite{Davidson:2014:WPG}, the mask \cmd{Vector} represents the set of vertices that have seen their distances from the source vertex change in this iteration. The mask can thus be used to zero out currently active vertices from the next traversal, because their distance information has already been taken into account in earlier traversal iterations. The mask can be used to help keep the active vertices \cmd{Vector} sparse throughout the SSSP; otherwise, it would be increasingly densifying.
  \item In adaptive PageRank (also known as PageRankDelta)~\cite{Kamwar:2004:AMC,Shun:2013:LLG}, the mask \cmd{Vector} represents the set of vertices that has converged already. The PageRank value for this set of vertices does not need to be updated in future iterations.
  \item In triangle counting~\cite{Azad:2015:PTC,Wolf:2017:FLA}, the mask \cmd{Matrix} represents the adjacency matrix where a value 1 at $\mM(i,j)$ indicates the presence of edge $i \rightarrow j$, and 0 indicates a lack of an edge. Performing a dot product $\mM \mM$ corresponds to finding for each index pair $(i, j)$ the number of wedges $i \rightarrow k \rightarrow j$ that can be formed for all $k \in V$. Thus applying the mask \cmd{Matrix} to the dot product will yield $\mM \mM\,.\!* \mM$, which indicates the set of wedges that are also triangles by virtue of the presence of edge $i \rightarrow j$. Here the $.*$ operation indicates element-wise operation. To get the number of wedges from the set of wedges, a further reduction is required. The algorithm described was to explain the purpose of the mask in triangle counting rather than for giving an optimal algorithm. For a better algorithm that does 6$\times$ less work, see Section~\ref{sec:tricount}.
\end{enumerate}

\subsection{Microbenchmarks}

Similar to our earlier microbenchmark (Section~\ref{sec:spmv-spmspv-input-sparsity-benchmark}), we benchmark how using masked SpMV and SpMSpV variants of \cmd{graphblas::mxv} performed compared with unmasked SpMV and SpMSpV as a function of mask \cmd{Vector} sparsity for a synthetic undirected Kronecker graph with 2M vertices and 182M edges. For more details of the experiment setup, see Section~\ref{sec:results}.

\begin{figure}[t]
  \centering
  \begin{subfigure}[t]{0.45\textwidth}
    \centering
    \includegraphics[width=1\linewidth]{fig/spmv2}
    \caption{SpMV not fused with mask (20 loads and stores).\label{fig:spmv3old}}
  \end{subfigure}
  \hfill
  \begin{subfigure}[t]{0.45\textwidth}
    \centering
    \includegraphics[width=1\textwidth]{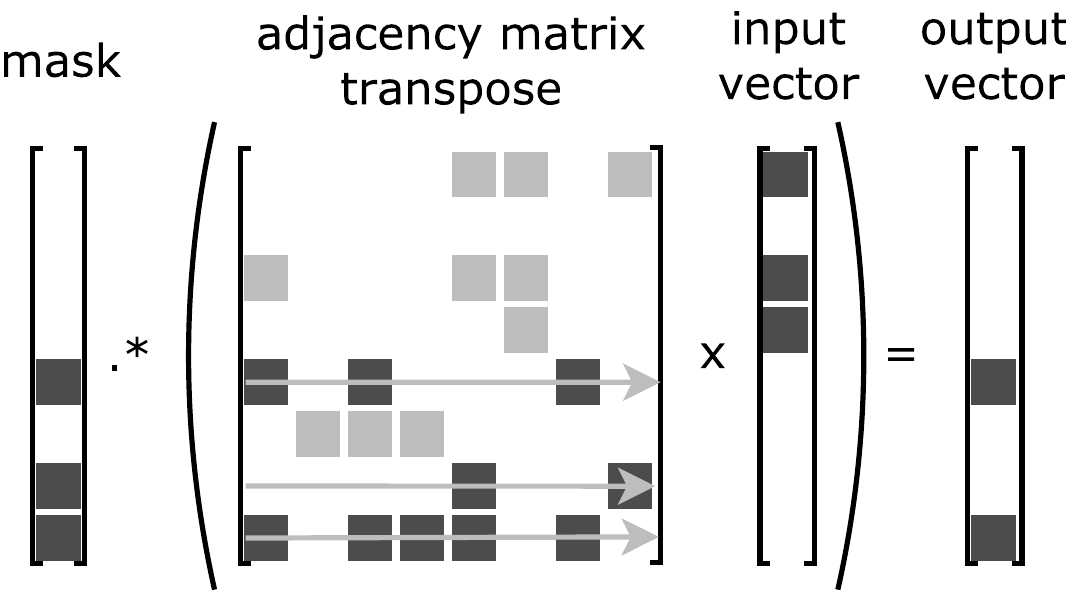}
    \caption{SpMV fused with mask (10 loads and stores).\label{fig:spmv3}}
  \end{subfigure}
  \begin{subfigure}[t]{0.45\textwidth}
    \centering
    \includegraphics[width=1\linewidth]{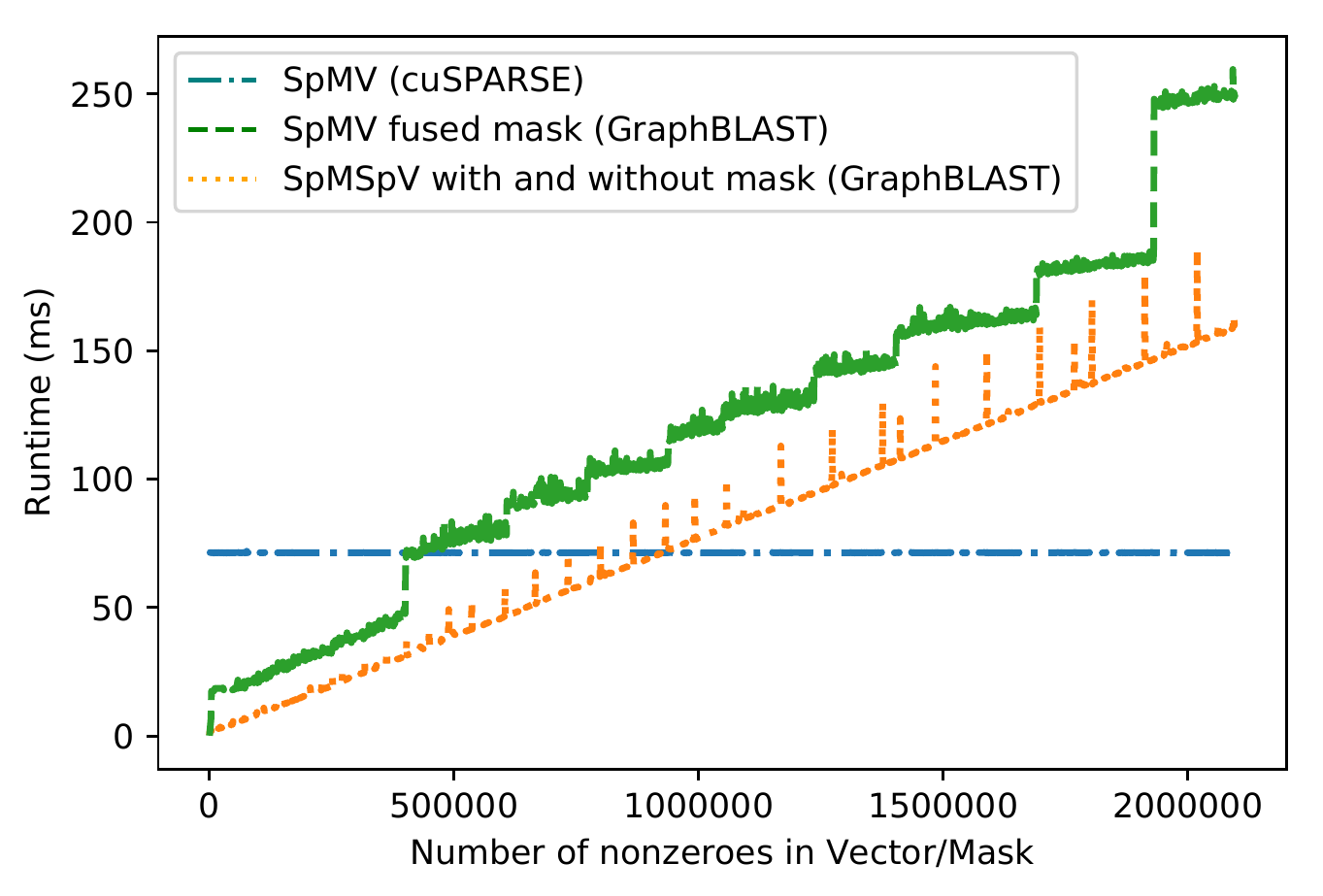}
    \caption{Algorithmic complexity as a function of vector and mask sparsity: (i)~SpMV without mask, (ii)~SpMV with mask, (iii)~SpMSpV without mask, (iv)~SpMSpV with mask. Note: Since SpMSpV with and without mask produced the same runtime, we lump them as ``SpMSpV with and without mask''.\label{fig:mask-complexity}}
  \end{subfigure}
  \hfill
  \begin{subfigure}[t]{0.45\textwidth}
    \centering
    \includegraphics[width=1\textwidth]{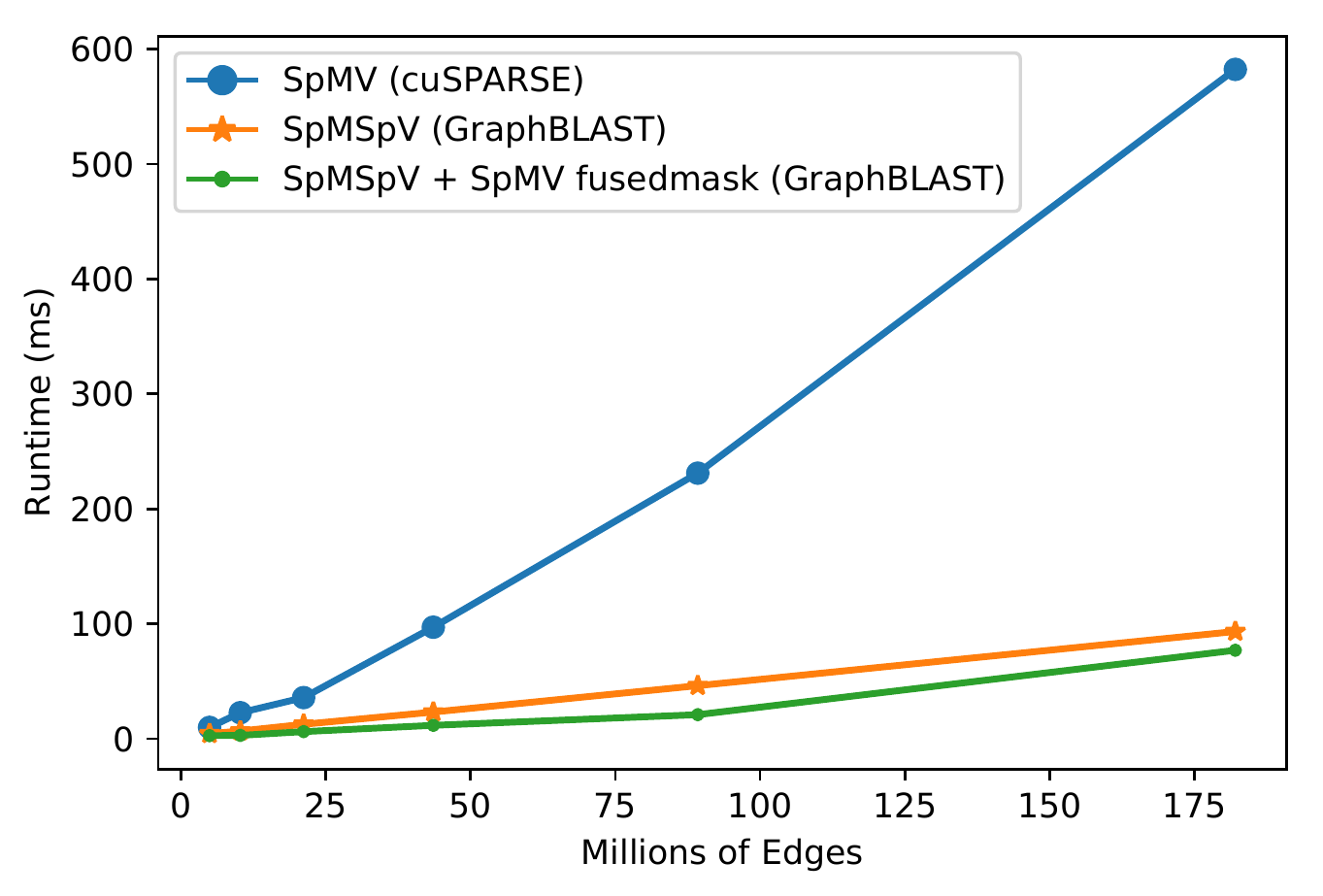}
    \caption{SpMV and SpMSpV runtime on Kronecker scale-\{16--21\} graphs.\label{fig:vs_nomask}}
  \end{subfigure}
  \caption{Comparison with and without fused mask.\label{fig:maskvsnomask}}
\end{figure}

As our microbenchmark in Figure~\ref{fig:maskvsnomask} illustrates, the masked SpMV variant of \cmd{graphblas::mxv} scales with the sparsity of the mask \cmd{Vector}. However, the masked SpMSpV is unchanged from the unmasked SpMSpV\@. This too matches the theoretical complexity shown in Table~\ref{tab:complexity-journal}, which shows that masked SpMV scales with $O(d\text{ } nnz(\mathbf{m}))$, where $\mathbf{m}$ is the mask \cmd{Vector}. However in our implementation, masked SpMSpV only performs the elementwise multiply with the mask after the SpMSpV operation, so it is unable to benefit from the mask's sparsity. After the columns of the sparse matrix are loaded, it may be possible to use binary search into the mask vector to reduce the number of operations required for the multiway merge, but this still does not asymptotically reduce the number of \revision{load and store instructions} into the sparse matrix as in the masked SpMV case.

In the running example, recall in Figure~\ref{fig:spmv3old} that standard SpMV, which performs the matrix-vector multiply followed by the elementwise product with the mask, took 20 \revision{load and store instructions}. However, when we reverse the sequence of operations by first loading the mask, seeing which elements of the mask are nonzero, and then only doing the matrix-vector multiply for those rows that map to a nonzero mask element, we see that the number of \revision{loads and stores} drops significantly from 20 down to 10.

\subsection{Masking insights}

One simple implementation of masking is to perform the matrix multiplication, and then apply the mask to the output. This approach has the benefit of being straightforward and easy to implement. However, we identify two scenarios in which accessing the mask ahead of the matrix multiplication is beneficial:

\begin{table}
  \small
  \centering
  \setlength{\tabcolsep}{3pt}
  \rowcolors{2}{lightgray}{white}
  \begin{tabular}{*{8}{c}} \toprule
    & \multicolumn{2}{c}{mxm first} && \multicolumn{2}{c}{mask first} & & \\
    \cmidrule{2-3}\cmidrule{5-6}
    \rowcolor{white}
    Dataset & Nonzeroes & Runtime (s) && Nonzeroes & Runtime (s) & Memory savings & Speedup \\
    \midrule
    coAuthorsCiteseer & 2.03M & 458.3 && 814K & 5.96 & 2.49$\times$ & 76.9$\times$ \\
    coPapersDBLP & 81.3M & 3869 && 15.2M & 78.66 & 5.35$\times$ & 13.2$\times$ \\
    road\_central & 29.0M & 3254  && 16.9M & 246.4 & 1.72$\times$ & 49.2$\times$ \\ \bottomrule
  \end{tabular}
  \caption[]{Runtime in milliseconds and speedup of accessing mask before \cmd{mxm} and after \cmd{mxm} on three datasets. Nonzeroes means how many nonzero elements are in the output of the \cmd{mxm}.\label{tab:maskvsmxm}}
\end{table}

\begin{enumerate}
  \item Masked \cmd{mxv}: As Figure~\ref{fig:maskvsnomask} illustrates, the masked SpMV is advantageous and to be preferred when the input vector nonzero count surpasses some threshold. Table~\ref{tab:switchpoint} is a good starting point at finding the optimal threshold for given hardware.
  \item Masked \cmd{mxm}: Table~\ref{tab:maskvsmxm} shows two benefits of accessing the mask before doing the \cmd{mxm}. The first benefit is lower memory consumption. Typically, \cmd{mxm} generates an order of magnitude more nonzeroes in the output matrix compared with the two input matrices, which in the absence of kernel fusion must be saved and typically causes out-of-memory errors. By accessing the mask first, this order of magnitude blow-up in nonzeroes can be avoided. Using the mask as an oracle, the mask yields an upper bound in where nonzeroes can be generated. Therefore, an order of magnitude less computation can be done by accessing the mask to determine nonzeroes $i, j$ s.t.\ $\mathbf{M}(i,j) \neq 0$, then loading only $\mathbf{A}(i,:)$ and $\mathbf{B}(:,j)$, performing the dot product between the two, and writing the result to $\mathbf{C}(i,j)$. Therefore, the second benefit is from avoiding computation.
\end{enumerate}

\section{GPU Implementation}
\label{sec:loadbalancing}
In this section, we discuss implementation details specific to the GPU\@.

\subsection{Memory management}
\revision{
  GBTL uses the Thrust template library to provide wrappers around GPU memory that automatically handle CPU-to-GPU and GPU-to-CPU communication. We instead decided to manually manage GPU pointers ourselves. This offers graph algorithm developers (i.e., the users) the same seamless experience of not having to concern themselves with whether the data is on the CPU or the GPU\@.

  For the concrete implementation of each object---SparseVector, DenseVector, SparseMatrix, DenseMatrix, etc.---we keep both a canonical GPU copy and a CPU copy that is allowed to go out-of-date. Upon initialization, we maintain the canonical copy on the GPU at all times and use a flag to keep track whether or not the CPU version differs from the canonical GPU copy and is stale. When operations mutating the GPU copy are run, this flag will get set to \cmd{true} indicating the CPU copy is now stale. For operations that interact with the world external to GraphBLAS such as \cmd{extract} and \cmd{extractTuples}, we will copy data back to the CPU depending on whether the flag tells us the CPU copy is stale or not. If it is not stale, we can return the CPU copy directly. If it is stale, we will copy data back to the CPU and reset the flag \cmd{false} indicating the GPU and CPU copy have equal values.

  On GPUs, memory allocation time can be a significant fraction of runtime. Since some operations require temporary memory, we keep a memory pool of already-allocated GPU memory. Currently, we associate this memory pool with the \cmd{Descriptor} object, because we do not find we often require more than one \cmd{Descriptor}. In the future, we will consider changing this to use the Factory pattern~\cite{Kircher:2002:PPM}, which is considered standard design for memory pools.
}

\subsection{Operators}
\label{subsec:operators}
\revision{
  One of the biggest challenges of implementing GraphBLAS is solving the problem of supporting the large cross product of possible functionalities. Consider, for example, \cmd{Semiring}. This operator needs to support 11 built-in GraphBLAS types (and any user-defined type), 22 built-in binary operations, and 8 built-in monoid operations. Even without user-defined types, this comes to a total of 1459 operators for the 3 GraphBLAS methods that take a \cmd{Semiring}, \cmd{mxv}, \cmd{vxm}, and \cmd{mxm}.

  In the literature~\cite{Davis:2018:SGG, Brock:2020:ARF, Zhang:2016:GCG}, two ways have been proposed to tackle this problem:

  \begin{enumerate}
    \item Using code generation tools and macros.
    \item Using C++ templates and functors. This is the approach taken by GBTL, which we adopt here as well.
  \end{enumerate}

  The first method is the approach taken by SuiteSparse~\cite{Davis:2018:SGG}. This has the advantage of being a shared object library, which can be linked to using frontends written in interpreted languages. We instead adopt the second method, which has the advantage of not needing to maintain code generation tools and macros, which can be challenging.

  To express monoids and semirings, we use \cmd{\_\_host\_\_} and \cmd{\_\_device\_\_} functors that overload the function call operator (i.e. \cmd{operator()}). We use a macro that constructs structs composed of one and two of these functors for monoids and semirings respectively. The macro also takes an identity-element input.

  \subsection{Implementation of key primitives and load balancing}
  What follows are the implementation details and load balancing techniques behind the four key primitives SpMSpV, SpMV, SpMM and masked SpGEMM\@. Load balancing attempts to distribute work equally across the GPU's processors (threads, warps and blocks). To motivate the need for load balance, consider an implementation that assigns each matrix row to a different processor. Because the number of nonzeroes per row may vary greatly, the amount of work per processor may also vary greatly, leading to inefficient execution. We use the following strategies to implement load-balanced kernels:

  \begin{enumerate}
    \item Multiple kernel approach: SpMSpV
    \item Merge-based: SpMV
    \item Merge-based and Row split: SpMM
    \item Row split: masked SpGEMM
  \end{enumerate}
}

\subsubsection{SpMSpV}
\label{subsubsec:spmspv}
\revision{
  Based on our earlier work~\cite{Yang:2015:FSM,Yang:2018:IPE}, our current SpMSpV implementation is composed of several steps. It heavily relies on scan primitives and in particular the \cmd{IntervalExpand} and \cmd{IntervalGather} operations of ModernGPU~\cite{Baxter:2015:MGL}.
  Recall that in SpMSpV, each column $i$ is multiplied by $\mX(i)$ if and only if $\mX(i)\neq 0$. This operation costs $\flops(\mA,\mX) = \sum_{i | \mX(i) \neq 0}{\dnnz(\mA(:,i))}$. \cmd{IntervalExpand} creates a temporary vector of length $\flops(\mA,\mX)$. This allows data-parallel multiplication of all the nonzeros that will contribute to the final output with their corresponding vector values. Note that the extra memory consumption for this temporary workspace is smaller than the input matrix size (unless the vector is completely dense, in which case we would be calling SpMV instead of SpMSpV). These intermediate values are then sorted using RadixSort in linear time to bring identical indices next to each other. Finally, a segmented reduction (ModernGPU's \cmd{ReduceByKey}) creates the desired output vector.
  Thus, SpMSpV has $O(\flops(\mA,\mX))$ work and $O(\lg(\flops(\mA,\mX)))$ depth.}

\begin{figure}
  \centering
  \captionsetup[subfigure]{justification=centering}
  \begin{subfigure}[t]{0.3\textwidth}
        \centering
      \includegraphics[width=\textwidth]{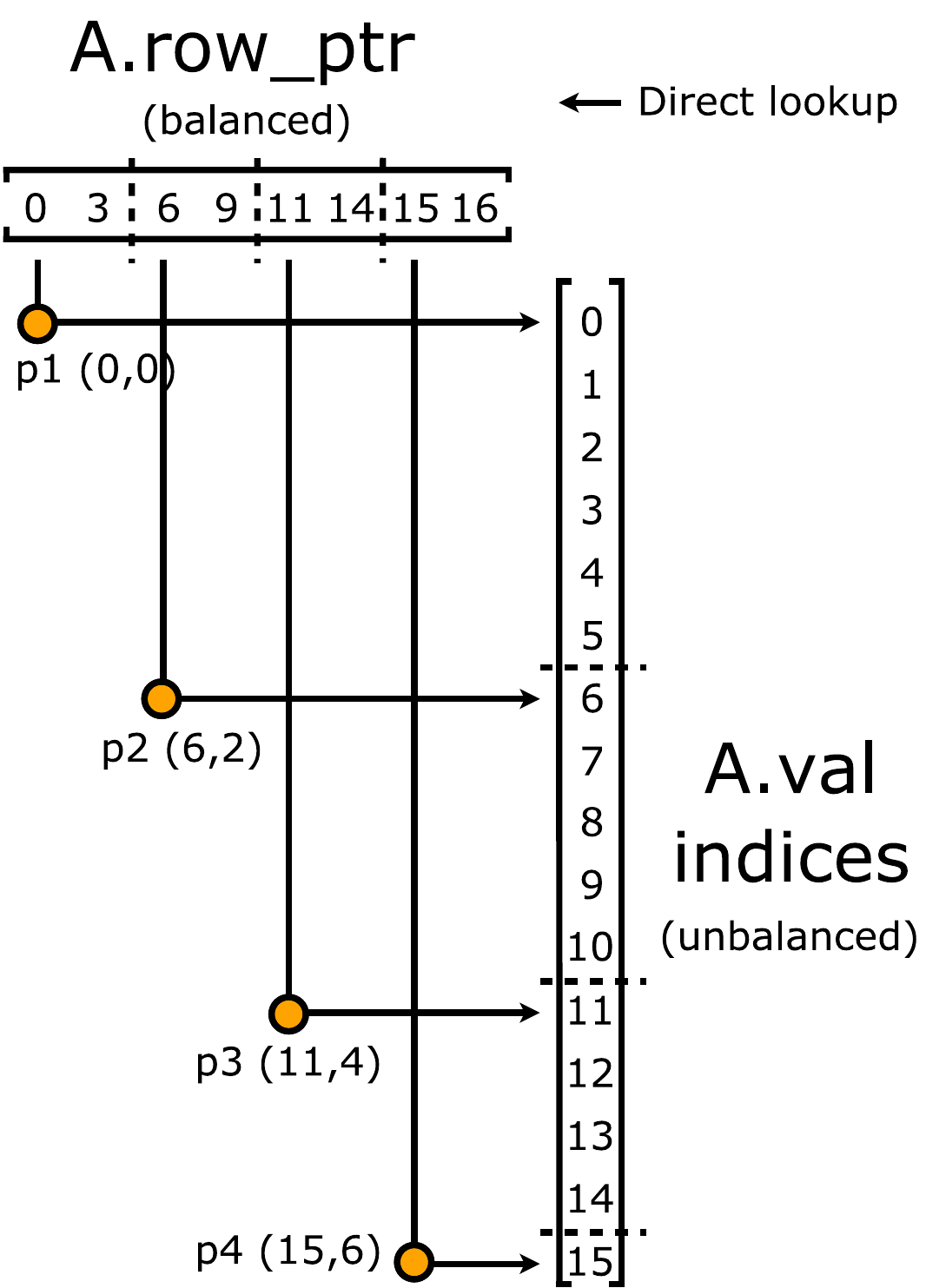}
      \caption{Row split\label{fig:rowsplit-dia}}
  \end{subfigure}
  \hfill
  \begin{subfigure}[t]{0.3\textwidth}
        \centering
      \includegraphics[width=\textwidth]{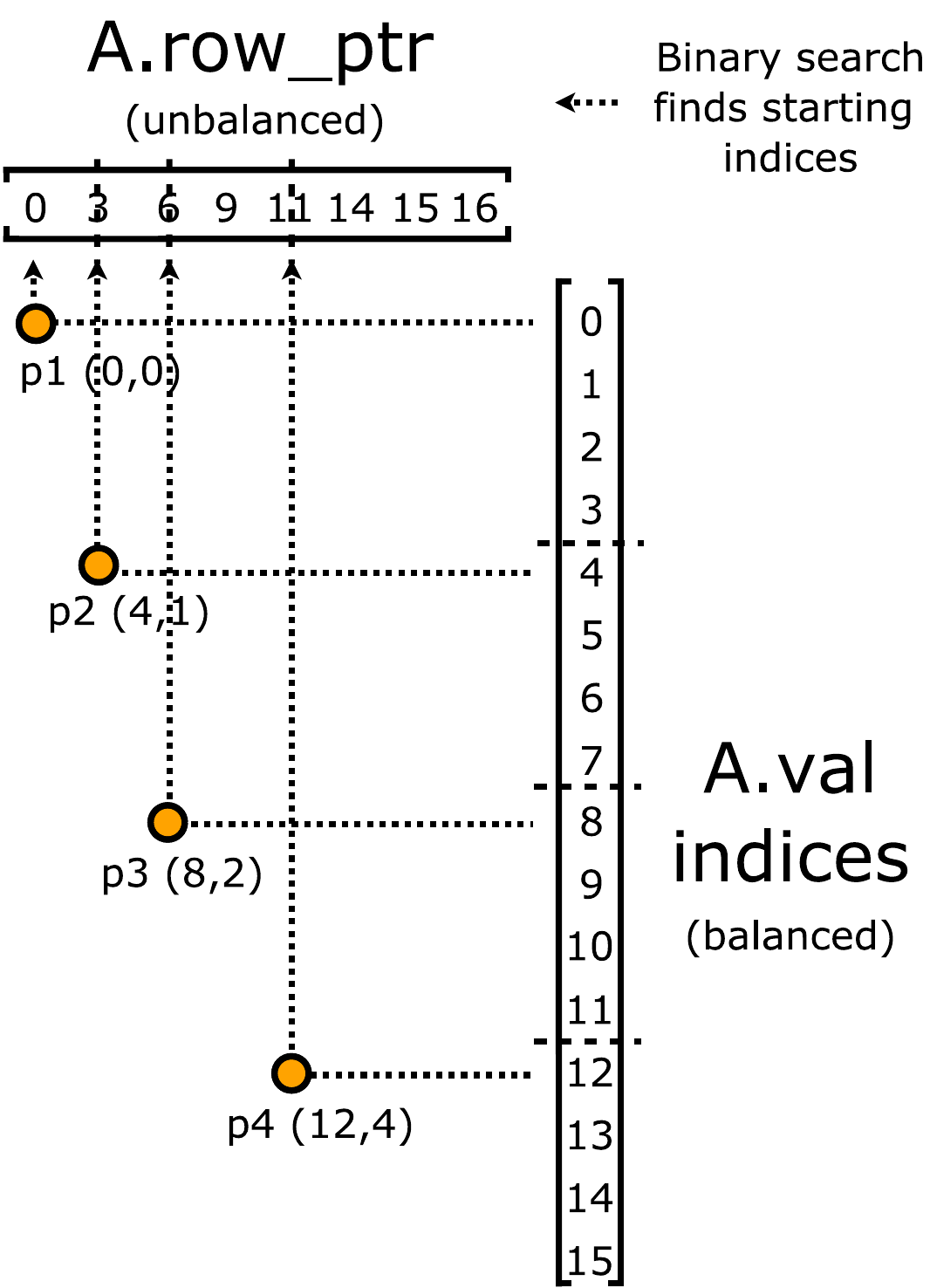}
      \caption{Merge-based\\ (nonzero split)\label{fig:nnzsplit-dia}}
  \end{subfigure}
  \hfill
  \begin{subfigure}[t]{0.3\textwidth}
        \centering
      \includegraphics[width=\textwidth]{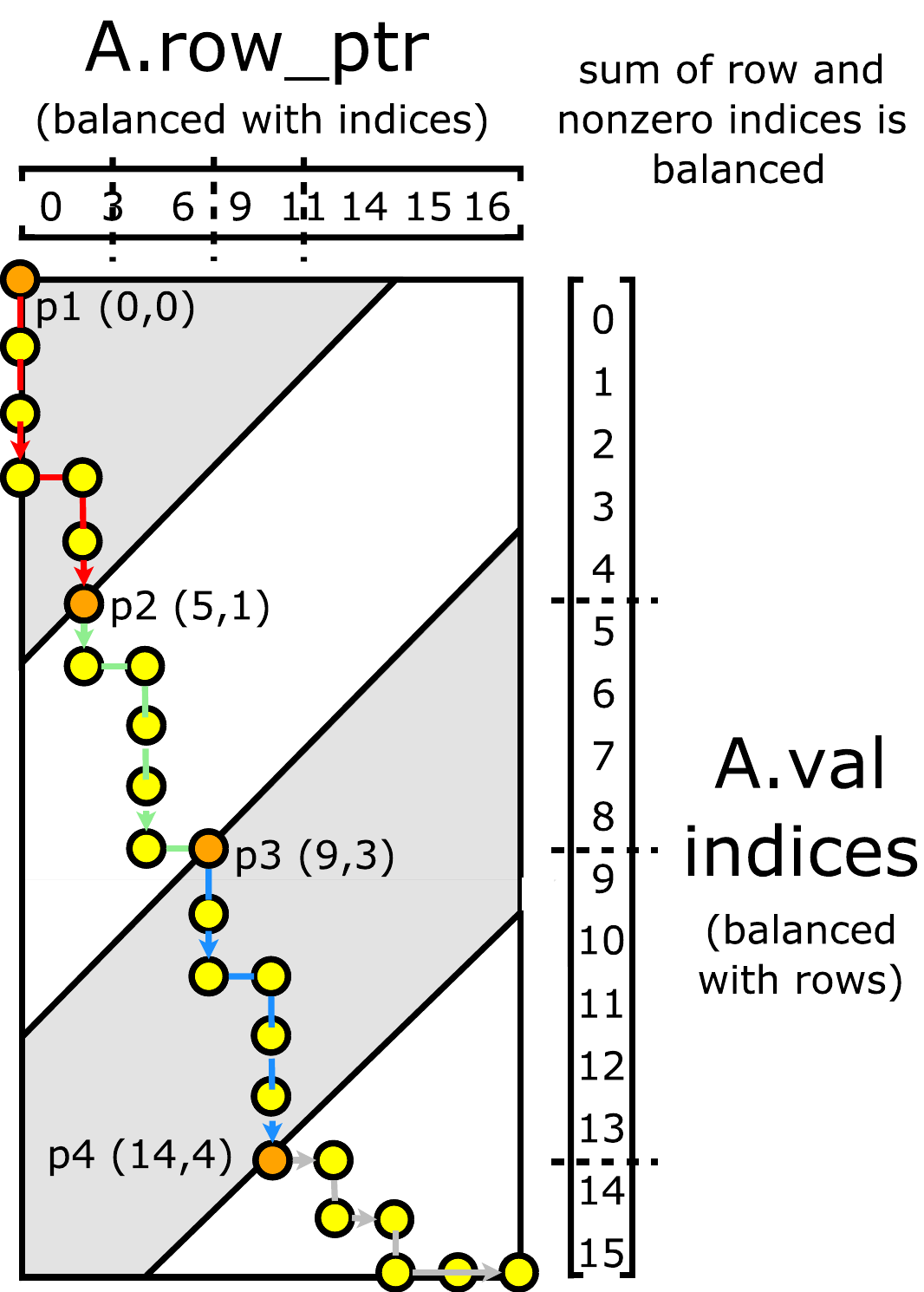}
      \caption{Merge-based\\(merge path)\label{fig:mergepath-dia}}
  \end{subfigure}
  \caption{The three parallelizations for CSR SpMV and SpMM on matrix $\mA$\@. The orange markers indicate the segment start for each processor ($P = 4$).\label{fig:loadbalance}}
\end{figure}

\subsubsection{SpMV}
\label{subsubsec:spmv}
\revision{
  For SpMV, we use the nonzero-split implementation of the merge-based algorithm provided by ModernGPU~\cite{Baxter:2015:MGL}. In benchmarking, we find that performance is similar to the merge path algorithm by Merrill and Garland~\cite{Merrill:2016:MPS}. The difference between the two algorithms is shown in Figure~\ref{fig:loadbalance} and may be described as follows:

  \begin{enumerate}
    \item Row split~\cite{Bell:2009:ISM}: Assigns an equal number of rows to each processor.
    \item Merge-based: Performs two-phase decomposition---the first kernel divides work evenly amongst CTAs, then the second kernel processes the work.
    \begin{enumerate}
      \item Nonzero-split~\cite{Dalton:2015:OSM,Baxter:2015:MGL}: Assign an equal number of nonzeros per processor. Then do a 1-D (1-dimensional) binary search on \emph{row offsets} to determine at which row to start.
      \item Merge path~\cite{Merrill:2016:MPS}: Assign an equal number of \{nonzeros and rows\} per processor. This is done by doing a 2-D binary search (i.e., on the diagonal line in Figure~\ref{fig:mergepath-dia}) over \emph{row offsets} and \emph{nonzero indices} of matrix $\mA$.
    \end{enumerate}
  \end{enumerate}

  The merge path algorithm has the advantage of doing well in the pathological case of arbitrarily many empty matrix rows, which can cause an arbitrarily large amount of load imbalance for ModernGPU's nonzero-split. However, we find that this does not happen often in practice. Hence, GraphBLAST currently uses ModernGPU's SpMV implementation that relies on the segmented-scan primitive. In theory, segmented-scan-based SpMV has depth logarithmic on the number of nonzeros involved~\cite{blelloch_prefix_1990}, and it can be implemented on a GPU in a work-efficient way~\cite{scanprimitives2007}. While in practice one has to choose a block size on the GPU, Sengupta et al.~\cite{sengupta2008efficient} showed that this does not increase work and depth of the segmented scan implementation asymptotically.

  We show the results of a microbenchmark comparing the merge-based algorithm against cuSPARSE's implementation of SpMV
  in Figure~\ref{fig:spmv-compare}, which we suspect relies on the row split algorithm. The experimental setup is described in Section~\ref{sec:results}. The right side of the $x$-axis represents load imbalance where long matrix rows are not divided enough, resulting in some computation resources on the GPU remaining idle while others are overburdened. The left size of the $x$-axis represents load imbalance where too many computational resources are allocated to each row, so some remain idle. From this figure, it is clear that the merge-based algorithm is superior to the row split algorithm at addressing these two types of load imbalance, despite there being two configurations 512 and 2048 for which row split is faster.
}

\subsubsection{SpMM}
\label{subsubsec:spmm}
\begin{figure}
  \centering
  \captionsetup[subfigure]{justification=centering}
  \begin{subfigure}[t]{0.45\textwidth}
        \centering
      \includegraphics[width=\textwidth]{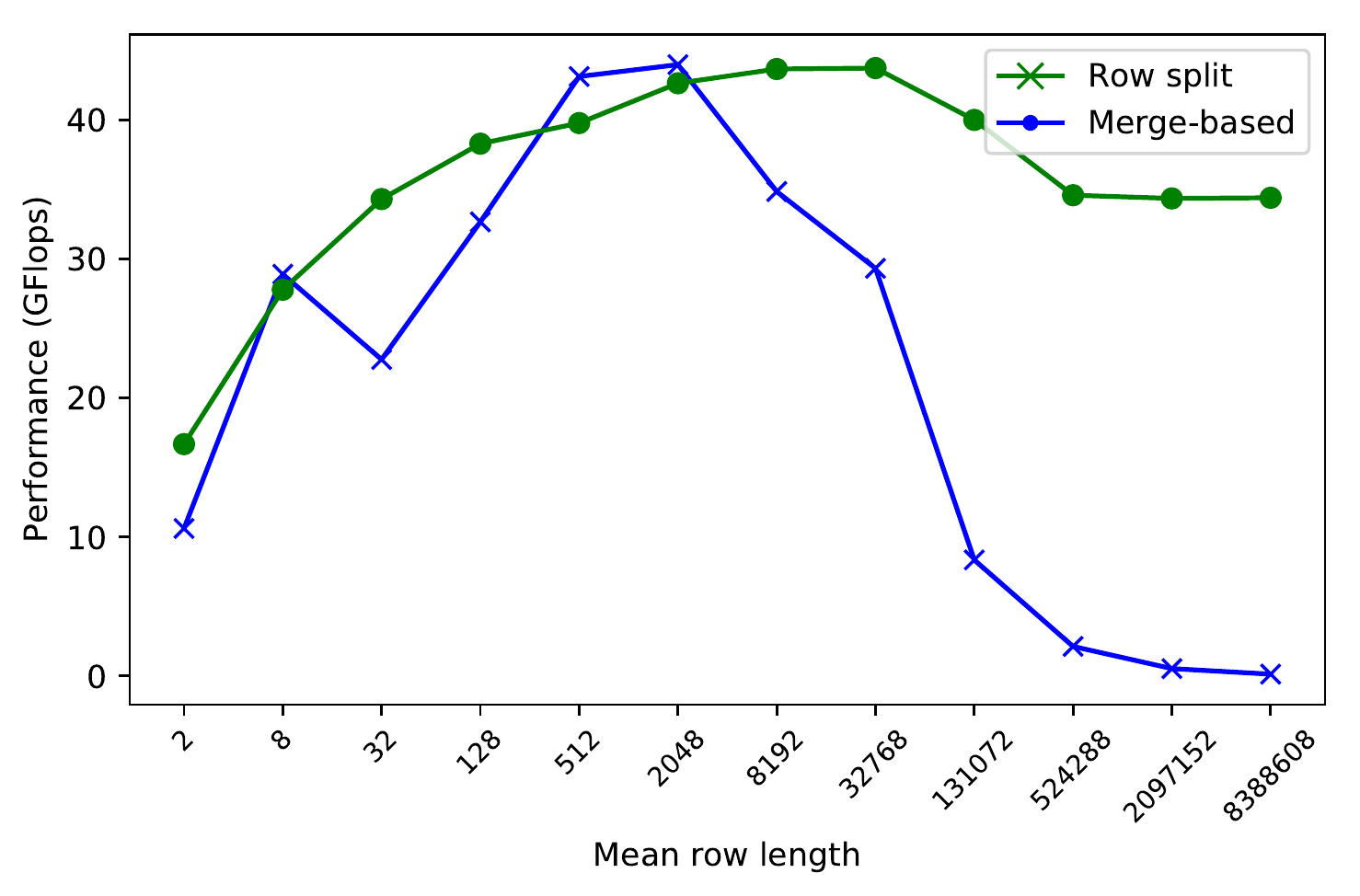}
      \caption{SpMV\label{fig:spmv-compare}}
  \end{subfigure}
  \hfill
  \begin{subfigure}[t]{0.45\textwidth}
        \centering
      \includegraphics[width=\textwidth]{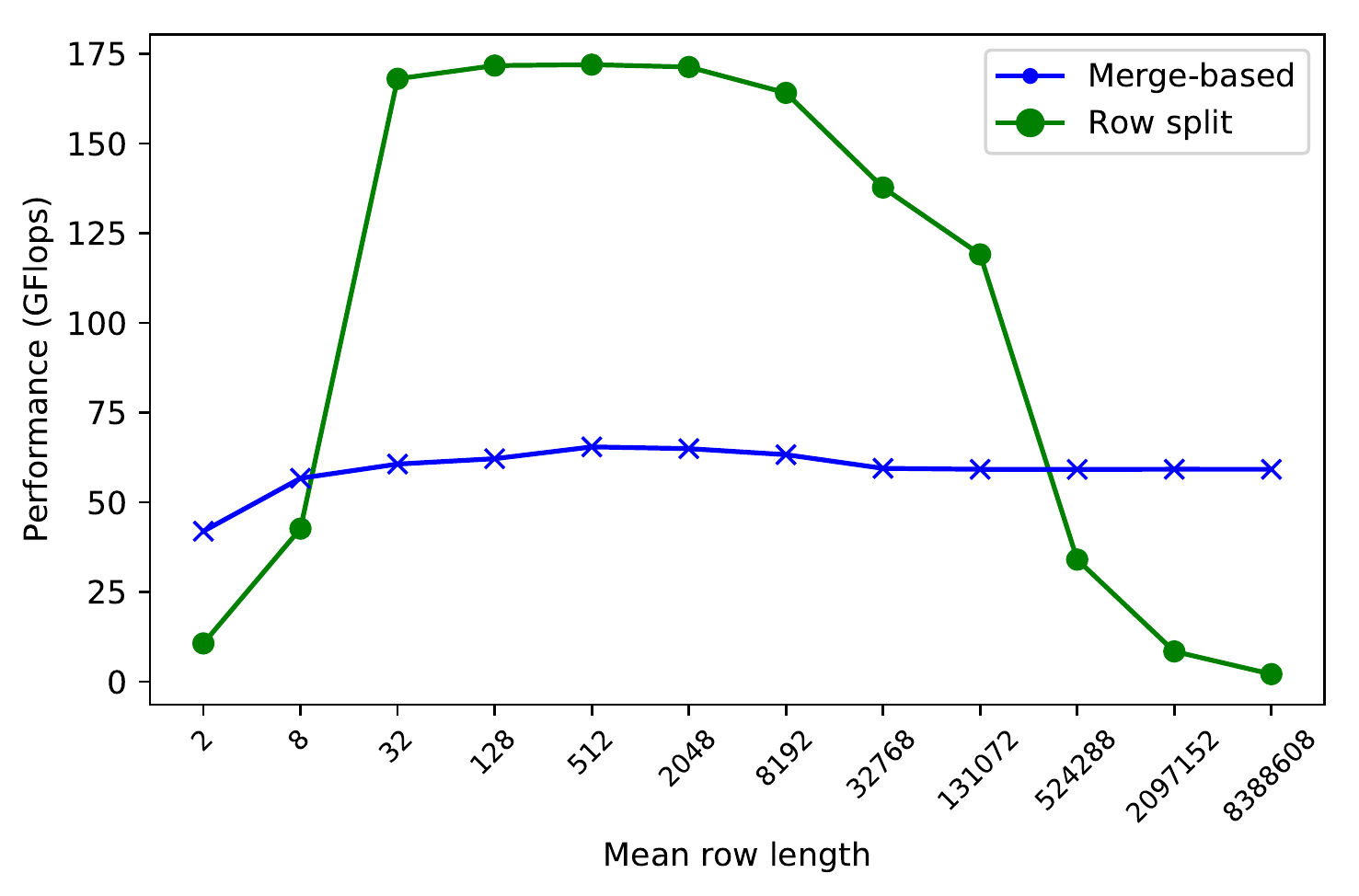}
      \caption{SpMM\label{fig:spmm-compare}}
  \end{subfigure}
  \caption{Microbenchmark showing performance of the merge-based algorithm compared against the row-split based algorithm for SpMV and SpMM\@.\label{fig:spmvspmm}}
\end{figure}

\revision{
  For SpMM, we have in our previous work~\cite{Yang:2018:DPF} extended the approach taken by SpMV\@. As
  Figure~\ref{fig:spmm-compare} shows, our conclusions from earlier work show that while merge-based SpMM does help with solving the two types of load balance as in SpMV, we have identified problems scaling the algorithm when there are many columns in the right-hand-side matrix. Therefore, based on this benchmark and experimentation using 157 SuiteSparse matrices, we developed a multi-algorithm:

  \begin{enumerate}
    \item When $\dnnz/M < 9.35$: Use merge-based algorithm.
    \item When $\dnnz/M \ge 9.35$: Use row split algorithm.
  \end{enumerate}

  \noindent
  The reason that this heuristic does not capture the right side of Figure~\ref{fig:spmm-compare}, where merge-based is superior, is that in practice we did not encounter any matrices that had a mean row length of greater than 524,288.
}

\subsubsection{Masked SpGEMM}
\label{subsubsec:maskedspgemm}
\revision{
  In our implementation, we use a generalization of this primitive where we assume we are solving the problem for three distinct matrices $\mC = \mA \mB\,.\!* \mathbf{M}$. We use a straightforward row split where we assign a warp per row of the mask $\mathbf{M}$, and for every nonzero $\mathbf{M}(i, j)$ in the mask, each warp loads the row $\mA(i,:)$ in order to perform the dot product $\mA(i,:)\mB(:,j)$. Using their $\mA$-elements, each thread in the warp performs binary search on column $\mB(:, j)$ and accumulates the result of the multiplication. After the row is finished, we perform a warp reduction and write the output to $\mC(i,j)$.

  We also experimented with first computing the SpGEMM and then applying the mask. However, as Table~\ref{tab:maskvsmxm} shows, our direct Masked SpGEMM implementation was 13--79$\times$ faster and used significantly less memory.
}

\section{Graph Algorithms}
\label{sec:app}
\begin{figure}[t]
        \centering
        \begin{subfigure}[t]{0.4\textwidth}
                \centering
                \includegraphics[width=1\linewidth]{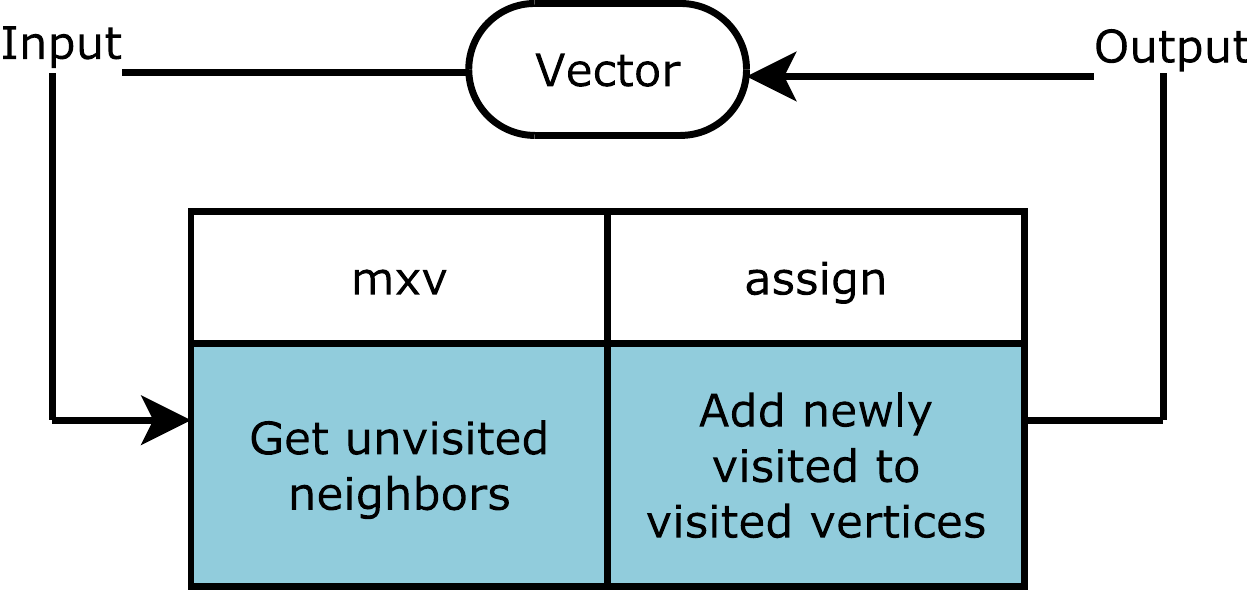}
                \caption{BFS\label{fig:app-bfs}}
                \vspace{0.5em}
        \end{subfigure}
        \hfill
        \begin{subfigure}[t]{0.4\textwidth}
                \centering
                \includegraphics[width=1\textwidth]{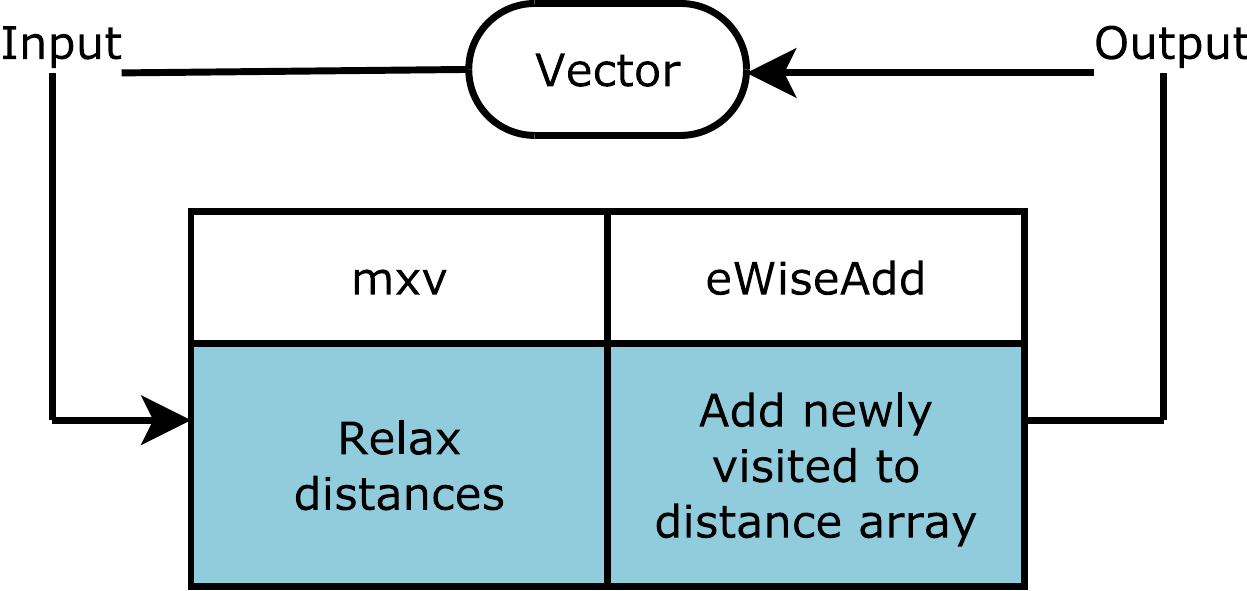}
                \caption{SSSP (Bellman-Ford)\label{fig:app-bf}}
                \vspace{0.5em}
        \end{subfigure}
        \begin{subfigure}[t]{0.4\textwidth}
                \centering
                \includegraphics[width=1\linewidth]{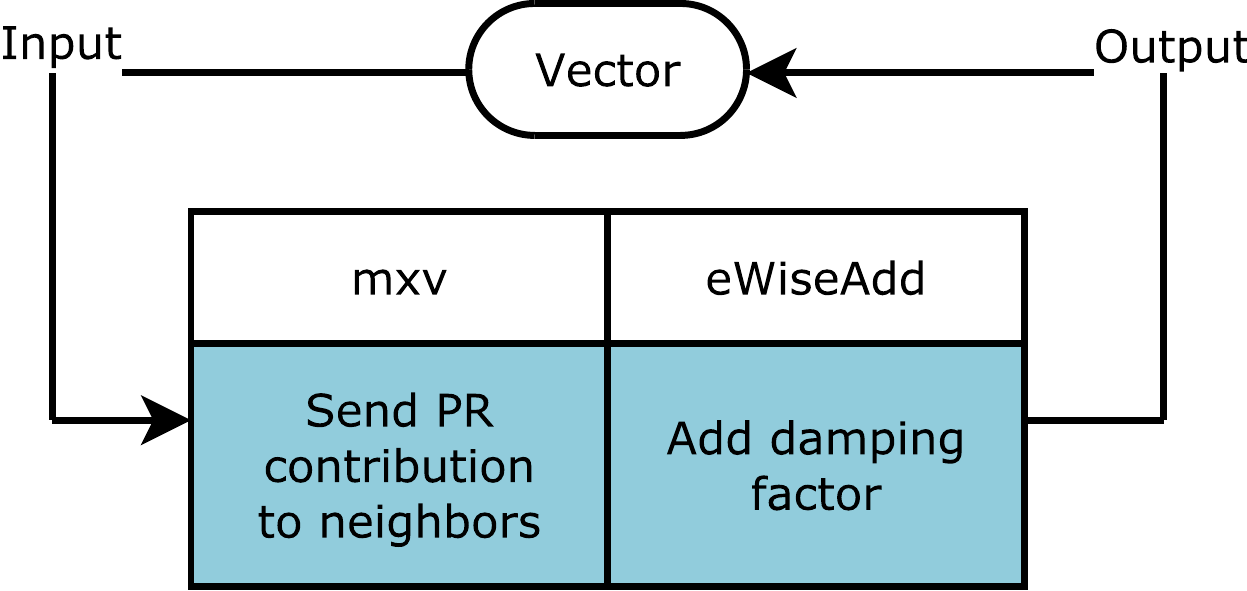}
                \caption{PR\label{fig:app-pr}}
        \end{subfigure}
        \hfill
        \begin{subfigure}[t]{0.4\textwidth}
                \centering
                \includegraphics[width=1\textwidth]{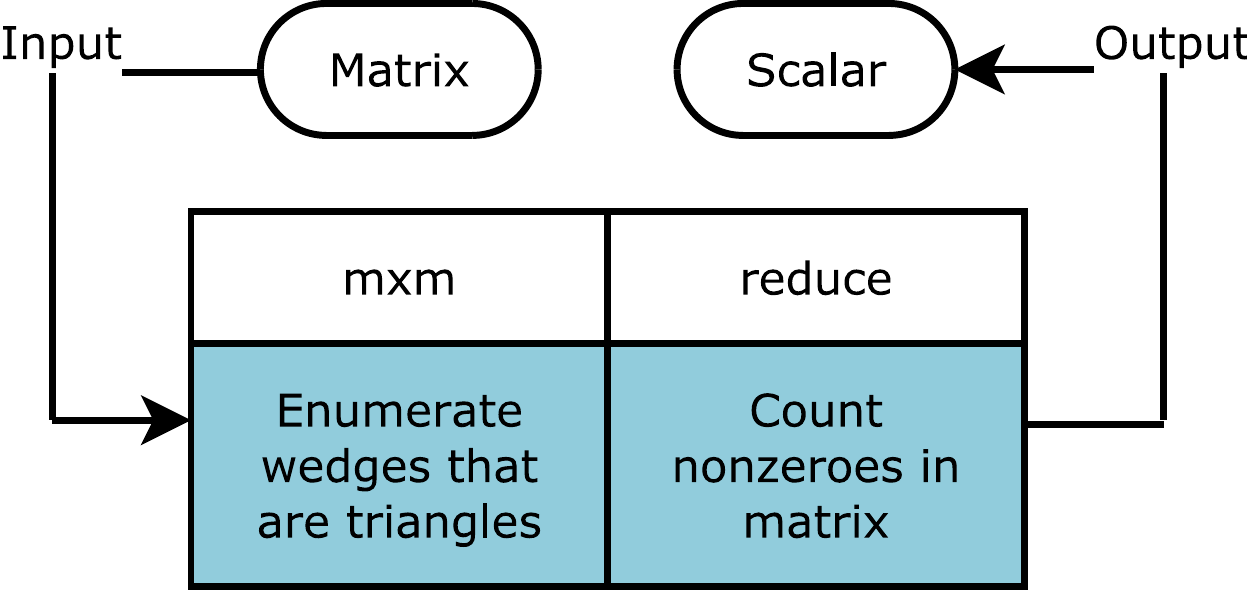}
                \caption{TC\label{fig:app-tc}}
        \end{subfigure}
        \begin{subfigure}[t]{0.72\textwidth}
                \centering
                \includegraphics[width=1\textwidth]{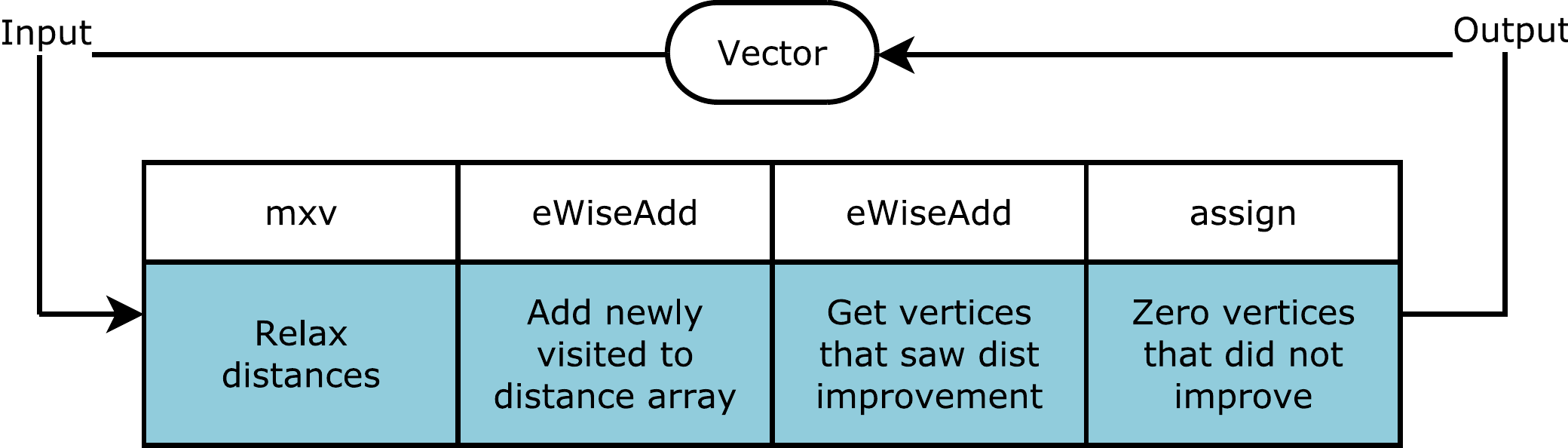}
                \caption{SSSP (Bellman-Ford with sparsification)\label{fig:app-sssp}}
        \end{subfigure}
        \begin{subfigure}[t]{\textwidth}
                \centering
                \includegraphics[width=1\textwidth]{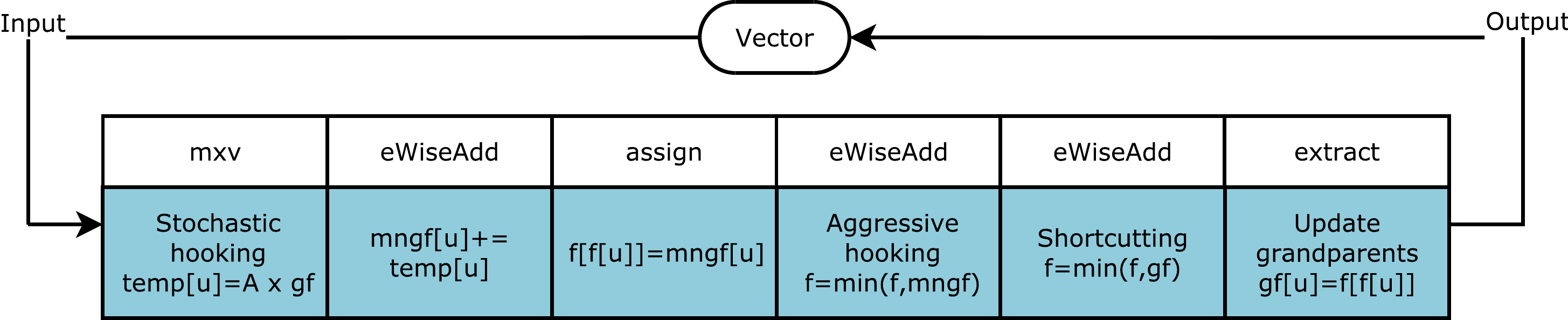}
                \caption{CC (FastSV)\label{fig:app-cc}}
        \end{subfigure}
        \caption{Operation flowchart for different algorithms expressed in GraphBLAS\@. A loop indicates a while-loop that runs until the Vector is empty.\label{fig:app}}
\end{figure}

One of the main advantages of GraphBLAS is that its operations can be composed to develop new graph algorithms. For each graph algorithm in this section, we describe the hardwired GPU implementation of that algorithm and how our implementation can be expressed using GraphBLAS\@. Then the next section will compare performance between hardwired and GraphBLAS implementations. Figure~\ref{fig:app} shows the GraphBLAS algorithms required to implement each algorithm.

We chose the five graph algorithms BFS, SSSP, PR, CC, and TC\@. Based on Beamer's thorough survey of graph processing frameworks in his Ph.D.\ dissertation~\cite{Beamer:2016:UIG}, they represent all five of the most commonly evaluated graph algorithms. In addition, they stress different components of graph frameworks. BFS stresses the importance of masking and being able to quickly filter out nonzeros that don't have an associated value. SSSP stresses masking and being able to run SpMV on nonzeros with an associated value representing distance. PR stresses having a well-load-balanced SpMV\@. CC tests expressibility and random memory accesses from hooking and pointer-jumping. TC stresses having a masked SpGEMM implementation.

\subsection{Breadth-first-search}
Given a source vertex $s \in V$, a BFS is a full exploration of a graph $G$ that produces a spanning tree of the graph, containing all the edges that can be reached from $s$, and the shortest path from $s$ to each one of them. We define the depth of a vertex as the number of hops it takes to reach this vertex from the root in the spanning tree. The visit proceeds in steps, examining one BFS level at a time. It uses three sets of vertices to keep track of the state of the visit: the \emph{frontier} contains the vertices that are being explored at the current depth, \emph{next} has the vertices that can be reached from \emph{frontier}, and \emph{visited} has the vertices reached so far. BFS is one of the most fundamental graph algorithms and serves as the basis of several other graph algorithms.

\begin{description}
  \item[Hardwired GPU implementation] The best-known BFS implementation of Merrill et al.~\cite{Merrill:2012:SGG} achieves its high performance through careful load-balancing, avoidance of atomics, and heuristics for avoiding redundant vertex discovery. Its chief operations are expand (to generate a new frontier) and contract (to remove redundant vertices) phases. Enterprise~\cite{Liu:2015:EBG}, a GPU-based BFS system, introduces a very efficient implementation that combines the benefits of the direction optimization of Beamer, Asanovi\'{c} and Patterson~\cite{Beamer:2012:DOB}, leverages the adaptive load-balancing workload mapping strategy of Merrill et al., and  chooses to not synchronize each BFS iteration, which addresses the kernel launch overhead problem (Section~\ref{sec:kernel-launch}).
  \item[GraphBLAST implementation] Merrill et al.'s expand and contract maps nicely to GraphBLAST's \cmd{mxv} operator with a mask using a Boolean semiring. Like Enterprise, we implement efficient load-balancing (Section~\ref{sec:loadbalancing}) and direction optimization, which was described in greater detail in Section~\ref{sec:diropt}. We do not use Enterprise's method of skipping synchronization between BFS iterations, but we use two optimizations: \emph{early-exit} and \emph{structure-only}, which are consequences of the Boolean semiring that is associated with BFS\@. We also use \emph{operand reuse}, which avoids having to convert from sparse to dense during direction optimization. These optimizations are inspired by Gunrock and are described in detail by the authors in an earlier work~\cite{Yang:2018:IPE}.
\revision{If our implementation were to always choose the top-down direction, it would have work complexity of $O(\dnnz(\mA)) = O(\vert E \vert)$ and depth $O(D \cdot \log{d_\text{max}})$ where $D$ is the graph diameter and $d_\text{max}$ is the maximum vertex degree.
This bound follows from the SpMSpV complexity described in Section~\ref{subsubsec:spmspv}. We are not aware of a worst-case analysis of the direction-optimized search but most switching heuristics are conservative and only switch to the bottom-up direction when it decreases work. Given that analysis in Section~\ref{subsubsec:spmv} shows that masked-SpMV depth is no larger than SpMSpV depth, we conclude that our implementation has worst-case work $O(\dnnz(\mA)) = O(\vert E \vert))$ and depth $O(D \cdot \log d_\text{max})$.}
\end{description}

\subsection{Single-source shortest-path}
Given a source vertex $s \in V$, a SSSP is a full exploration of weighted graph $G$ that produces a distance array of all vertices $v$ reachable from $s$, representing paths from $s$ to each $v$ such that the path distances are minimized.

\begin{description}
  \item[Hardwired GPU implementation] Currently the highest-performing
  SSSP algorithm implementation on the GPU is the work from Davidson
  et al.~\cite{Davidson:2014:WPG}. They provide two key
  optimizations in their SSSP implementation: (1)~a load balanced graph
  traversal method, and (2)~a priority queue implementation that
  reorganizes the workload.
  \item[GraphBLAST implementation] We take a different approach from Davidson et al.\ to solve SSSP\@. We show that our approach both avoids the need for ad hoc data structures such as priority queues and wins in performance. The optimizations we use are: (1)~\emph{generalized direction optimization}, which is handled automatically within the \cmd{mxv} operation rather than inside the user's graph algorithm code, and (2)~sparsifying the set of active vertices after each iteration by comparing each active vertex to see whether or not it improved over the stored distance in the distance array. The second phase introduces two additional steps (compare Figures~\ref{fig:app-bf} and~\ref{fig:app-sssp}). \revision{These facts make our SSSP implementation an adaptive variant of Bellman-Ford. While the worst-case work complexity is $O(\lvert E \rvert \, \lvert V \rvert))$, it is much faster in practice due to (a)~convergence being achieved significantly before $\vert V \vert$ iterations, and (b)~direction optimization. Each iteration has only $O(\log{d_\text{max}})$ depth.}
\end{description}

\subsection{PageRank}

The PageRank link analysis algorithm assigns a numerical weighting to each element of a hyperlinked set of documents, such as the World Wide Web, with the purpose of quantifying its relative importance within the set. The iterative method of computing PageRank gives each vertex an initial PageRank value and updates it based on the PageRank of its neighbors\revision{~(this is the ``pull'' formulation of PageRank)}, until the PageRank value for each vertex converges. There are variants of the PageRank algorithm that stop computing PageRank for vertices that have converged already and also remove it from the set of active vertices. This is called adaptive PageRank~\cite{Kamwar:2004:AMC} (also known as PageRankDelta). In this paper, we do not implement or compare against this variant of PageRank.
\revision{We also acknowledge that different kinds of iterative solvers can be used for computing PageRank~\cite{Gleich:2004:FPP}.}

\begin{description}
  \item[Hardwired GPU implementation] One of the highest-performing implementations of PageRank is written by Khorasani, Vora, and Gupta~\cite{Khorasani:2014:CVG}. In their system, they use solve the load imbalance and GPU underutilization problem with a GPU adoption of GraphChi's Parallel Sliding Window scheme~\cite{Kyrola:2012:GLG}. They call this preprocessing step ``G-Shard'' and combine it with a concatenated window method to group edges from the same source IDs. We realize that due to G-Shard's preprocessing this comparison is not exactly fair to GraphBLAS, but include the comparison, because they are one of the leaders in PageRank performance and despite their preprocessing, our dynamic load-balancing is sufficient to make our implementation faster in the geomean (geometric mean).
  \item[Gunrock implementation] \revision{Gunrock supports both pull- and push-based PageRank; its push-based implementation is in general faster than its pull-based implementation. For a fair algorithmic comparison, we measure against Gunrock's pull-based implementation.}
  \item[GraphBLAST implementation] In GraphBLAST, we rely on the merge-based load-balancing scheme discussed in Section~\ref{sec:loadbalancing}. The advantage of the merge-based scheme is that unlike Khorasani, Vora, and Gupta, we do not need any specialized storage format; the GPU is efficient enough to do the load-balancing on the fly. In terms of exploiting input sparsity, we demonstrate that our system is intelligent enough to determine that we are doing repeated matrix-vector multiplication and because the vector does not get any sparser, it is more efficient to use SpMV rather than SpMSpV\@. \revision{As described in Section~\ref{subsubsec:spmv}, our SpMV implementation is based on ModernGPU~\cite{Baxter:2015:MGL}, which uses the segmented-scan primitive with logarithmic depth and linear work. Consequently, an iteration of PageRank has $O(\tilde{e})$  work and $O(\lg{\tilde{e}})$ depth, where $\tilde{e} = \sum_{i \vert \mathbf{m}(i) \neq 0} {\dnnz(\mA(i,:))}$ is the number of nonzeros that need to be touched in that iteration. Note that $\tilde{e}$ is at most $\dnnz(\mA)$ but is often smaller due to already converged vertices.}

\end{description}

\revision{\subsection{Connected components}
Weakly connected components (abbreviated as connected components) is the problem of: (1)~identifying all subgraphs (or components) in an undirected graph such that every pair of vertices in the subgraph are connected by edges and no edges connect vertices in different subgraphs, and (2)~labeling each vertex with its component ID\@.

\begin{description}
  \item[Hardwired GPU implementation] Soman, Kishore and Narayanan~\cite{Soman:2010:AFG} base their GPU implementation on two algorithms from the PRAM literature: hooking and pointer-jumping.  Hooking takes an edge as the input and tries to set the component IDs of the two end vertices of that edge to the same value. In odd-numbered iterations, the lower vertex writes its value to the higher vertex, and vice versa in the even-numbered iteration. This strategy is used to increase the rate of convergence over a more naive approach such as a breadth-first-search.  Pointer-jumping reduces a multi-level tree in the graph to a one-level tree (star).  By repeating these two operators until no component ID changes for any node in the graph, the algorithm will compute the number of connected components for the graph and the connected component to which each node belongs.
  \item[Gunrock implementation] Gunrock uses a filter operator on an edge frontier to implement hooking. The frontier starts with all edges and during each iteration, one end vertex of each edge in the frontier tries to assign its component ID to the other vertex, and the filter step removes the edge whose two end vertices have the same component ID\@.  We repeat hooking until no vertex's component ID changes and then proceed to pointer-jumping, where a filter operator on vertices assigns the component ID of each vertex to its parent's component ID until it reaches the root.
 Then a filter step removes the node whose component ID equals its own node ID\@. The pointer-jumping phase also ends when no vertex's component ID changes.
  \item[GraphBLAST implementation] GraphBLAST's implementation of CC is based on the FastSV algorithm~\cite{zhang2020fastsv}. FastSV is a linear-algebraic connected components algorithm~\cite{Zhang:2020:PAF} that is based on the classic PRAM algorithm of Shiloach and Vishkin~\cite{Shiloach:1982:AOL} that uses hooking and pointer-jumping. We make two interesting observations. (1)~We include a \emph{sparsification} optimization that is discussed in the FastSV paper~\cite{zhang2020fastsv}. With 3 lines of code that zero out the redundant values, our push-pull design is able to handle this optimization automatically. (2)~We avoid unnecessary GPU-to-CPU and CPU-to-GPU memory copies that would otherwise be required in the original FastSV GraphBLAS implementation using 2 new variants of \cmd{assign} and \cmd{extract}. Instead of using the \cmd{Index *} found with the standard variants, we use a GraphBLAS \cmd{Vector} that will implicitly have its values treated as the indices. Since \cmd{Vector} will always have the most recent data in GPU memory, this solves the problem of not being able to deduce whether the \cmd{Index} pointer is located in CPU or GPU memory. Unlike the other linear-algebraic CC implementation (LACC), FastSV does not guarantee $O(\log{\lvert V \rvert})$ worst-case iteration bound because it avoids unconditional hooking for higher performance~\cite{Zhang:2020:PAF}. We chose FastSV to include in GraphBLAST because it is consistently faster than LACC in practice, despite having worse complexity bounds. Consequently, our CC implementation also takes worst-case $O(\lvert E \rvert \, \lvert V \rvert)$ time and $O(\lvert V \rvert)$ depth. However, it is much faster in practice, similar to other quadratic time and linear depth CC algorithms such as multistep-CC~\cite{Slota:2014:BAC}.

    \end{description}}

\subsection{Triangle counting}
\label{sec:tricount}
Triangle counting is the problem of counting the number of unique triplets $u, v, w$ in an undirected graph such that $\{(u, v), (u,w), (v, w)\} \in E$. Many important measures of a graph are triangle-based, such as clustering coefficient and transitivity ratio.

\begin{description}
  \item[Hardwired GPU implementation]
  One of the best-performing implementation of triangle counting is by Bisson and Fatica~\cite{Bisson:2017:HPE}. In their work, they demonstrate an effective use of a static workload mapping of thread, warp, block per matrix row together with using bitmaps.
  \item[GraphBLAST implementation]
  In GraphBLAST, we follow Azad and Bulu\c{c}~\cite{Azad:2015:PTC} and Wolf et al.~\cite{Wolf:2017:FLA} in modeling the TC problem as a masked matrix-matrix multiplication problem. Given an adjacency matrix of an undirected graph $\mA$, and taking the lower triangular component $\mathbf{L}$, the number of triangles is the reduction of the matrix $\mathbf{B} = \mathbf{L}\mathbf{L}^T \,.\!* \mathbf{L}$ to a scalar. \revision{Rows of $\mathbf{L}$ are sorted by increasing number of nonzeros, following the literature that demonstrates the benefits of sorting by degree prior to triangle counting~\cite{cohen2009graph}.} In our implementation, we use a generalization of this algorithm where we assume we are solving the problem for three distinct matrices $\mA$, $\mB$, and $\mathbf{M}$ by computing $\mC = \mA \mB\,.\!* \mathbf{M}$.
\revision{We use the masked SpGEMM primitive whose implementation is detailed in Section~\ref{subsubsec:maskedspgemm}. This is followed by a reduction of matrix $\mC$ to a scalar, returning the number of triangles in graph $\mA$.
For each nonzero entry in the mask $\mathbf{M}(i,j) \neq 0$, our masked matrix-matrix multiplication performs an intersection of the nonzeros in the $i$-th row of $\mA$ with the $j$-th column of $\mB$.
Using a straightforward merge-based set intersection would have yielded an implementation with $O(\lvert E \rvert^{3/2})$ work and $O(\lg^{3/2} \lvert V \rvert)$ depth~\cite{shun2015multicore}. Instead of performing the set intersection using merging or hash tables, we use repeated binary searches from the elements of the shorter list to the larger list. Multiple publications concluded that the method of repeated binary searches was either competitive with or faster than the merge-based method on GPUs~\cite{fox2018fast, hu2018tricore}. Theoretically, it increases work marginally by a factor $\log{d}$ on average where $d$ is the average degree of a vertex. On the positive side, it has more parallelism.}

\end{description}

\section{Experimental Results}
\label{sec:results}
We first show overall performance analysis of GraphBLAST on nine datasets including both real-world and synthetically generated graphs; the topology of these datasets spans from regular to scale-free. Five additional datasets are used specifically for triangle counting, because they are the ones typically used for comparison of triangle counting~\cite{Wang:2016:ACS,Bisson:2017:HPE}.

\begin{table}
  \small
  \centering
  \setlength{\tabcolsep}{3pt}
  \rowcolors{2}{lightgray}{white}
  \begin{tabular}{*{6}{c}} \toprule Dataset &Vertices&Edges&Max Degree& Diameter &Type \\
    \midrule
    soc-orkut & 3M & 212.7M & 27,466 & 9 & rs
    \\ soc-Livejournal1 & 4.8M & 85.7M & 20,333 & 16 & rs
    \\ hollywood-09 & 1.1M & 112.8M & 11,467 & 11 & rs
    \\ indochina-04 & 7.4M & 302M & 256,425 & 26 & rs
    \\ rmat\_s22\_e64 & 4.2M & 483M & 421,607 & 5 & gs
    \\ rmat\_s23\_e32 & 8.4M & 505.6M & 440,396 & 6 & gs
    \\ rmat\_s24\_e16 & 16.8M & 519.7M & 432,152 & 6 & gs
    \\ rgg\_n\_24 & 16.8M & 265.1M & 40 & 2622 & gm
    \\ roadnet\_USA & 23.9M & 577.1M & 9 & 6809 & rm \\ \midrule
    coAuthorsCiteseer & 227K & 1.63M & 1372 & 31* & rs \\
    coPapersDBLP & 540K & 30.6M & 3299 & 18* & rs \\
    cit-Patents & 3.77M & 33M & 793 & 24* & rs \\
    com-Orkut & 3.07M & 234M & 33313 & 8* & rs \\
    road\_central & 14.1M & 33.9M & 8 & 4343* & rm \\ \midrule
    Journals & 124 & 12K & 123 & 2 & rs \\
    G43 & 1K & 20K & 36 & 4 & gs \\
    ship\_003 & 122K & 3.8M & 143 & 58* & rs \\
    belgium\_osm & 1.4M & 3.1M & 10 & 1923* & rm \\
    roadNet-CA & 2M & 5.5M & 12 & 617* & rm \\
    delaunay\_24 & 16.8M & 101M & 26 & 1720* & rm
    \\ \bottomrule
  \end{tabular}
  \caption[Dataset description table.]{Dataset Description Table. Graph types are: r: real-world, g: generated, s: scale-free, and m: mesh-like. All datasets
    have been converted to undirected graphs. Self-loops and duplicated edges are removed. Datasets in the top segment are used for BFS, SSSP and PR\@. Datasets in the middle segment are used for TC\@. Datasets in the bottom segment are used for comparison with GBTL~\cite{Zhang:2016:GCG}. An asterisk indicates the diameter is estimated using samples from 10,000 vertices. \label{tab:dataset-journal}}
\end{table}

\paragraph{Measurement methodology}
\revision{We report both runtime and traversed edges per second (TEPS) as our performance metrics. In general we report runtimes in milliseconds and TEPS as millions of traversals per second [MTEPS]. Runtime is measured by measuring the GPU kernel running time and TEPS is computed by the number of edges in the undirected graph divided by the runtime. We do not compute TEPS for CC, because it is not well-defined for this algorithm due to the hooking and pointer-jumping.}

\paragraph{Hardware characteristics}
We ran all experiments in this paper on a Linux workstation with 2$\times$3.50~GHz Intel 4-core, hyperthreaded E5-2637 v2 Xeon CPUs, 528~GB of main memory, and an NVIDIA K40c GPU with 12~GB on-board memory. GPU programs were compiled with NVIDIA's nvcc compiler (version~8.0.44) with the -O3 flag. CuSha was compiled using commit e753734 on their GitHub page. Galois was compiled using v2.2.1 (r0). Ligra was compiled using icpc 15.0.1 with CilkPlus at v1.5. Mapgraph was compiled at v0.3.3. SuiteSparse was compiled at v3.0.1 (beta1). Enterprise was compiled at commit 426846f on their GitHub page. Gunrock was compiled at v0.4 for the BFS, SSSP, PR and TC comparisons, and at v0.5 for the CC comparison. All results ignore transfer time (both disk-to-memory and CPU-to-GPU)\@. All Gunrock and GraphBLAST tests were run 10 times with the average runtime and MTEPS used for results.

\revision{\paragraph{Datasets} We summarize the datasets in Table~\ref{tab:dataset-journal}. soc-orkut (soc-ork), com-orkut (com-ork), soc-Livejournal1 (soc-lj), and hollywood-09 (h09) are social graphs; indochina-04 (i04) is a crawled hyperlink graph from indochina web domains; coAuthorsCiteseer (coauthor), coPapersDBLP (copaper), and cit-Patents (cit-pat) are academic citation and patent citation networks; Journals (journal) is a graph indicating common readership across Slovenian magazines and journals; rmat\_s22\_e64 (rmat-22), rmat\_s23\_e32 (rmat-23), and rmat\_s24\_e16 (rmat-24) are three generated R-MAT graphs; and G43 (g43) is a random graph. All twelve datasets are scale-free graphs with diameters of less than 30 and unevenly distributed node degrees (80\% of nodes have degree less than 64). ship-003 is a graph of a finite element model. The following datasets---rgg\_n\_24 (rgg), road\_central (road\_cent), roadnet\_USA (road\_usa), belgium\_osm (belgium), roadNet-CA (road\_ca), and delaunay\_n24 (delaunay)---have large diameters with small and evenly distributed node degrees (most nodes have degree less than 12). soc-ork and com-Ork are from the Network Repository~\cite{Rossi:2015:TND}; soc-lj, i04, h09, road\_central, road\_usa, coauthor, copaper, and cit-pat are from the University of Florida Sparse Matrix Collection~\cite{Davis:2011:TUF}; rmat-22, rmat-23, rmat-24, and rgg are R-MAT and random geometric graphs we generated. The R-MAT graphs were generated with the following parameters: $a= 0.57$, $b= 0.19$, $c= 0.19$, $d= 0.05$. The edge weight values (used in SSSP) for each dataset are uniformly random integer values between 1 and 64.}

\begin{table}
  \small
  \centering
  \renewcommand{\arraystretch}{0.5} %
  \resizebox{\linewidth}{!}{
    \begin{NiceTabular}{*{13}{c}}[code-before = \rowcolors{4}{lightgray}{}]
      \toprule
      && \multicolumn{5}{c}{Runtime (ms) [lower is better]} && \multicolumn{5}{c}{Edge throughput (MTEPS) [higher is better]} \\
      \cmidrule{3-7}\cmidrule{9-13}
      & & SuiteSparse & Hardwired & & & && SuiteSparse & Hardwired & & & \\
      Alg. & Dataset & GraphBLAS & GPU & Ligra & Gunrock & GraphBLAST && GraphBLAS & GPU & Ligra & Gunrock & GraphBLAST \\ \midrule
      \Block[color=white]{9-1}{\rotate BFS}
      & soc-ork & 2542 & 25.81      & 26.1  & \textbf{5.573}   & 7.230   && 83.66 & 12360      & 8149  & \textbf{38165} & 29217  \\
      & soc-lj     & 2218 & 36.29    & 42.4  & \textbf{14.05} & 14.16  && 38.61 & 5661     & 2021  & \textbf{6097} & 6049   \\
      & h09        & 1013 & 11.37    & 12.8  & \textbf{5.835} & 7.138  && 111.1 & 14866 & 8798  & \textbf{19299} & 15775 \\
      & i04         & 2646 & \textbf{67.7}      & 157  & 77.21   & 80.37 && 112.7 & \textbf{8491}    & 1899  & 3861 & 3709    \\
      & rmat-22 & 5401 & 41.81      & 22.6 & \textbf{3.943}  & 4.781 && 89.44  & 17930     & 21374  & \textbf{122516} & 101038  \\
      & rmat-23 & 8628  & 59.71      & 45.6 & \textbf{7.997}  & 8.655 && 58.61  & 12971    & 11089  & \textbf{63227} & 58417   \\
      & rmat-24 & 21032  & 270.6      & 89.6  & 16.74   & \textbf{16.59} && 24.71 & 1920    & 5800  & 31042  & \textbf{31327}  \\
      & rgg         & 230602 & \textbf{138.6}  & 918  & 593.9  & 2991 && 1.201  & \textbf{2868}    & 288.8   & 466.4 & 92.59   \\
      & road\_usa& 9413 & \textbf{141}    & 978  & 676.2  & 7155  && 6.131  & \textbf{1228}    & 59.01  & 85.34 & 8.065   \\ \midrule
      \Block[color=white]{9-1}{\rotate SSSP}
      & soc-ork & 7223 & 807.2   & \textbf{595}   & 981.6 & 676.7  && 29.45   & 263.5 & \textbf{357.5}  & 216.7 & 314.3 \\
      & soc-lj  & 3599 & 369   & 368   & 393.2 & \textbf{256.3} && 23.81   & 232.2      & 232.8   & 217.9 & \textbf{334.2} \\
      & h09     & 2585 & 143.8   & 164   & \textbf{83.2} & 109.123  && 43.58   & 783.4      & 686.9   & \textbf{1354} & 1032  \\
      & i04     & 1087 & ---   & 397   & \textbf{371.8} & 414.5 && 274.22   & ---   & 750.8   & \textbf{801.7} & 719.2 \\
      & rmat-22 & 30688 & ---   & 774   & 583.9 & \textbf{477.5}  && 15.74   & --- & 624.1  & 827.3 & \textbf{1011.7} \\
      & rmat-23 & 25268 & ---   & 1110  & 739.1 & \textbf{680.0} && 20.01   & ---   & 455.5   & 684.1 & \textbf{743.6} \\
      & rmat-24 & 39105 & ---   & 1560  & \textbf{884.5} & 905.2 && 13.29   & ---   & 333.1  & \textbf{587.5} & 574.0 \\
      & rgg     & 1649653 & ---   & \textbf{80800} & 115554 & 144291 && 0.161   & ---  & \textbf{3.28}   & 2.294 & 1.84 \\
      & road\_usa & 801311 & \textbf{4860}   & 29200 & 11037 & 144962  && 0.072   & \textbf{11.87} & 1.98   & 5.229 & 0.398 \\ \midrule
      \Block[color=white]{9-1}{\rotate PR (pull)}
      & soc-ork & 942.3 & \textbf{52.54} & 476   & 173.1 & 64.22  && 225.7 & \textbf{4048} & 446.8 & 1229 & 3312 \\
      & soc-lj  & 741.8 & 33.61 & 200   & 54.1 & \textbf{21.54}    && 115.5 & 2550 & 428.5 & 1584 & \textbf{3978} \\
      & h09     & 306.2 & 34.71 & 77.4  & 20.05 & \textbf{8.12}  && 41.81 & 368.8 & 165.4 & 638.4 & \textbf{1577} \\
      & i04     & 1154 & 164.6 & 210   & 41.59 & \textbf{19.16}  && 261.6 & 1835 & 1438 & 7261 & \textbf{15763} \\
      & rmat-22 & 5328 & 188.5 & 1250  & 304.5 & \textbf{115.6}  && 90.65 & 2562 & 386.4 & 1586 & \textbf{4178} \\
      & rmat-23 & 7087 & \textbf{147}   & 1770  & 397.2 & 161.3  && 71.34 & \textbf{3439} & 285.6 & 1273 & 3134 \\
      & rmat-24 & 9033 & \textbf{128}   & 2180  & 493.2 & 211.5  && 57.54 & \textbf{4060} & 238.4 & 1054 & 2457 \\
      & rgg     & 2233 & 53.93 & 247   & 181.3 & \textbf{34.58}  && 118.7 & 4916 & 1073 & 1462 & \textbf{7665} \\
      & road\_usa & 3030 & --- & 209   & \textbf{24.11} & 26.91  && 190.4 & --- & 2761 & \textbf{23936} & 21449 \\ \midrule
      \Block[color=white]{9-1}{\rotate CC}
      & soc-ork   & 34813 & \textbf{46.97} & 260 & 179.24 & 275.87    && --- & \textbf{---} & --- & --- & --- \\
      & soc-lj      & 32051 & \textbf{43.51} & 184 & 81.24 & 185.06    && --- & \textbf{---} & --- & --- & --- \\
      & h09          & 15551 & \textbf{24.63} & 90.8 & 92.1 & 64.44     && --- & \textbf{---} & --- & --- & --- \\
      & i04           & 57211 & \textbf{130.3} & 315 & 786.83 & 343.64    && --- & \textbf{---} & --- & --- & --- \\
      & rmat-22   & 81021 & \textbf{149.4} & 563 & 369.23 & 365.75    && --- & \textbf{---} & --- & --- & --- \\
      & rmat-23   & 89421 & \textbf{212} & 1140 & 498.01 & 705.45     && --- & \textbf{---} & --- & --- & --- \\
      & rmat-24   & 102867 & \textbf{245.7} & 1730 & 560.9 & 978.64  && --- & \textbf{---} & --- & --- & --- \\
      & rgg            & 190083 & \textbf{103.9} & 6000 & 353.41 & 5602    && --- & \textbf{---} & --- & --- & --- \\
      & road\_usa & 264328 & \textbf{124.9} & 50500 & 212.62 & 26880 && --- & \textbf{---} & --- & --- & --- \\
      \midrule
      \Block[color=white]{6-1}{\rotate TC}
      & coauthor & 6.337   & \textbf{2.2}       & ---      & 4.51   & 5.96   && 128.5   & \textbf{370}  & ---     & 181  & 137    \\
      & copaper  & \textbf{50.93}   & 64.4     & ---     & 197   & 246      && \textbf{630}   & 498      & ---    & 163  & 130    \\
      & soc-lj     & 1221   & \textbf{295}      & 490      & 896   & 1125    && 56.5   & \textbf{234}      & 141     & 77.0  & 61.3    \\
      & cit-pat    & 380.9   & \textbf{34.5}      & 79.5     & 156   & 137     && 43.3   & \textbf{478}      & 208    & 105  & 121 \\
      & com-ork & 5100   & \textbf{1626}    & 1920  & 6636   & 5367 && 23   & \textbf{72.1}      & 61.0     & 17.7  & 21.8    \\
      & road\_cent & 231.3   & \textbf{5.6}   & ---     & 61.4  & 78.7     && 73.1   & \textbf{3018}     & ---     & 275  & 215    \\
      \bottomrule
    \end{NiceTabular}}
  \caption[GraphBLAST's performance comparison with other graph libraries.]
  {GraphBLAST's performance comparison for runtime and edge throughput with other graph libraries (SuiteSparse, Ligra, Gunrock) and hardwired GPU implementations on a Tesla K40c GPU\@. All PageRank times are normalized to one iteration. Hardwired GPU implementations for each primitive are Enterprise (BFS)~\protect\cite{Liu:2015:EBG}, delta-stepping SSSP~\protect\cite{Davidson:2014:WPG}, pull-based PR~\protect\cite{Khorasani:2014:CVG}, \revision{hooking and pointer-jumping CC~\cite{Soman:2010:AFG},} and triangle counting~\protect\cite{Bisson:2017:HPE}. A missing data entry means there is a runtime error.%
    \label{tab:exp_largetable}}
\end{table}

\begin{figure}[t]
  \includegraphics[width=\textwidth]{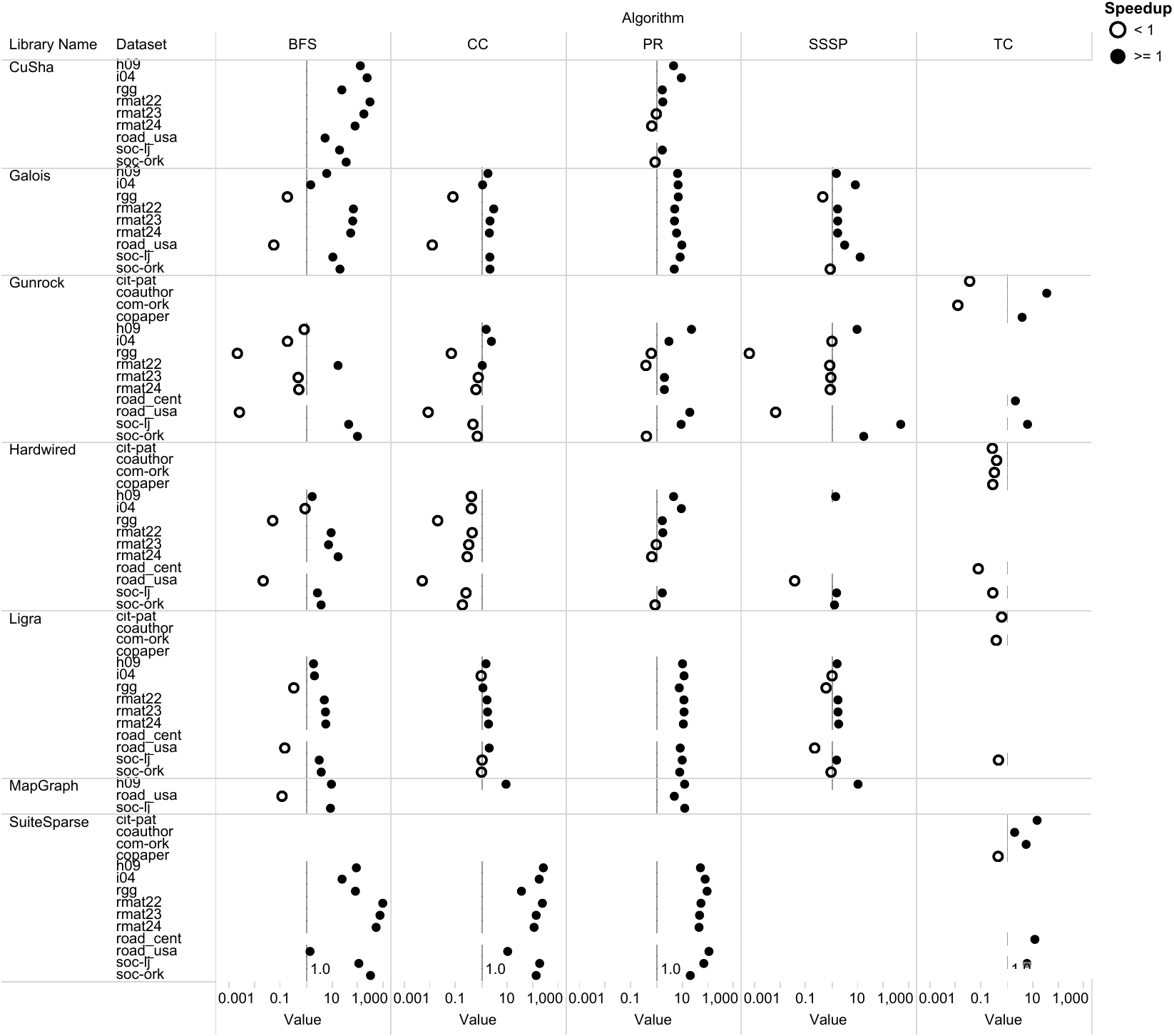}
  \centering
  \caption{Speedup of GraphBLAST over seven other graph processing libraries/hardwired algorithms on different graph inputs. Black dots indicate GraphBLAST is faster, white dots slower.\label{fig:speedup}}
\end{figure}

\subsection{Performance summary}

Table~\ref{tab:exp_largetable} and Figure~\ref{fig:speedup} compare GraphBLAST's performance against several other graph libraries and hardwired GPU implementations. In general, GraphBLAST's performance on traversal-based algorithms (BFS and SSSP) is better on the seven scale-free graphs (soc-orkut, soc-lj, h09, i04, and rmats) than on the small-degree large-diameter graphs (rgg and road\_usa). The main reason is our load-balancing strategy during traversal and particularly our emphasis on high performance for highly-skewed-distribution irregular graphs. Therefore, we incur a certain amount of overhead for our merge-based load-balancing and our requirement of a kernel launch in every iteration. For these types of graphs, asynchronous approaches, pioneered by Enterprise~\cite{Liu:2015:EBG}, that do not require exiting the kernel until the breakpoint has been met is a way to address the kernel launch problem. However, this does not work for non-BFS solutions, so asynchronous approaches in this area remain an open problem. In addition, graphs with uniformly low degree expose less parallelism and would tend to show smaller gains in comparison to CPU-based methods.

\subsection{Comparison with CPU graph frameworks}

We compare GraphBLAST's performance with three CPU graph libraries: the SuiteSparse GraphBLAS library, the first GraphBLAS implementation for multi-threaded CPU~\cite{Davis:2018:SGG}; and Galois~\cite{Nguyen:2013:ALI} and Ligra~\cite{Shun:2013:LLG}, both among the highest-performing multi-core shared-memory graph libraries. Against SuiteSparse, the speedup of GraphBLAST on average on all algorithms is geomean $27.9\times$ ($1268\times$ peak) and geomean $43.51\times$ ($1268\times$ peak) on scale-free graphs. Compared to Galois, GraphBLAST's performance is generally faster. We are 2.6$\times$ geomean ($64.2\times$ peak) faster across all algorithms. We get the greatest speedup on BFS, because we implement direction optimization. We get the next greatest speedup on PR, where the amount of computation tends to be greater than for BFS or SSSP\@.

Compared to Ligra, GraphBLAST's performance is generally comparable on most tested graph algorithms; note Ligra results are on a 2-CPU machine of the same timeframe as the K40c GPU we used to test. We are $3.38\times$ ($1.35\times$ peak) faster for BFS vs.\ Ligra for scale-free graphs, because we incorporate some BFS-specific optimizations such as \emph{masking}, \emph{early-exit}, and \emph{operand reuse}, as discussed in Section~\ref{sec:app}. However, we are $4.88\times$ slower on the road network graphs. For SSSP, a similar picture emerges. Compared to Ligra for scale-free graphs, we get $1.35\times$ ($1.72\times$ peak) speed-up, but are $2.98\times$ slower on the road networks. We believe this is because our Bellman-Ford with sparsification means we can do less work on scale-free graphs, but our framework is not optimized for road networks. For PR, we are $9.23\times$ ($10.96\times$ peak) faster, because we use a highly-optimized merge-based load balancer that is suitable for this SpMV-based problem. \revision{For CC, we are $1.30\times$ ($1.88\times$ peak) faster. With regards to TC, we are $2.80\times$ slower, because we have a simple algorithm for the masked matrix-matrix multiply.}

\subsection{Comparison with GPU graph frameworks and GPU hardwired}
Compared to hardwired GPU implementations, depending on the dataset, GraphBLAST's performance is comparable or better on BFS, SSSP, and PR\@. \revision{For CC, GraphBLAST is $3.1\times$ slower (geometric mean) on scale-free graphs and $107.7\times$ slower on road network graphs. We think the reason is that we are strictly following Shiloach-Vishkin in doing only one level of pointer jumping per iteration, whereas the hardwired implementation is doing pointer jumping until each tree has been reduced to a star. This is an advantage on the GPU, because it allows their implementation to significantly reduce kernel launch overheads (see Section~\ref{sec:kernel-launch}), which become significant especially for graphs with high diameter such as road networks. If kernel fusion is added to GraphBLAST, we would be able to take advantage of this optimization.}

For TC, GraphBLAST is $3.3\times$ slower (geometric mean) than the hardwired GPU implementation due to fusing of the matrix-multiply and the reduce, which lets the hardwired implementation avoid the step of writing out the output to the matrix-multiply. The alternative is having a specialized kernel that does a fused matrix-multiply and reduce. This tradeoff is not typical of our other algorithms. While still achieving high performance, GraphBLAST's application code is smaller in size and clearer in logic compared to other GPU graph libraries.

Compared to CuSha and MapGraph, GraphBLAST's performance is quite a bit faster. We get geomean speedups of $8.40\times$ and $3.97\times$ respectively ($420\times$ and $64.2\times$ peak). The speedup comes from direction optimization. CuSha only does the equivalent of pull-traversal, so their performance is most comparable to ours in PR\@. MapGraph is push-only.

\revision{Compared to Gunrock, the fastest GPU graph framework, GraphBLAST's performance is comparable on BFS, CC and TC with Gunrock being 11.8\%, 14.8\% and 11.1\% faster in the geomean respectively. On SSSP, GraphBLAST is faster by $1.1\times$ ($1.53\times$ peak). This can be attributed to GraphBLAST using \emph{generalized direction optimization} and Gunrock only doing push-based advance. On PR, GraphBLAST is significantly faster and gets speedups of $2.39\times$ ($5.24\times$ peak). For PR, the speed-up again can be attributed to GraphBLAST automatically using \emph{generalized direction optimization} to select the right direction, which is SpMV in this case. Gunrock does push-based advance.}

\subsection{Comparison with Gunrock on latest GPU architecture}

In Table~\ref{tab:gunrock-titan}, we compare against Gunrock on BFS, SSSP, CC and PR using the latest generation GPU, Titan V\@. As the result shows, we have a 3.13$\times$ slowdown compared to Gunrock on BFS, which indicates that we do worse on Titan V\@. On SSSP, we are $1.08\times$ ($2.39\times$ peak) faster when not including the road network datasets, and $0.48\times$ slower when including them. On CC, we have a 4.00$\times$ slowdown compared to Gunrock. On PR, we are $2.90\times$ ($7.67\times$ peak) faster in the geomean.

\begin{table}
  \small
  \centering
  \setlength{\tabcolsep}{3pt}

  \begin{NiceTabular}{*{11}{c}}[code-before = \rowcolors{4}{lightgray}{}\columncolor{white}{1-2}\columncolor{white}{11}]
    \toprule
    &&& \multicolumn{2}{c}{Runtime (ms)} && \multicolumn{2}{c}{Edge throughput (GTEPS)} &&& \\
    &&& \multicolumn{2}{c}{[lower is better]} && \multicolumn{2}{c}{[higher is better]} &&& \\
    \cmidrule{4-5}\cmidrule{7-8}
    Alg. & Type & Dataset & Gunrock & GraphBLAST && Gunrock & GraphBLAST && Speedup & \multirow{2}{*}[4.08mm]{\parbox[t]{1.2cm}{Geomean Speedup}} \\
    \midrule\parbox[t]{2mm}{\multirow{9}{*}{\rotatebox[origin=c]{90}{BFS}}}
    &
      \multirow{7}{*}{Scale-free} &
                                    soc-ork    & 1.61    & 4.02   && 132.3 & 52.93 && 0.40$\times$ & \\
    && soc-lj        & 2.95    & 8.27   && 29.05 & 10.36 && 0.36$\times$ &\\
    && h09           & 1.60    & 5.16   && 70.51 & 21.83 && 0.31$\times$ &\\
    && i04            & 14.72  & 32.09 && 20.25 & 9.29   && 0.46$\times$ & 0.44$\times$\\
    && rmat-22   & 1.13    & 2.36   && 425.7 & 204.6 && 0.48$\times$ &\\
    && rmat-23   & 2.04   & 3.64    && 247.3 & 138.8 && 0.56$\times$ &\\
    && rmat-24   & 3.91   & 6.79    && 132.8 & 76.47 && 0.58$\times$ &\\
    \cmidrule[0.05mm]{2-11}
    &Road & rgg           & 321.0 & 3333   && 0.863 & 0.083 && 0.096$\times$ & \multirow{2}{*}{0.10$\times$}\\
    &network& road\_usa& 782.5 & 7467  && 0.074 & 0.0077 && 0.10$\times$ &\\
    \midrule\parbox[t]{2mm}{\multirow{9}{*}{\rotatebox[origin=c]{90}{SSSP}}}
    &\multirow{7}{*}{Scale-free}
      & soc-ork    & 263.0 & 110.2 && 0.809 & 1.93 && \textbf{2.39$\times$} &\\
    && soc-lj        & 122.5 & 72.41 && 0.699 & 1.18 && \textbf{1.69$\times$} &\\
    && h09           & 15.46 & 43.87 && 7.29   & 2.57 && 0.35$\times$ &\\
    && i04            & 79.01 & 150.8 && 3.77   & 1.98 && 0.52$\times$ & 1.08$\times$\\
    && rmat-22   & 103.6 & 76.23 && 4.66   & 6.34 && \textbf{1.35$\times$} &\\
    && rmat-23   & 175.5 & 122.8 && 2.88   & 4.12 && \textbf{1.43$\times$} &\\
    && rmat-24   & 254.0 & 209.5 && 2.05   & 2.48 && \textbf{1.21$\times$} &\\
    \cmidrule[0.05mm]{2-11}
    &Road& rgg           & 349.6 & 91704 && 0.792& 0.0030 && 0.0038$\times$  & \multirow{2}{*}{0.025$\times$}\\
    &network& road\_usa& 2928 & 17994 && 0.020 & 0.0032 && 0.16$\times$ &\\
    \midrule\parbox[t]{2mm}{\multirow{9}{*}{\rotatebox[origin=c]{90}{PR (pull)}}}
    &\multirow{7}{*}{Scale-free}
      & soc-ork    & 51.76 & 9.29   && 4.11   & 22.88 && \textbf{5.57$\times$} &\\
    && soc-lj        & 14.66 & 4.41   && 5.84   & 19.45 && \textbf{3.33$\times$} &\\
    && h09           & 3.99   & 2.11   && 28.23 & 53.40 && \textbf{1.89$\times$} &\\
    && i04            & 6.54   & 5.38   && 45.58 & 55.37 && \textbf{1.21$\times$} & 2.86$\times$\\
    && rmat-22   & 49.26 & 15.86 && 9.81   & 30.46 && \textbf{3.11$\times$} &\\
    && rmat-23   & 85.62 & 24.23 && 5.91   & 20.87 && \textbf{3.53$\times$} &\\
    && rmat-24   & 130.1 & 39.14 && 3.99   & 13.28 && \textbf{3.32$\times$} &\\
    \cmidrule[0.05mm]{2-11}
    &Road& rgg            & 62.17 & 8.10   && 4.46   & 34.18 && \textbf{7.67$\times$} & \multirow{2}{*}{3.06$\times$}\\
    &network& road\_usa& 9.17   & 7.51    && 6.29   & 7.68   && \textbf{1.22$\times$} & \\
    \midrule\parbox[t]{2mm}{\multirow{9}{*}{\rotatebox[origin=c]{90}{CC}}}
    &\multirow{7}{*}{Scale-free}
      & soc-ork    & 30.14 & 45.54  && N/A & N/A && 0.66$\times$ &\\
    && soc-lj        & 15.91 & 47.7   && N/A & N/A && 0.33$\times$ &\\
    && h09           & 18.05 & 18.3    && N/A & N/A && 0.99$\times$ &\\
    && i04            & 65.55 & 105.87 && N/A & N/A && 0.62$\times$ & 0.68$\times$\\
    && rmat-22   & 65.33 & 56.68  && N/A & N/A && \textbf{1.15$\times$} &\\
    && rmat-23   & 87.9 & 117.79   && N/A & N/A && 0.76$\times$ &\\
    && rmat-24   & 126.84 & 226.36 && N/A & N/A && 0.56$\times$ &\\
    \cmidrule[0.05mm]{2-11}
    &Road& rgg           & 64.25 & 4826.8 && N/A & N/A && 0.013 $\times$  & \multirow{2}{*}{0.0077$\times$}\\
    &network& road\_usa& 39.78 & 9016.04 && N/A & N/A && 0.0044$\times$ &
    \\ \bottomrule
  \end{NiceTabular}
  \caption{GraphBLAST's performance comparison for runtime and edge throughput with Gunrock~\cite{Wang:2017:GGG} for four graph algorithms on a Titan V GPU\@. Datasets where this work is faster are shown in bold.\label{tab:gunrock-titan}}
\end{table}

\begin{figure}
  \centering
  \captionsetup[subfigure]{justification=centering}
  \begin{subfigure}[b]{0.25\textwidth}
        \centering
      \includegraphics[width=\textwidth]{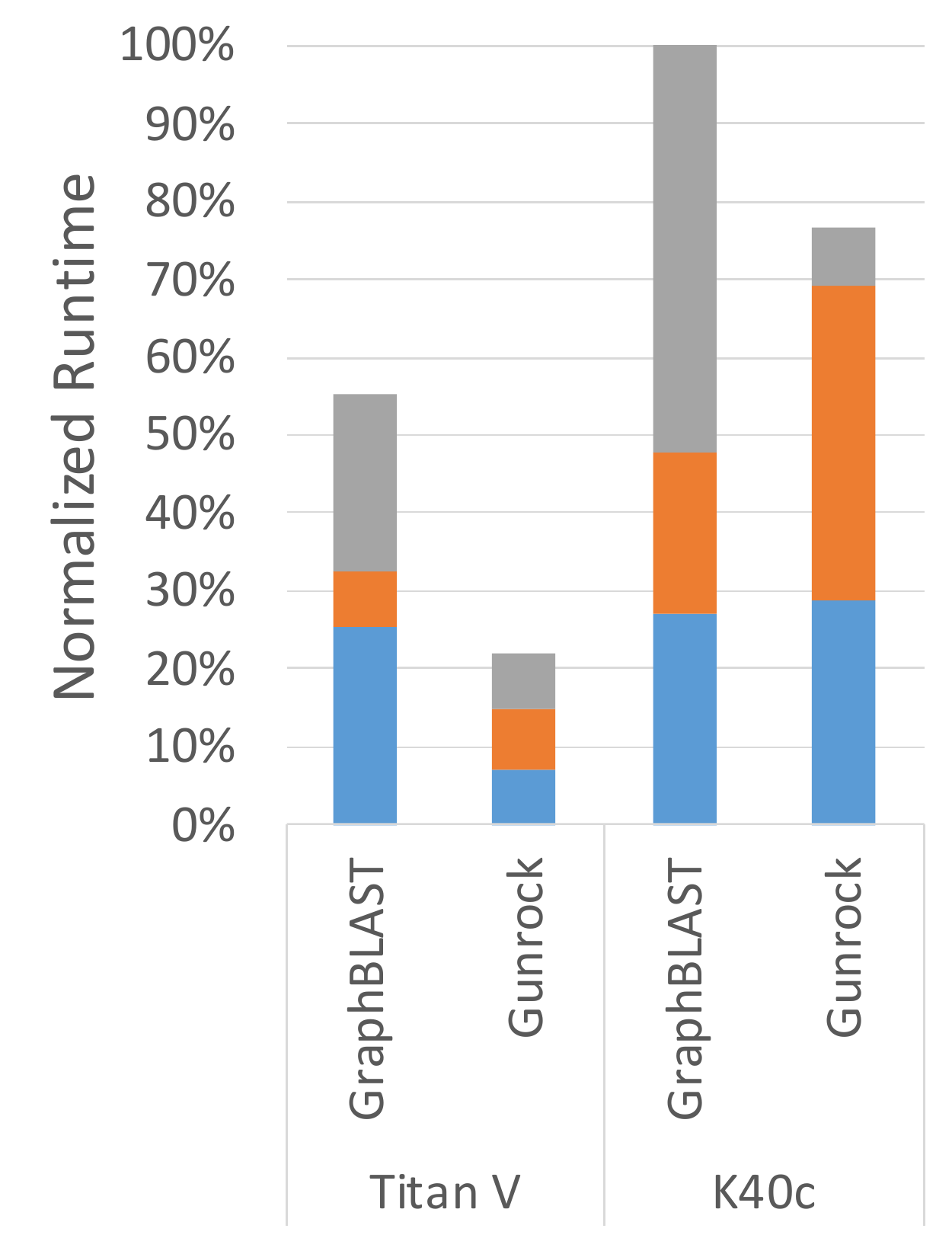}
      \caption{BFS}
  \end{subfigure}
  \begin{subfigure}[b]{0.20\textwidth}
        \centering
      \includegraphics[width=\textwidth]{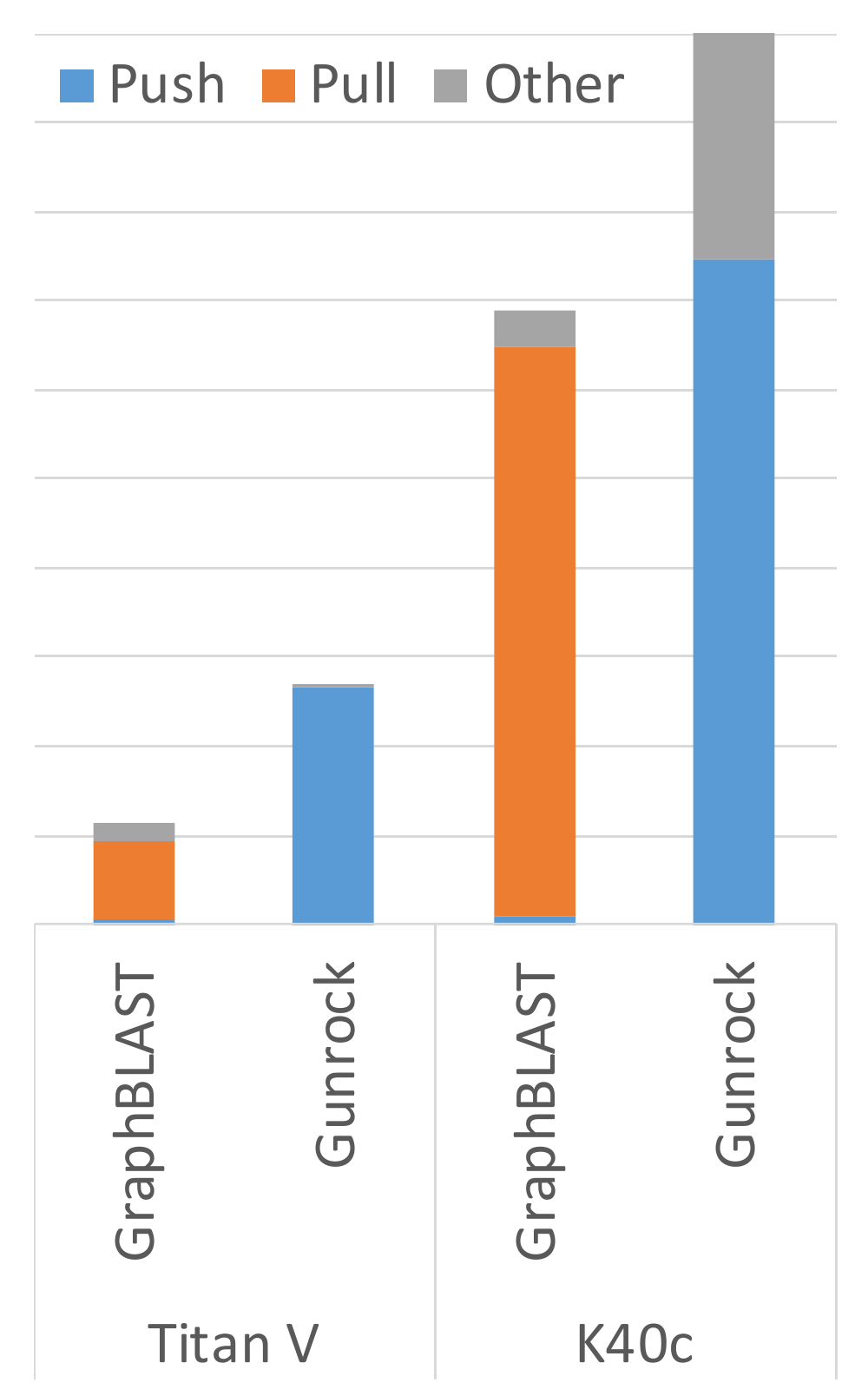}
      \caption{SSSP}
  \end{subfigure}
  \begin{subfigure}[b]{0.20\textwidth}
        \centering
      \includegraphics[width=\textwidth]{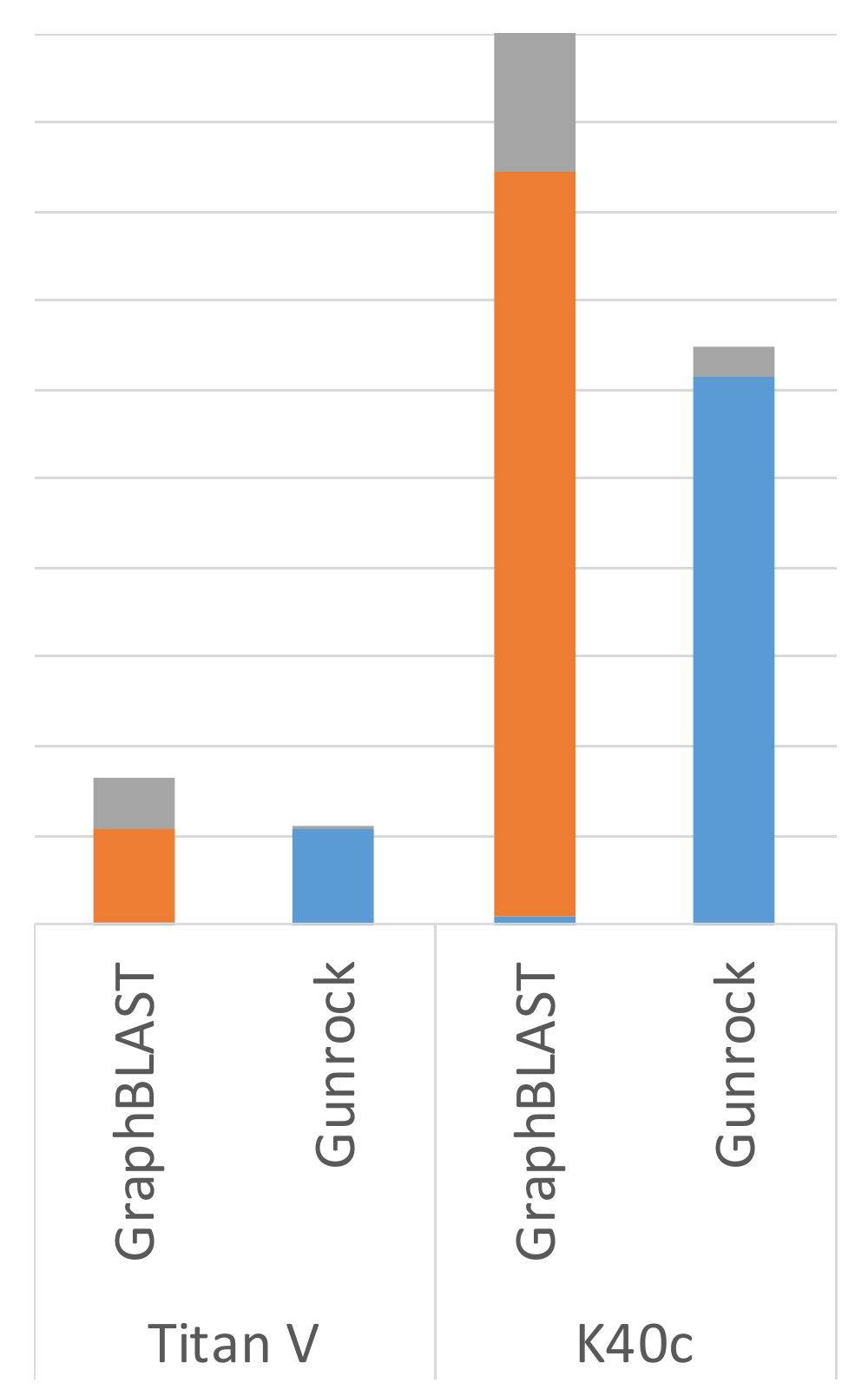}
      \caption{CC}
  \end{subfigure}
  \begin{subfigure}[b]{0.20\textwidth}
        \centering
      \includegraphics[width=\textwidth]{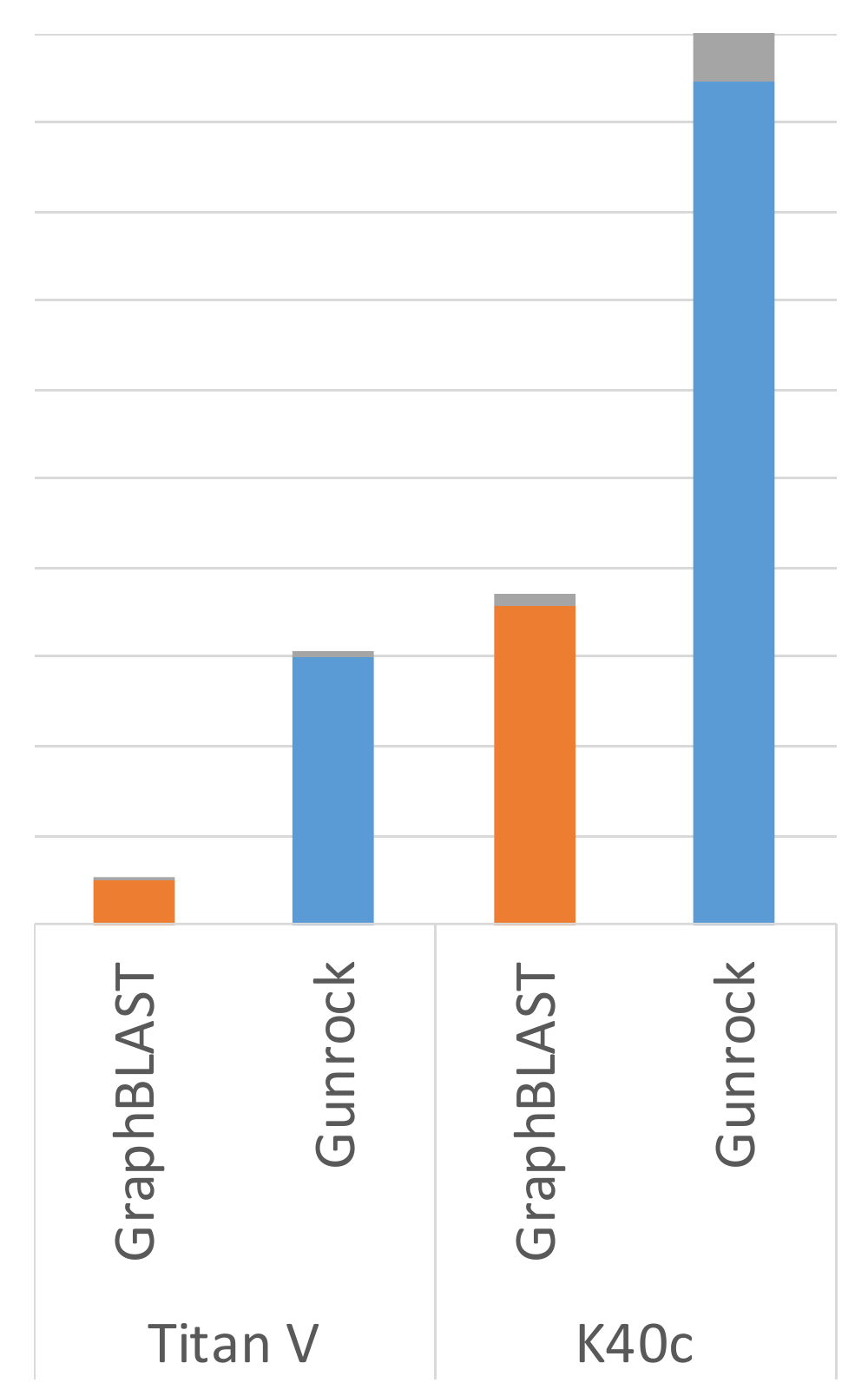}
      \caption{PR}
  \end{subfigure}
  \caption{Runtime breakdown of GraphBLAST and Gunrock migrating from K40c to Titan V GPU for BFS, SSSP, CC and PR on `soc-ork', normalized to the slowest combination per graph algorithm.\label{fig:gunrock-titan}}
\end{figure}

Taking a closer look at this in Figure~\ref{fig:gunrock-titan}, we can see that both push and pull components of Gunrock's BFS benefit due to moving from K40c to Titan V, but ``other'' does not. However for GraphBLAST, only the ``other'' and pull benefit from moving from K40c to Titan V\@. We hypothesize the reason for this is the GraphBLAST push is implemented using a radixsort to perform a multiway merge, and radixsort does not see a noticeable improvement in performance from K40c to Titan V on the problem sizes typical of BFS\@. On the other hand, Gunrock uses uses a series of inexpensive heuristics~\cite{Wang:2017:GGG} to reduce but not eliminate redundant entries in the output frontier. These heuristics include a global bitmask, a block-level history hashtable, and a warp-level hashtable. The size of each hashtable is adjustable to achieve the optimal tradeoff between performance and redundancy reduction rate. However, this approach may not be suitable for GraphBLAS, because such an optimization may be too BFS-focused and would generalize poorly.

\revision{For SSSP, GraphBLAST has the advantage of the direction optimization automatically choosing the optimal direction, which for SSSP happens to be pull. Gunrock's SSSP is written to use push, so it is unable to take advantage of this feature. For PR, the situation is similar to SSSP in that Gunrock uses push when pull is better. In CC, Gunrock's implementation has the advantage of following the approach of the hardwired implementation by Soman, Kishore and Narayanan~\cite{Soman:2010:AFG}. This algorithm performs pointer-jumping until the tree has become a star, whereas GraphBLAST follows Shiloach-Vishkin strictly and only does one level of pointer-jumping. We believe Gunrock's primary performance advantage, then, is faster convergence. More research is required to understand whether this optimization can be used to improve the linear algebraic implementation of FastSV in the presence and absence of kernel fusion (see Section~\ref{sec:kernel-fusion}).}

\subsection{Comparison with GraphBLAS-like framework on GPU}
In Table~\ref{tab:gbtl}, we compare against GBTL~\cite{Zhang:2016:GCG}, the first GraphBLAS-like implementation for the GPU on BFS\@. Our implementation is $31.8\times$ ($58.5\times$ peak) faster in the geomean. We attribute this speed-up to several factors: (1)~they use the Thrust library~\cite{Bell:2012:TPO} to manage the CPU-to-GPU memory traffic that works for all GPU applications, while we use a domain-specific memory allocator that only copies from CPU to GPU when necessary; (2)~they specialize the CUSP library's \cmd{mxm} operation~\cite{Dalton:2014:CGP,Dalton:2015:OSM} for a matrix with a single column to mimic the \cmd{mxv} required by the BFS, while we have a specialized \cmd{mxv} operation that is more efficient; and (3)~we utilize the design principles of exploiting input and output sparsity, as well as proper load balancing, none of which are in GBTL\@.

\begin{table}
  \small
  \centering
  \setlength{\tabcolsep}{3pt}
  \rowcolors{2}{lightgray}{white}
  \begin{tabular}{*{8}{c}} \toprule
    & \multicolumn{2}{c}{Runtime (ms)} && \multicolumn{2}{c}{Edge throughput (MTEPS)} && \\
    & \multicolumn{2}{c}{[lower is better]} && \multicolumn{2}{c}{[higher is better]} && \\
    \cmidrule{2-3}\cmidrule{5-6}
    \rowcolor{white}
    Dataset & GBTL & GraphBLAST && GBTL & GraphBLAST && Speedup\\
    \midrule
    Journals     & 5.76    & 0.147 && 2.074 & 80.98 && 39.05$\times$\\
    G43          & 14.61   & 0.503 && 1.368 & 39.72 && 29.04$\times$\\
    ship\_003    & 559.0  & 9.562 && 14.25 & 832.9 && 58.46$\times$\\
    belgium\_osm & 10502 & 476.3 && 0.295 & 6.508 && 22.05$\times$\\
    roadNet-CA   & 4726 & 259.2 && 1.168 & 21.30  && 18.23$\times$\\
    delaunay\_24 & 65508 & 1677  && 1.537 & 60.02 && 39.06$\times$
    \\ \bottomrule
  \end{tabular}
  \caption{GraphBLAST's performance comparison for runtime and edge throughput with GBTL~\cite{Zhang:2016:GCG} for BFS on a Tesla K40c GPU\@. \label{tab:gbtl}}
\end{table}

\begin{figure}
  \centering
  \includegraphics[width=0.7\textwidth]{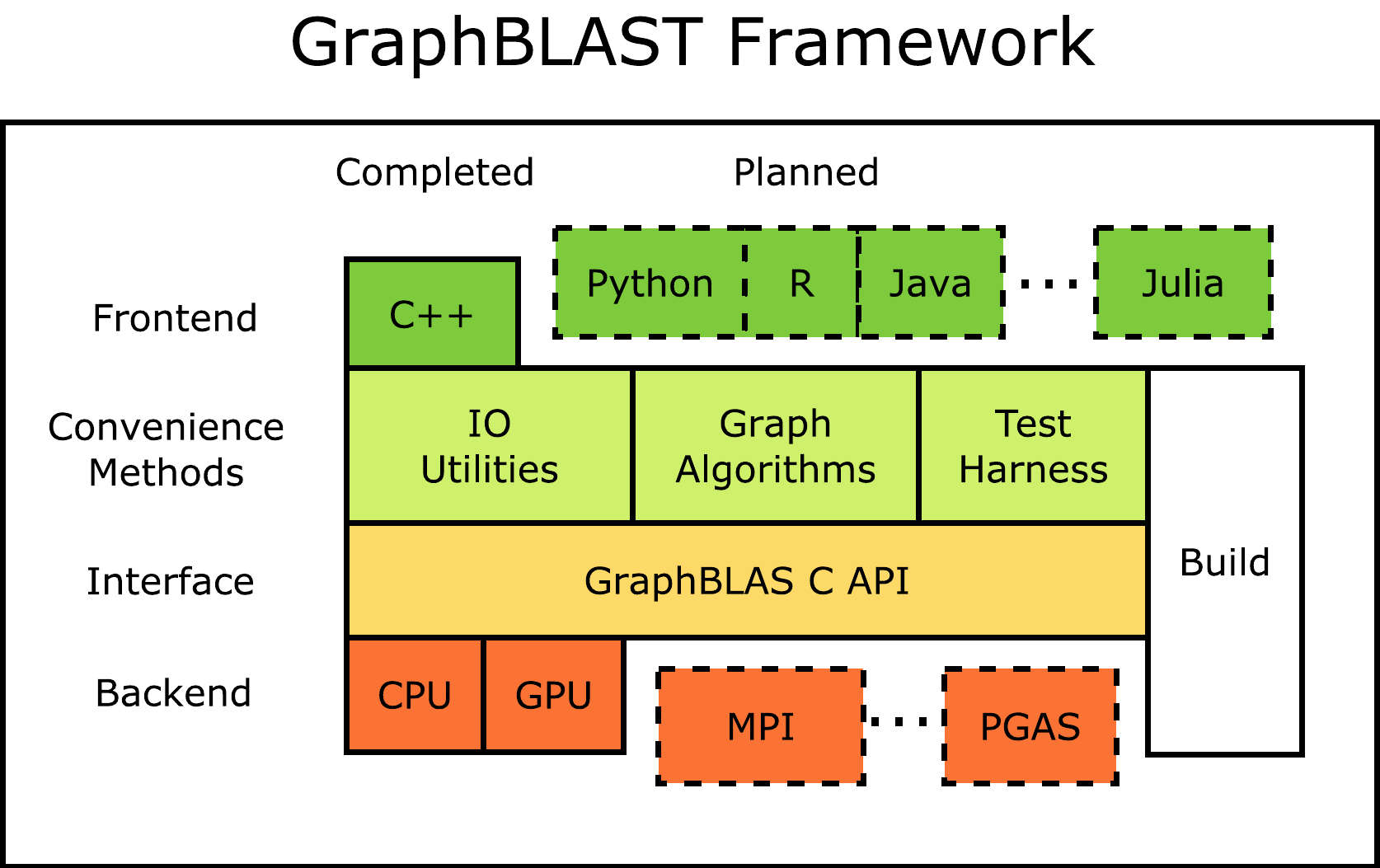}
  \centering
  \caption[Hourglass design of GraphBLAST\@.]{Design of GraphBLAST: Completed and planned components, and how open standard GraphBLAS API fits into the framework.\label{fig:hourglass}}
\end{figure}

In addition to getting comparable or faster performance, GraphBLAST has the advantage of being concise, as shown in Table~\ref{tab:loc}. Developing new graph algorithms in GraphBLAST requires modifying a single file and writing straightforward C++ code. Currently, we are working on a Python frontend interface too, to allow users to build new graph algorithms without having to recompile. Additional language bindings are being planned as well (see Figure~\ref{fig:hourglass}). Similar to working with machine learning frameworks, writing GraphBLAST code does not require any parallel programming knowledge of OpenMP, OpenCL or CUDA, or even performance optimization experience.

\section{Conclusion}
\label{sec:conclusion}
In this paper, we set out to answer the question: What is needed for a high-performance graph framework that is based on linear algebra? The answer we conclude is that it must: (1)~exploit input sparsity through direction optimization, (2)~exploit output sparsity through masking, and (3)~\revision{address the GPU considerations of avoiding CPU-to-GPU copies, supporting generalized semiring operators, and load-balancing}. In order to give empirical evidence for this hypothesis, using the above design principles we built a framework called GraphBLAST based on the GraphBLAS open standard. Testing GraphBLAST on \revision{five} graph algorithms, we were able to obtain $36\times$ geomean $892\times$ peak) over SuiteSparse GraphBLAS (sequential CPU) and $2.14\times$ geomean ($10.97\times$ peak) and $1.01\times$ ($5.24\times$ peak) speed-up over Ligra and Gunrock respectively, which are state-of-the-art graph frameworks on the CPU and GPU\@.

\revision{What follows are the main limitations of GraphBLAST\@: \emph{multi-GPU and multi-node scaling}, \emph{kernel fusion}, \emph{asynchronous execution}, and \emph{matrix-matrix generalization of direction optimization}. As evidenced by other algorithms listed in its code repository, GraphBLAST supports a much larger set of algorithms than the ones described in this paper. In fact, given that our work implements the GraphBLAS API with only minor modifications, it potentially supports the growing list of algorithms that are implemented using GraphBLAS, a majority of which are included in the LAGraph repository\footnote{www.github.com/GraphBLAS/LAGraph}. The only graph algorithms that do not map to the GraphBLAS API efficiently are those that are inherently sequential such as depth-first search and algorithms that use priority queues such as Dijkstra's algorithm.

\paragraph{Multi-GPU and multi-node scalability} By construction, the GraphBLAS open standard establishes its first two goals---\emph{portable performance} and \emph{conciseness}. Portable performance is from making implementers adhere to the same standard interface; conciseness, by basing the interface design around the language of mathematics, which is one of the most concise forms of expression. In this paper, we set out to meet the third goal of \emph{high performance}, which is a prerequisite towards \emph{scalability}. Our work does not address scalability. While we have demonstrated that GraphBLAS is  effective at the scale of a single GPU, we have not addressed the issues associated with scaling across multiple GPUs much less multiple nodes.

The largest challenge for the scale-up direction (more capable nodes) is the limited size of GPU main memory. The NVIDIA K40c (and V100) we used in our experiments, for instance, only support 12~GB (32~GB) of main memory, and all of the datasets we have used in our experiments fit into a 12~GB memory allocation. CPUs support a much larger main memory, allowing CPU-based systems like Ligra to scale their performance to much larger datasets on a single processor. For graph applications that run on GPUs but use CPU memory to store a (larger) graph, the low bandwidth between CPU and GPU and the generally irregular access patterns into the graph data structure would likely make using the CPU for backing storage noncompetitive vs.\ datasets that fit into GPU memory.

The largest challenge for the scale-out direction (more nodes) is effectively and quickly partitioning scale-free graphs. Road-network-like graphs partition well, resulting in a tractable amount of communication between partitions, and thus would generally be amenable to scaling across multiple GPUs. But, because of their high connectivity, scale-free graphs are much more difficult to partition. The resulting high communication volume makes scalability much more difficult.}

\begin{figure}
  \captionsetup[figure]{skip=0pt}
  \centering
  \includegraphics[width=\textwidth,trim={0.1cm 0 0 0.2cm},clip]{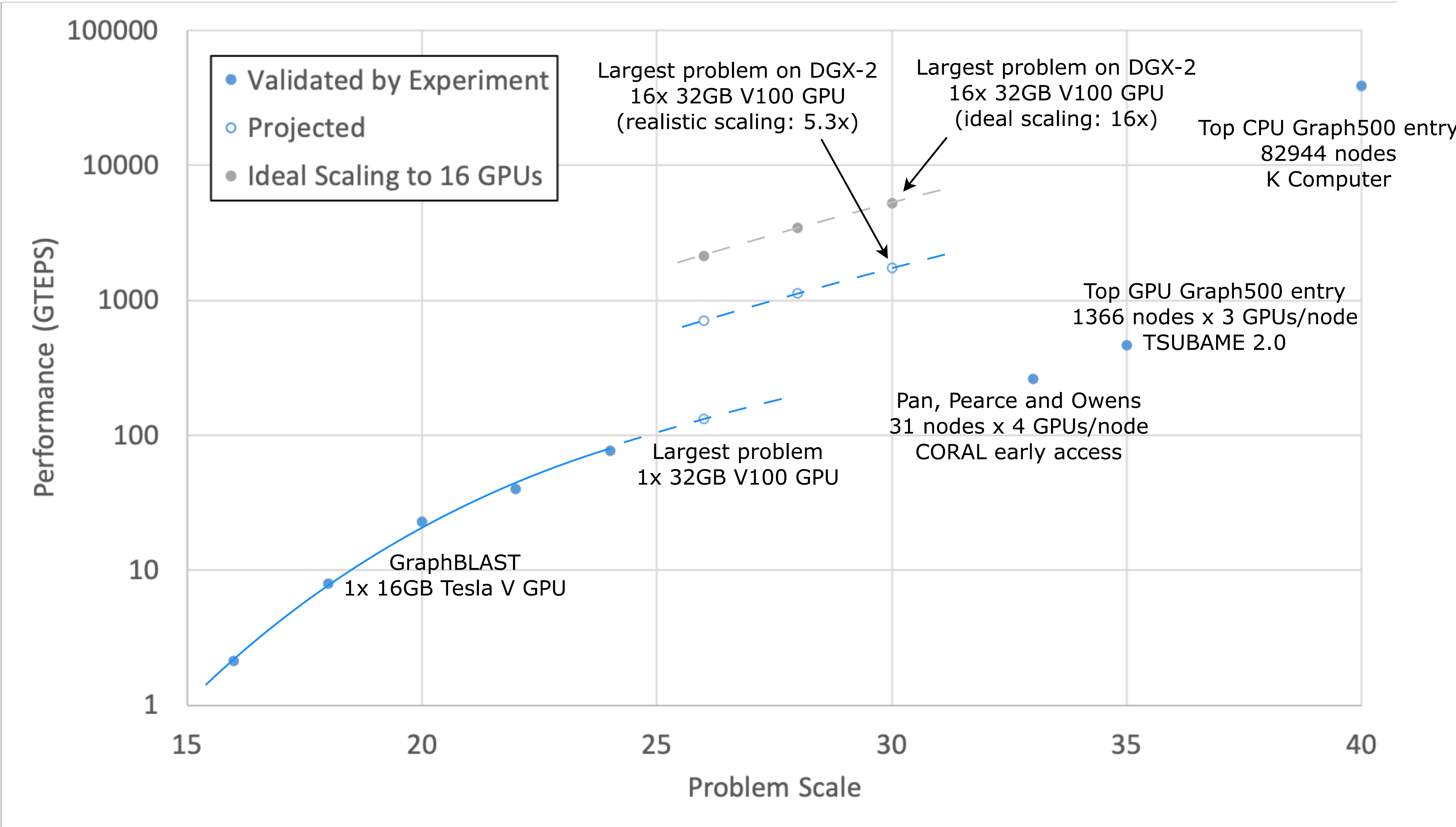}
  \centering
  \caption[Scalability]{Data points from GraphBLAST and points representative of the state-of-the-art in distributed BFS\@. The Dashed line indicates projected performance assuming ideal scaling and realistic scaling accounting for bisection bandwidth for 16 GPUs. In random graph generation for each problem scale $\mathit{SCALE}$, the graph will have $2^{\mathit{SCALE}}$ vertices and $16\times2^{\mathit{SCALE}}$ edges according to Graph500 rules. We acknowledge that the R-MAT generator used by Graph500 has known densification issues~\cite{Seshadhri:2011:AID} that make weak scaling studies problematic. However, Graph500 remains the community standard for benchmarking weakly scaled graph studies. \label{fig:scalability}}
\end{figure}

GPU-based implementations have found difficulty in scaling to as many nodes as CPU-based implementations. This is partly due to GPUs speeding up each node's local computation phase, thus increasing the algorithm's sensitivity to any latency from inter-node communication; and partly because each GPU has very limited main memory compared to CPUs. New GPU-based fat nodes such as the DGX-2 may offer an interesting solution to both problems. By offering $16\times$ GPUs with 32~GB memory each and by being connected using NVSwitch technology that offers a bisection bandwidth of 2.4~TB/s, the DGX-2 may be a contender for multi-GPU top BFS performance. \revision{For example, in Figure~\ref{fig:scalability}, the dashed line and hollow point indicate the potential performance of a DGX-2 system, assuming linear scalability from the $1\times$ GPU GraphBLAST BFS and realistic scalability given the bisection bandwidth computation as follows:

The V100 has a memory bandwidth of 900~GB/s. The BFS on scale-free graphs that are challenging to partition have $O(|E|/P)$ bandwidth cost, where $P$ is the number of devices (such as GPUs), regardless of the algorithm used~\cite{Buluc:2011:PBF}. Given half of the processors will be on each side of the bisection, $O(\vert E \vert/2)$ data will need to be exchanged. If a single V100 was used, per-edge bandwidth would be $900/\vert E \vert$, because we need to touch each edge at some point. With 16 V100s, it is $4800/\vert E \vert$, so a more realistic speedup is $5.3\times$ faster on the DGX-2 compared to a single V100.}

\paragraph{Kernel fusion} In this paper, we hinted at several open problems as potential directions of research. One open problem is the problem of kernel fusion (Section~\ref{sec:kernel-fusion}). In the present situation, a GraphBLAS-based triangle counting algorithm in \emph{blocking mode} (i.e. where operations are required to complete before the next operation begins) can never be as efficient as a hardwired GPU implementation, because it requires a matrix-matrix multiply followed by a reduce. This bulk-synchronous approach forces the computer to write the output of the matrix-matrix multiply to main memory before reading from main memory again in the reduce. A worthwhile area of programming language research would be to use a computation graph to store the operations that must happen, do a pass over the computation graph to identify profitable kernels to fuse, generate the CUDA kernel code at runtime, just-in-time (JIT) compile the code to machine code, and execute the fused kernel. This may be possible in GraphBLAS's non-blocking mode where operations are not required to return immediately after each operation, but only when the user requests an output or an explicit \cmd{wait}.

Such an approach is what is done in machine learning, but with graph algorithms the researcher is faced with additional challenges. One such challenge is that the runtime of graph kernels is dependent on the input data, so in a multiple iteration algorithm such as BFS, SSSP or PR, it may be profitable to fuse two kernels in one iteration and two different kernels in a different iteration. Another challenge is the problem of load balancing. Typically code that is automatically generated is not as efficient as hand-tuned kernels, and may not load-balance well enough to be efficient.

\paragraph{Asynchronous execution model} For road network graphs, asynchronous approaches pioneered by Enterprise~\cite{Liu:2015:EBG} that do not require exiting the kernel until the breakpoint has been met is a way to address the kernel launch problem. This opens the door to two avenues of research: (1)~How can one detect whether one is dealing with a road network that will require thousands of iterations to converge rather than tens of iterations? (2)~How can such an asynchronous execution model be reconciled with GraphBLAS, which is based on the bulk-synchronous parallel model? The first problem requires a system that detects whether the graph is chordal, planar, bipartite, etc. before running the graph traversal algorithm, while the latter problem may also have implications when scaling to distributed implementations.

\paragraph{Matrix-matrix generalization of direction optimization} Currently, direction optimization is only applied for matrix-vector multiplication. However, in the future, the optimization can be extended to matrix-matrix multiplication. The analogue is thinking of the matrix on the right as not a single vector, but as composed of many column vectors, each representing a graph traversal from a different source node. Applications include batched betweenness centrality and all-pairs shortest-path. Instead of switching between SpMV and SpMSpV, we could be switching between SpMM (sparse matrix-dense matrix) and SpGEMM (sparse matrix-sparse matrix). %

\revision{
Finally, several statistical algorithms can estimate the size and structure of SpGEMM output~\cite{cohen1998structure, amossen2010better}, which can be used to choose the right algorithm when implementing direction-optimizing matrix-matrix multiplication.}

\section*{Acknowledgments}
\label{sec:acks}
We thank Yuechao Pan for valuable insight into BFS optimizations. We would like to acknowledge Scott McMillan for important feedback on early drafts of the paper. Thanks to Olivier Beaumont for the historical background on push-pull terminology. We thank Muhammad Osama and Charles Rozhon for collecting valuable experimental results for Gunrock. We would like to acknowledge Collin McCarthy for maintaining the servers the experiments were run on in top shape. We appreciate funding from the National Science Foundation (Award \# CCF-1629657 and OAC-1740333), the DARPA XDATA, HIVE and SDH programs, and for support from an NVIDIA AI Laboratory (UC Davis Center for GPU Graph Analytics). This material is based on research sponsored by Air Force Research Lab (AFRL), the Army Research Lab (ARL), and the Defense Advanced Research Projects Agency (DARPA) under agreements FA8650-18-2-7836, W911QX-12-C-0059 and HR0011-18-3-0007.

This research was supported in part by the Applied Mathematics program of the DOE Office of Advanced Scientific Computing Research under Contract No.\ DE-AC02-05CH11231, and in part the Exascale Computing Project (17-SC-20-SC), a collaborative effort of the U.S.\ Department of Energy Office of Science and the National Nuclear Security Administration.

The U.S.\ Government is authorized to reproduce and distribute reprints for Governmental purposes notwithstanding any copyright notation thereon. The views and conclusions contained herein are those of the authors and should not be interpreted as necessarily representing the official policies or endorsements, either expressed or implied, of AFRL, ARL, DARPA, or the U.S.\ Government.

\bibliographystyle{plain}
\bibliography{graphblast}

\appendix

\section{Lines of Code}
\label{sec:appendix}
For the purposes of counting lines of code, we only compare the code specific to each algorithm, and we assume the graph data structure is already stored in the preferred format by each framework. We compare the lines of code by manually deleting the include and namespace statements, running the open-source code formatting tool \cmd{clang-format} using the default options and then using the open-source tool \cmd{cloc} to count the numbers of lines of code.  This process is used to have as fair a comparison as possible, because some frameworks require many include and namespace statements, while others have only a few. Similarly, some codebases exceed the 80 character line limit, whereas others respect it. By using \cmd{clang-format} and \cmd{cloc}, we can sidestep many of these issues. In this section, we compare our framework against the non-GraphBLAS-based framework with the fewest lines of code, which happens to be Ligra~\cite{Shun:2013:LLG} for all five algorithms.

\subsection{Breadth-first-search}
\lstset{
  basicstyle=\ttfamily,
  columns=fullflexible,
  keepspaces=true,
  numbers=left
}
{
\miniscule
\begin{minipage}[t]{0.45\textwidth}
\begin{lstlisting}
void bfs(Vector<float> *v, const Matrix<float> *A, Index s, Descriptor *desc) {
  Index A_nrows;
  CHECK(A->nrows(&A_nrows));
  CHECK(v->fill(0.f));
  Vector<float> q1(A_nrows);
  Vector<float> q2(A_nrows);
  std::vector<Index> indices(1, s);
  std::vector<float> values(1, 1.f);
  CHECK(q1.build(&indices, &values, 1, GrB_NULL));
  float iter = 1;
  float succ = 0.f;
  do {
    assign<float, float>(v, &q1, GrB_NULL, iter, GrB_ALL, A_nrows, desc);
    CHECK(desc->toggle(GrB_MASK));
    vxm<float, float, float, float>(
        &q2, v, GrB_NULL, LogicalOrAndSemiring<float>(), &q1, A, desc);
    CHECK(desc->toggle(GrB_MASK));
    CHECK(q2.swap(&q1));
    reduce<float, float>(&succ, GrB_NULL, PlusMonoid<float>(), &q1, desc);
    iter++;
  } while (succ > 0);
}
\end{lstlisting}
\end{minipage}
\hfill
\begin{minipage}[t]{0.45\textwidth}
\begin{lstlisting}
template <class ET> inline void writeOr(ET *a, ET b) {
  volatile ET newV, oldV;
  do {
    oldV = *a;
    newV = oldV | b;
  } while ((oldV != newV) && !CAS(a, oldV, newV));
}
struct BFS_F {
  uintE *Parents;
  long *Visited;
  BFS_F(uintE *_Parents, long *_Visited)
      : Parents(_Parents), Visited(_Visited) {}
  inline bool update(uintE s, uintE d) {
    writeOr(&Visited[d / 64], Visited[d / 64] | ((long)1 << (d \% 64)));
    Parents[d] = s;
    return 1;
  }
  inline bool updateAtomic(uintE s, uintE d) {
    writeOr(&Visited[d / 64], Visited[d / 64] | ((long)1 << (d \% 64)));
    return Parents[d] == UINT_E_MAX && CAS(&Parents[d], UINT_E_MAX, s);
  }
  inline bool cond(uintE d) {
    return !(Visited[d / 64] & ((long)1 << (d \% 64)));
  }
};
template <class vertex> void Compute(graph<vertex> &GA, commandLine P) {
  long start = P.getOptionLongValue("-r", 0);
  long n = GA.n;
  uintE *Parents = newA(uintE, n);
  parallel_for(long i = 0; i < n; i++) Parents[i] = UINT_E_MAX;
  Parents[start] = start;
  long numWords = (n + 63) / 64;
  long *Visited = newA(long, numWords);
  { parallel_for(long i = 0; i < numWords; i++) Visited[i] = 0; }
  Visited[start / 64] = (long)1 << (start \% 64);
  vertexSubset Frontier(n, start);
  while (!Frontier.isEmpty()) {
    vertexSubset output = edgeMap(GA, Frontier, BFS_F(Parents, Visited));
    Frontier.del();
    Frontier = output;
  }
  Frontier.del();
  free(Parents);
  free(Visited);
}
\end{lstlisting}
\end{minipage}
\hspace{50pt}
\newline
\centerline{%
  \begin{minipage}{\textwidth}
    \captionof{algorithm}{BFS code for GraphBLAST (left), Ligra~\cite{Shun:2013:LLG} (right)}
\end{minipage}}
}

\subsection{Single-source shortest-path}
\lstset{
  basicstyle=\ttfamily,
  columns=fullflexible,
  keepspaces=true,
  numbers=left
}
{\miniscule
\begin{minipage}[t]{0.45\textwidth}
\begin{lstlisting}
void sssp(Vector<float> *v, const Matrix<float> *A, Index s, Descriptor *desc) {
  Index A_nrows;
  A->nrows(&A_nrows);
  std::vector<graphblas::Index> indices(1, s);
  std::vector<float> values(1, 0.f);
  v->build(&indices, &values, 1, GrB_NULL);
  Vector<float> w(A_nrows);
  Vector<float> zero(A_nrows);
  zero.fill(std::numeric_limits<float>::max());
  Index iter = 1;
  float succ_last = 0.f;
  float succ = 1.f;
  do {
    succ_last = succ;
    vxm<float, float, float, float>(
        &w, GrB_NULL, GrB_NULL, MinimumPlusSemiring<float>(), v, A, desc);
    eWiseAdd<float, float, float, float>(
        v, GrB_NULL, GrB_NULL, MinimumPlusSemiring<float>(), v, &w, desc);
    eWiseMult<float, float, float, float>(
        &w, GrB_NULL, GrB_NULL, PlusLessSemiring<float>(), v, &zero, desc);
    reduce<float, float>(&succ, GrB_NULL, PlusMonoid<float>(), &w, desc);
    iter++;
  } while (succ_last != succ);
}
\end{lstlisting}
\end{minipage}
\hfill
\begin{minipage}[t]{0.45\textwidth}
\begin{lstlisting}
#define WEIGHTED 1
struct BF_F {
  intE *ShortestPathLen;
  int *Visited;
  BF_F(intE *_ShortestPathLen, int *_Visited)
      : ShortestPathLen(_ShortestPathLen), Visited(_Visited) {}
  inline bool update(uintE s, uintE d, intE edgeLen) {
    intE newDist = ShortestPathLen[s] + edgeLen;
    if (ShortestPathLen[d] > newDist) {
      ShortestPathLen[d] = newDist;
      if (Visited[d] == 0) {
        Visited[d] = 1;
        return 1;
      }
    }
    return 0;
  }
  inline bool updateAtomic(uintE s, uintE d, intE edgeLen) {
    intE newDist = ShortestPathLen[s] + edgeLen;
    return (writeMin(&ShortestPathLen[d], newDist) && CAS(&Visited[d], 0, 1));
  }
  inline bool cond(uintE d) { return cond_true(d); }
};
struct BF_Vertex_F {
  int *Visited;
  BF_Vertex_F(int *_Visited) : Visited(_Visited) {}
  inline bool operator()(uintE i) {
    Visited[i] = 0;
    return 1;
  }
};
template <class vertex> void Compute(graph<vertex> &GA, commandLine P) {
  long start = P.getOptionLongValue("-r", 0);
  long n = GA.n;
  intE *ShortestPathLen = newA(intE, n);
  { parallel_for(long i = 0; i < n; i++) ShortestPathLen[i] = INT_MAX / 2; }
  ShortestPathLen[start] = 0;
  int *Visited = newA(int, n);
  { parallel_for(long i = 0; i < n; i++) Visited[i] = 0; }
  vertexSubset Frontier(n, start);
  long round = 0;
  while (!Frontier.isEmpty()) {
    if (round == n) {
      {
        parallel_for(long i = 0; i < n; i++) ShortestPathLen[i] =
            -(INT_E_MAX / 2);
      }
      break;
    }
    vertexSubset output = edgeMap(GA, Frontier, BF_F(ShortestPathLen, Visited),
                                  GA.m / 20, dense_forward);
    vertexMap(output, BF_Vertex_F(Visited));
    Frontier.del();
    Frontier = output;
    round++;
  }
  Frontier.del();
  free(Visited);
  free(ShortestPathLen);
}
\end{lstlisting}
\end{minipage}
\hspace{50pt}
\newline
\centerline{%
  \begin{minipage}{\textwidth}
    \captionof{algorithm}{SSSP code for GraphBLAST (left), Ligra~\cite{Shun:2013:LLG} (right)}
  \end{minipage}}
}

\subsection{PageRank}
\lstset{
  basicstyle=\ttfamily,
  columns=fullflexible,
  keepspaces=true,
  numbers=left
}
{\miniscule
\hfill
\begin{minipage}[t]{0.45\textwidth}
\begin{lstlisting}
void pr(Vector<float> *p, const Matrix<float> *A, float alpha, float eps,
        Descriptor *desc) {
  Index A_nrows;
  CHECK(A->nrows(&A_nrows));
  CHECK(p->clear());
  CHECK(p->fill(1.f / A_nrows));
  Vector<float> p_prev(A_nrows);
  Vector<float> p_swap(A_nrows);
  Vector<float> r(A_nrows);
  r.fill(1.f);
  Vector<float> r_temp(A_nrows);
  float error_last = 0.f;
  float error = 1.f;
  for (int iter = 1; error > eps && iter <= desc->descriptor_.max_niter_;
       ++iter) {
    error_last = error;
    p_prev = *p;
    vxm<float, float, float, float>(&p_swap, GrB_NULL, GrB_NULL,
                                    PlusMultipliesSemiring<float>(), &p_prev, A,
                                    desc);
    eWiseAdd<float, float, float, float>(
        p, GrB_NULL, GrB_NULL, PlusMultipliesSemiring<float>(), &p_swap,
        (1.f - alpha) / A_nrows, desc);
    eWiseMult<float, float, float, float>(
        &r, GrB_NULL, GrB_NULL, PlusMinusSemiring<float>(), p, &p_prev, desc);
    eWiseAdd<float, float, float, float>(&r_temp, GrB_NULL, GrB_NULL,
                                         MultipliesMultipliesSemiring<float>(),
                                         &r, &r, desc);
    reduce<float, float>(&error, GrB_NULL, PlusMonoid<float>(), &r_temp, desc);
    error = sqrt(error);
  }
}
\end{lstlisting}
\end{minipage}
\hfill
\begin{minipage}[t]{0.45\textwidth}
\begin{lstlisting}
template <class vertex> struct PR_F {
  double *p_curr, *p_next;
  vertex *V;
  PR_F(double *_p_curr, double *_p_next, vertex *_V)
      : p_curr(_p_curr), p_next(_p_next), V(_V) {}
  inline bool update(uintE s,
                     uintE d) {
    p_next[d] += p_curr[s] / V[s].getOutDegree();
    return 1;
  }
  inline bool updateAtomic(uintE s, uintE d) {
    writeAdd(&p_next[d], p_curr[s] / V[s].getOutDegree());
    return 1;
  }
  inline bool cond(intT d) { return cond_true(d); }
};
struct PR_Vertex_F {
  double damping;
  double addedConstant;
  double *p_curr;
  double *p_next;
  PR_Vertex_F(double *_p_curr, double *_p_next, double _damping, intE n)
      : p_curr(_p_curr), p_next(_p_next), damping(_damping),
        addedConstant((1 - _damping) * (1 / (double)n)) {}
  inline bool operator()(uintE i) {
    p_next[i] = damping * p_next[i] + addedConstant;
    return 1;
  }
};
struct PR_Vertex_Reset {
  double *p_curr;
  PR_Vertex_Reset(double *_p_curr) : p_curr(_p_curr) {}
  inline bool operator()(uintE i) {
    p_curr[i] = 0.0;
    return 1;
  }
};
template <class vertex> void Compute(graph<vertex> &GA, commandLine P) {
  long maxIters = P.getOptionLongValue("-maxiters", 100);
  const intE n = GA.n;
  const double damping = 0.85, epsilon = 0.0000001;
  double one_over_n = 1 / (double)n;
  double *p_curr = newA(double, n);
  { parallel_for(long i = 0; i < n; i++) p_curr[i] = one_over_n; }
  double *p_next = newA(double, n);
  { parallel_for(long i = 0; i < n; i++) p_next[i] = 0; }
  bool *frontier = newA(bool, n);
  { parallel_for(long i = 0; i < n; i++) frontier[i] = 1; }
  vertexSubset Frontier(n, n, frontier);
  long iter = 0;
  while (iter++ < maxIters) {
    edgeMap(GA, Frontier, PR_F<vertex>(p_curr, p_next, GA.V), 0, no_output);
    vertexMap(Frontier, PR_Vertex_F(p_curr, p_next, damping, n));
    {
      parallel_for(long i = 0; i < n; i++) {
        p_curr[i] = fabs(p_curr[i] - p_next[i]);
      }
    }
    double L1_norm = sequence::plusReduce(p_curr, n);
    if (L1_norm < epsilon)
      break;
    vertexMap(Frontier, PR_Vertex_Reset(p_curr));
    swap(p_curr, p_next);
  }
  Frontier.del();
  free(p_curr);
  free(p_next);
}
\end{lstlisting}
\end{minipage}
\hfill
\newline
\centerline{%
  \begin{minipage}{\textwidth}
    \captionof{algorithm}{PR code for GraphBLAST (left), Ligra~\cite{Shun:2013:LLG} (center right)}
  \end{minipage}}
}

\subsection{Connected components}
\lstset{
  basicstyle=\ttfamily,
  columns=fullflexible,
  keepspaces=true,
  numbers=left
}
{\miniscule
\hfill
\begin{minipage}[t]{0.45\textwidth}
\begin{lstlisting}
void cc(Vector<int> *v, const Matrix<int> *A, int seed, Descriptor *desc) {
  Index A_nrows;
  CHECK(A->nrows(&A_nrows));
  Vector<bool> diff(A_nrows);
  Vector<int> parent(A_nrows);
  Vector<int> parent_temp(A_nrows);
  Vector<int> grandparent(A_nrows);
  Vector<int> grandparent_temp(A_nrows);
  Vector<int> min_neighbor_parent(A_nrows);
  Vector<int> min_neighbor_parent_temp(A_nrows);
  CHECK(parent.fillAscending(A_nrows));
  CHECK(min_neighbor_parent.dup(&parent));
  CHECK(min_neighbor_parent_temp.dup(&parent));
  CHECK(grandparent.dup(&parent));
  CHECK(grandparent_temp.dup(&parent));
  int succ = 0;
  for (int iter = 1; iter <= desc->descriptor_.max_niter_; ++iter) {
    CHECK(parent_temp.dup(&parent));
    mxv<int, int, int, int>(&min_neighbor_parent_temp, GrB_NULL, GrB_NULL,
                            MinimumSelectSecondSemiring<int>(), A, &grandparent,
                            desc);
    eWiseAdd<int, bool, int, int>(&min_neighbor_parent, GrB_NULL, GrB_NULL,
                                  MinimumSelectSecondSemiring<int>(),
                                  &min_neighbor_parent,
                                  &min_neighbor_parent_temp, desc);
    assignScatter<int, bool, int, int>(
        &parent, GrB_NULL, GrB_NULL, &min_neighbor_parent, &parent_temp, desc);
    eWiseAdd<int, bool, int, int>(&parent, GrB_NULL, GrB_NULL,
                                  MinimumPlusSemiring<int>(), &parent,
                                  &min_neighbor_parent, desc);
    eWiseAdd<int, bool, int, int>(&parent, GrB_NULL, GrB_NULL,
                                  MinimumPlusSemiring<int>(), &parent,
                                  &parent_temp, desc);
    extractGather<int, bool, int, int>(&grandparent, GrB_NULL, GrB_NULL,
                                       &parent, &parent, desc);
    eWiseMult<bool, bool, int, int>(&diff, GrB_NULL, GrB_NULL,
                                    MinimumNotEqualToSemiring<int, int, bool>(),
                                    &grandparent_temp, &grandparent, desc);
    reduce<int, bool>(&succ, GrB_NULL, PlusMonoid<int>(), &diff, desc);
    if (succ == 0)
      break;
    CHECK(grandparent_temp.dup(&grandparent));
    CHECK(desc->toggle(GrB_MASK));
    assign<int, bool, int, int>(&grandparent, &diff, GrB_NULL,
                                std::numeric_limits<int>::max(), GrB_ALL,
                                A_nrows, desc);
    CHECK(desc->toggle(GrB_MASK));
  }
  CHECK(v->dup(&parent));
}
\end{lstlisting}
\end{minipage}
\hfill
\begin{minipage}[t]{0.45\textwidth}
\begin{lstlisting}
struct CC_Shortcut {
  uintE *IDs;
  *prevIDs;
  CC_Shortcut(uintE *_IDs, uintE *_prevIDs) : IDs(_IDs), prevIDs(_prevIDs) {}
  inline bool operator()(uintE i) {
    uintE l = IDs[IDs[i]];
    if (IDs[i] != l)
      IDs[i] = l;
    if (prevIDs[i] != IDs[i]) {
      prevIDs[i] = IDs[i];
      return 1;
    } else
      return 0;
  }
};
struct CC_F {
  uintE *IDs;
  *prevIDs;
  CC_F(uintE *_IDs, uintE *_prevIDs) : IDs(_IDs), prevIDs(_prevIDs) {}
  inline bool update(uintE s, uintE d) {
    uintE origID = IDs[d];
    if (IDs[s] < origID) {
      IDs[d] = min(origID, IDs[s]);
    }
    return 1;
  }
  inline bool updateAtomic(uintE s, uintE d) {
    uintE origID = IDs[d];
    writeMin(&IDs[d], IDs[s]);
    return 1;
  }
  inline bool cond(uintE d) { return cond_true(d); }
};
template <class vertex> void Compute(graph<vertex> &GA, commandLine P) {
  long n = GA.n;
  m = GA.m;
  uintE *IDs = newA(uintE, n);
  *prevIDs = newA(uintE, n);
  {
    parallel_for(long i = 0; i < n; i++) {
      prevIDs[i] = i;
      IDs[i] = i;
    }
  }
  bool *all = newA(bool, n);
  { parallel_for(long i = 0; i < n; i++) all[i] = 1; }
  vertexSubset All(n, n, all);
  bool *active = newA(bool, n);
  { parallel_for(long i = 0; i < n; i++) active[i] = 1; }
  vertexSubset Active(n, n, active);
  while (!Active.isEmpty()) {
    edgeMap(GA, Active, CC_F(IDs, prevIDs), m / 20, no_output);
    vertexSubset output = vertexFilter(All, CC_Shortcut(IDs, prevIDs));
    Active.del();
    Active = output;
  }
  Active.del();
  All.del();
  free(IDs);
  free(prevIDs);
}
\end{lstlisting}
\end{minipage}
\hfill
\newline
\centerline{%
  \begin{minipage}{\textwidth}
    \captionof{algorithm}{CC code for GraphBLAST (left), Ligra~\cite{Shun:2013:LLG} (right)}
  \end{minipage}}
}

\subsection{Triangle counting}
\lstset{
  basicstyle=\ttfamily,
  columns=fullflexible,
  keepspaces=true,
  numbers=left
}
{\miniscule
\hfill
\begin{minipage}[t]{0.45\textwidth}
\begin{lstlisting}
void tc(int *ntris, const Matrix<int> *A, Matrix<int> *B, Descriptor *desc) {
  Index A_nrows;
  CHECK(A->nrows(&A_nrows));
  CHECK(desc->toggle(graphblas::GrB_INP1));
  mxm<int, int, int, int>(B, A, GrB_NULL, PlusMultipliesSemiring<int>(), A, A,
                          desc);
  reduce<int, int>(ntris, GrB_NULL, PlusMonoid<int>(), B, desc);
}
\end{lstlisting}
\end{minipage}
\hfill
\begin{minipage}[t]{0.45\textwidth}
\begin{lstlisting}
template <class vertex>
long countCommon(vertex &A, vertex &B, uintE a, uintE b) {
  uintT i = 0, j = 0, nA = A.getOutDegree(), nB = B.getOutDegree();
  uintE *nghA = (uintE *)A.getOutNeighbors(),
        *nghB = (uintE *)B.getOutNeighbors();
  long ans = 0;
  while (i < nA && j < nB && nghA[i] < a &&
         nghB[j] < b) {
    if (nghA[i] == nghB[j])
      i++, j++, ans++;
    else if (nghA[i] < nghB[j])
      i++;
    else
      j++;
  }
  return ans;
}
template <class vertex> struct countF {
  vertex *V;
  long *counts;
  countF(vertex *_V, long *_counts) : V(_V), counts(_counts) {}
  inline bool update(uintE s, uintE d) {
    if (s > d)
      writeAdd(&counts[s], countCommon<vertex>(V[s], V[d], s, d));
    return 1;
  }
  inline bool updateAtomic(uintE s, uintE d) {
    if (s > d)
      writeAdd(&counts[s], countCommon<vertex>(V[s], V[d], s, d));
    return 1;
  }
  inline bool cond(uintE d) { return cond_true(d); }
};
struct intLT {
  bool operator()(uintT a, uintT b) { return a < b; };
};
template <class vertex>
struct initF {
  vertex *V;
  long *counts;
  initF(vertex *_V, long *_counts) : V(_V), counts(_counts) {}
  inline bool operator()(uintE i) {
    counts[i] = 0;
    quickSort(V[i].getOutNeighbors(), V[i].getOutDegree(), intLT());
    return 1;
  }
};
template <class vertex> void Compute(graph<vertex> &GA, commandLine P) {
  uintT n = GA.n;
  long *counts = newA(long, n);
  bool *frontier = newA(bool, n);
  { parallel_for(long i = 0; i < n; i++) frontier[i] = 1; }
  vertexSubset Frontier(n, n, frontier);
  vertexMap(Frontier, initF<vertex>(GA.V, counts));
  edgeMap(GA, Frontier, countF<vertex>(GA.V, counts), -1, no_output);
  long count = sequence::plusReduce(counts, n);
  cout << "triangle count = " << count << endl;
  Frontier.del();
  free(counts);
}
\end{lstlisting}
\end{minipage}
\hfill
\newline
\centerline{%
  \begin{minipage}{\textwidth}
    \captionof{algorithm}{TC code for GraphBLAST (left), Ligra~\cite{Shun:2013:LLG} (right)}
  \end{minipage}}
}

\end{document}